\documentclass[sigplan,screen]{acmart}

\makeatletter
\def\@ACM@checkaffil{% Only warnings
    \if@ACM@instpresent\else
    \ClassWarningNoLine{\@classname}{No institution present for an affiliation}%
    \fi
    \if@ACM@citypresent\else
    \ClassWarningNoLine{\@classname}{No city present for an affiliation}%
    \fi
    \if@ACM@countrypresent\else
        \ClassWarningNoLine{\@classname}{No country present for an affiliation}%
    \fi
}
\makeatother

\makeatletter
\def\@ACM@copyright@check@cc{}
\makeatother

\copyrightyear{2025}
\acmYear{2025}
\setcopyright{cc}
\setcctype{by}
\acmConference[ASPLOS '25]{Proceedings of the 30th ACM International Conference on Architectural Support for Programming Languages and Operating Systems, Volume 2}{March 30-April 3, 2025}{Rotterdam, Netherlands}
\acmBooktitle{Proceedings of the 30th ACM International Conference on Architectural Support for Programming Languages and Operating Systems, Volume 2 (ASPLOS '25), March 30-April 3, 2025, Rotterdam, Netherlands}\acmDOI{10.1145/3676641.3716277}
\acmISBN{979-8-4007-1079-7/2025/03}

\date{}

\newcommand{\myparagraph}[1]{\vspace{1.5pt} \noindent{\bf {#1}.}}

\newcommand{\out}[1] {}

\newcounter{codeLineCntr}

\reversemarginpar
\newif\ifnotes
\notestrue

\newcommand{\punt}[1]{}

\newcommand{\proc}[1]{\ifmmode\mbox{\textsc{#1}}\else\textsc{#1}\fi}

  \newcommand{\func}[1]{\ifmmode\mathrm{#1}\else\textrm{#1}fi} %
\newcounter{remark}[section]

\usepackage[inline]{enumitem}
\setlist{noitemsep,topsep=0pt,parsep=0pt,partopsep=0pt}

\usepackage[subtle]{savetrees}

\usepackage{setspace}
\usepackage[margin=5pt,font={stretch=0.9}]{caption}

\usepackage{grumble}
\usepackage{subfigure}
\usepackage{caption}
\usepackage{booktabs}
\usepackage{multirow}
\usepackage{listings}
\usepackage{xcolor}
\usepackage{float}
\usepackage{xspace}
\usepackage[binary-units=true]{siunitx}
\usepackage{tikz}

\usepackage{xcolor,colortbl}
\usepackage{algorithmic}
\usepackage[ruled,vlined,linesnumbered]{algorithm2e}
\usepackage{grumble}

\newcommand\blfootnote[1]{%
  \begingroup
  \renewcommand\thefootnote{}\footnote{#1}%
  \addtocounter{footnote}{-1}%
  \endgroup
}

\newcommand{\projecttitle}{\textsc{tnic}\xspace}

\newcommand{\trustedfpga}{\textsc{t-fpga}\xspace}

\definecolor{codegreen}{rgb}{0,0.6,0}
\definecolor{codegray}{rgb}{0.5,0.5,0.5}
\definecolor{codepurple}{rgb}{0.58,0,0.82}
\definecolor{backcolour}{rgb}{0.95,0.95,0.92}

\definecolor{lightGrey}{rgb}{0.9, 0.9, 0.9}
\definecolor{beaublue}{rgb}{0.74, 0.83, 0.9}
\definecolor{lightred}{RGB}{229, 220, 220}

\definecolor{burlywood}{rgb}{0.87, 0.72, 0.53}

\lstdefinestyle{customc}{
    backgroundcolor=\color{backcolour},   
    commentstyle=\color{codegreen},
    keywordstyle=\color{magenta},
    numberstyle=\tiny\color{codegray},
    stringstyle=\color{codepurple},
    basicstyle=\ttfamily\footnotesize,
    breaklines,
    tabsize=2,
    numbers=left,
    columns=fullflexible,
    keepspaces=true,
    frame=lines,
    numbersep=4pt,
    escapechar=@,
    mathescape=true,
    captionpos=b,
    language=c++,
    keywords = {auto, new, void, Raft_ctx, Msg, for}
}

\begin{document}
\author{Dimitra Giantsidi}
\affiliation{%
  \institution{The University of Edinburgh}
  % \country{Germany}
}
% \email{d.giantsidi@sms.ed.ac.uk}
\author{Julian Pritzi}
\affiliation{%
  \institution{Technical University of Munich}
  % \country{Germany}
}
% \email{julian.pritzi@tum.de}
\author{Felix Gust}
\affiliation{%
  \institution{Technical University of Munich}
  % \country{Germany}
}
% \email{gustf@in.tum.de}
\author{Antonios Katsarakis$^*$}
\affiliation{%
  \institution{Huawei Research}
  % \country{Germany}
}
% \email{antonios.katsarakis@huawei.com}
\author{Atsushi Koshiba}
\affiliation{%
  \institution{Technical University of Munich}
  % \country{Germany}
}
% \email{atsushi.koshiba@tum.de}
\author{Pramod Bhatotia}
\affiliation{%
  \institution{Technical University of Munich}
  % \country{Germany}
}
% \email{pramod.bhatotia@tum.de}
\renewcommand{\shortauthors}{Giantsidi, et al.}

%make title bold and 14 pt font (Latex default is non-bold, 16 pt)
% \title{{TNIC: A Trusted NIC Architecture}\\ \vspace{-2mm}
% {\large A hardware-network substrate for building high-performance trustworthy distributed systems}}
\title{{TNIC: A Trusted NIC Architecture}}
\subtitle{{\large \bf A hardware-network substrate for building high-performance trustworthy distributed systems}}
\renewcommand{\shorttitle}{TNIC: A Trusted NIC Architecture}

% \subsection*{Abstract}
\begin{abstract}
We introduce \projecttitle{}, a trusted NIC architecture for building trustworthy distributed systems deployed in heterogeneous, untrusted (Byzantine) cloud environments. \projecttitle{} builds a minimal, formally verified, silicon root-of-trust at the network interface level. We strive for three primary design goals: (1) a host CPU-agnostic unified security architecture by providing trustworthy network-level isolation; (2) a minimalistic and verifiable TCB based on a silicon root-of-trust by providing two core properties of transferable authentication and non-equivocation; and (3) a hardware-accelerated trustworthy network stack leveraging SmartNICs. Based on the \projecttitle{} architecture and associated network stack, we present a generic set of programming APIs and a recipe for building high-performance, trustworthy, distributed systems for Byzantine settings. We formally verify the safety and security properties of our \projecttitle{} while demonstrating its use by building four trustworthy distributed systems. Our evaluation of \projecttitle{} shows up to $6\times$ performance improvement compared to CPU-centric TEE systems.
\end{abstract}

% \ccsdesc[500]{Computer systems organization~Cloud computing}


\maketitle

% Use the following at camera-ready time to suppress page numbers.
% Comment it out when you first submit the paper for review.
%\thispagestyle{empty}

\section{Introduction}
\blfootnote{$^*$This work started when the author was at the University of Edinburgh.}
Distributed systems are integral to the third-party cloud infrastructure~\cite{amazon_ec2, microsoft_azure, rackspace, google_engine}. While these systems manifest in diverse forms (e.g., storage systems~\cite{dynamo, azure_storage, tao, spanner, 51, zippy, AmazonS3}, management services~\cite{Hunt:2010, Burns2016}, computing frameworks~\cite{aws_lambda, azure_functions, google_cloud_functions}) they all must be fast and remain correct upon failures. %when failures occur. 

Unfortunately, the widespread adoption of the cloud has drastically increased the surface area of attacks and faults~\cite{Gunawi_bugs-in-the-cloud, Shinde2016, high_resolution_side_channels} that are beyond the traditional fail-stop (or crash fault) model~\cite{delporte}. The modern (untrusted) third-party cloud infrastructure severely suffers from arbitrary  ({\em Byzantine}) \linebreak faults~\cite{Lamport:1982} that can range from malicious (network) attacks to configuration errors and bugs and are capable of irreversibly disrupting the correct execution of the system~\cite{Gunawi_bugs-in-the-cloud, Shinde2016, high_resolution_side_channels, Castro:2002}.
% ford2010availability, Mazieres2002b, Garay2000}.

A promising solution to build trustworthy distributed systems that can sustain Byzantine failures is based on the {\em silicon root of trust}---specifically, the Trusted Execution Environments (TEEs)~\cite{cryptoeprint:2016:086, arm-realm, amd-sev, riscv-multizone, intelTDX}. While the TEEs offer a (single-node) isolated Trusted Computing Base (TCB),  we have identified three core challenges ($\S$~\ref{subsec:challenges}) that complicate their adoption for building trustworthy distributed systems spanning multiple nodes in Byzantine cloud environments.

{\bf \em First, TEEs in heterogeneous cloud environments introduce programmability and security challenges}. A cloud environment offers diverse heterogeneous host-side CPUs with different TEEs (e.g., Intel SGX/TDX, AMD SEV-SNP, AWS Nitro Enclaves, Arm TrustZone/CCA, RISC-V Keystone). These heterogeneous host-side TEEs require different programming models and offer varying security properties. Therefore, they cannot (easily) provide a generic substrate for building trustworthy distributed systems. Our work overcomes this challenge by designing a {\em host CPU-agnostic} {\em silicon root of trust} at the network interface (NIC) level ($\S$~\ref{sec:t-nic-hardware}). We provide a generic programming API ($\S$~\ref{sec:t-nic-software}) and a {\em recipe} ($\S$~\ref{subsec:transformation}) for building high-performance, trustworthy distributed systems ($\S$~\ref{sec:use_cases}).
%, exposing a {\em unified trusted} network-level isolation 

{\bf \em Secondly, TEEs with a large TCB are plagued with security vulnerabilities, rendering them non-verifiable}. With hundreds of security bugs already uncovered~\cite{10.1145/3456631}, TEEs' large TCBs further increase their security vulnerabilities~\cite{10.1145/3379469, 10.5555/1756748.1756832}, impeding a formal verification of their security. We overcome this with a {\em minimalistic verifiable TCB} ($\S$~\ref{subsec:nic_attest_kernel}). Our TCB resides at the NIC hardware and is equipped with {\em the lower bound of security primitives};  we provide only two key security properties of non-equivocation and transferable authentication for building trustworthy distributed systems ($\S$~\ref{subsec:trustworthy_ds}). Since we strive for a minimal trusted interface, we can (and we did) formally verify the security properties of our TCB ($\S$~\ref{subsec::formal_verification_remote_attestation}). 

{\bf \em Thirdly, TEEs report significant performance bottlenecks.} TEEs syscalls execution for (network) I/O is extremely costly~\cite{hotcalls}, whereas even state-of-the-art network stacks showed a lower bound of 4$\times$ slowdown~\cite{avocado}. We attack this challenge based on two aspects. First, we build a scalable transformation with our minimal TCB's security properties ($\S$~\ref{subsec:transformation}) to transform Byzantine faults (3$f$+1) to much cheaper crash faults (2$f$+1) for tolerating $f$  (distributed) Byzantine nodes.  Secondly, we design hardware-accelerated offload of the security computation at the NIC level by extending the scope of SmartNICs with {\em the lower bound of security primitives} ($\S$~\ref{sec:t-nic-hardware}) while offering kernel-bypass networking ($\S$~\ref{sec:t-nic-network}).

To overcome these challenges, we present  \projecttitle{}, a trusted NIC architecture for building trustworthy distributed systems deployed in Byzantine cloud environments. \projecttitle{} realizes an abstraction of trustworthy network-level isolation by building a hardware-accelerated silicon root of trust at the NIC level. Overall, \projecttitle{} follows a layered design:
\begin{itemize}[leftmargin=*]
    \item {\bf Trusted NIC hardware architecture ($\S$~\ref{sec:t-nic-hardware}):}  We materialize a \underline{minimalistic}, \underline{verifiable}, and \underline{host-CPU-agnostic} TCB at the network interface level as the key component to design trusted distributed systems for Byzantine settings. Our TCB guarantees the security properties of non-equivocation and transferable authentication that suffice to implement an efficient transformation of systems for Byzantine settings. We build \projecttitle{} on top of FPGA-based SmartNICs~\cite{u280_smartnics}. We formally verify the safety and security guarantees of \projecttitle{} protocols using Tamarin Prover~\cite{tamarin-prover}. 

    \item \rev{(a)}{{\bf Network stack ($\S$~\ref{sec:t-nic-network}) and library ($\S$~\ref{sec:t-nic-software}):} Based on the \projecttitle{} architecture, we design a \underline{HW-accelerated} network stack to access the hardware bypassing kernel for performance. On top of \projecttitle{}'s network stack, we present a networking library that exposes a \underline{simplified} programming model. We show {\em how to use} \projecttitle{} APIs to construct a \underline{generic transformation} of a distributed system operating under the CFT model to target Byzantine settings.}
    % \item {\bf Trusted network stack ($\S$~\ref{sec:t-nic-network}) and library ($\S$~\ref{sec:t-nic-software}):} Based on the \projecttitle{} architecture, we design a trusted \underline{HW-accelerated} network stack to access the hardware bypassing kernel for performance. On top of \projecttitle{}'s network stack, we present a trusted networking library that exposes a \underline{simplified} programming model. We show {\em how to use} \projecttitle{} APIs to construct a \underline{generic transformation} of a distributed system operating under the CFT model to target Byzantine settings.
    
    \item {\bf Trusted distributed systems using \projecttitle{} ($\S$~\ref{sec:use_cases}):} We build with \projecttitle{} the following (distributed) systems for Byzantine environments: Attested Append-only Memory (A2M)~\cite{A2M}, Byzantine Fault Tolerance (BFT)~\cite{pbft}, Chain Replication~\cite{chain-replication}, and Accountability with PeerReview~\cite{peer-review}---showing the \underline{generality of our approach}.
\end{itemize}

% We build \projecttitle{} on top of Alveo U280 FPGA-based SmartNICs~\cite{u280_smartnics}.  extending the Coyote system~\cite{coyote}.
% Our core component is the \projecttitle{}'s {\em attestation kernel} ($\S$~\ref{subsec:nic_attest_kernel}), the minimal required hardware-assisted TCB that guarantees the lower bound of security properties for BFT across the network ($\S$~\ref{sec:requirements-ds}). 
% We formally verify the safety and correctness properties of all \projecttitle{}'s operations (i.e., from remote attestation to networking) using Tamarin prover~\cite{tamarin-prover} ($\S$~\ref{subsec::formal_verification_remote_attestation}). Our 
% \projecttitle{}'s attestation kernel resides on the network data path to optimize for latency while we further design a unified trusted network stack to implement user-space networking following the RDMA programming paradigm ($\S$~\ref{subsec:roce_protocol_kernel} and $\S$~\ref{sec:t-nic-network}). Lastly, we leverage our \projecttitle{} trusted network library($\S$~\ref{sec:t-nic-software}) to show a generic {\em recipe} ($\S$~\ref{subsec:transformation})to transform a distributed system operating under the fail-stop model for Byzantine settings ($\S$~\ref{subsec:transformation}). 

We evaluate \projecttitle{} with a  state-of-the-art software-based network stack, eRPC~\cite{erpc}, on top of RDMA~\cite{rdma}/DPDK~\cite{dpdk} with two different TEEs (Intel SGX~\cite{intel-sgx} and AMD-sev~\cite{amd-sev}). Our evaluation shows that \projecttitle{} offers 3$\times$---5$\times$ lower latency than the software-based approach with the CPU-based TEEs. For trusted distributed systems, \projecttitle{} improves throughput by up to $6\times$ compared to their TEE-based implementations.
 % We evaluate the \projecttitle{}-based versions of the four implemented systems against their TEEs-based versions. 

\section{Motivation and Background}
\label{sec:requirements-ds}

We first examine the design requirements for high-performance, trustworthy distributed systems for cloud environments. % hosted in the untrusted heterogeneous cloud infrastructure.

\subsection{Trustworthy Distributed Systems}\label{subsec:trustworthy_ds}
\myparagraph{Byzantine fault model} 
In the untrusted cloud infrastructure, arbitrary (Byzantine) faults are a frequent occurrence in the wild~\cite{Gunawi_bugs-in-the-cloud, Shinde2016, 10.1145/1189256.1189259, 10.5555/1267308.1267318}. To this end, system designers introduced Byzantine Fault Tolerant (BFT) systems that remain correct even under the presence of (a bounded number of) Byzantine failures~\cite{Lamport:1982}. \rev{(b)}{Traditional BFT protocols need \emph{at least} $3f+1$ nodes in order to provide consistent replication while tolerating up to $f$ Byzantine failures.} While BFT accurately captures the realistic security needs in the cloud~\cite{bft_made_practical}, it is rarely adopted in practice~\cite{bftForSkeptics} due to its complexity and limited performance~\cite{268273, 10.1145/2658994}. 

\myparagraph{Crash fault model} 
The vast majority of cloud applications operate under the fail-stop (crash fault) model~\cite{spanner, 27897, cockroachdb_raft, zippydb, foundationdb}, optimistically {\em assuming} that the entire cloud infrastructure is trusted and only fails by crashing~\cite{delporte}. \rev{(b)}{Compared to BFT replication, Crash Fault Tolerant (CFT) protocols~\cite{10.1145/279227.279229, raft, primary-backup, Hunt:2010}, require $2f+1$ replicas to tolerate $f$ (yet non-Byzantine) failures.} While CFT systems can offer performance and scalability~\cite{f04eb9b864204bab958e72055062748c}, they are fundamentally incapable of ensuring safety in the presence of non-benign faults, hence, are ill-suited for the modern cloud. 

% In the untrusted cloud infrastructure, arbitrary (Byzantine) faults are a frequent occurrence in the wild~\cite{Gunawi_bugs-in-the-cloud, Shinde2016, 10.1145/1189256.1189259, 10.5555/1267308.1267318}. To this end, system designers introduced Byzantine Fault Tolerant (BFT) systems that remain correct even under the presence of (a bounded number of) Byzantine failures~\cite{Lamport:1982}. While BFT accurately captures the realistic security needs in the cloud~\cite{bft_made_practical}, it is rarely adopted in practice~\cite{bftForSkeptics} due to its complexity and limited performance~\cite{268273, 10.1145/2658994}. In contrast, the vast majority of cloud applications operate under the fail-stop (crash fault) model, optimistically {\em assuming} that the entire cloud infrastructure is trusted and only fails by crashing~\cite{delporte}. While Crash Fault Tolerant (CFT) systems usually offer performance and scalability~\cite{f04eb9b864204bab958e72055062748c}, they are ill-suited for the modern cloud as they are fundamentally incapable of ensuring safety in the presence of non-benign faults. 
 
% \noindent{\bf{Security properties for BFT.}} 

\myparagraph{Security properties for BFT} 
\rev{(b), A2, A4}{
We seek to build BFT systems while reducing their programmability and performance overheads. Our approach, inspired by the theoretical findings of Clement et al.~\cite{clement2012}, {\em transforms} CFT systems into BFT systems by providing the {\em lower bound} of security properties, i.e., {\em transferable authentication} and {\em non-equivocation}.
}

% \rev{A2}{We next explain the properties:}
% , which are minimal security properties required to build trustworthy systems under the BFT model. 

% \myparagraph{Transferable authentication}
\revcont{
We next explain the two security properties. First, {\em transferable authentication} allows a node to verify the original sender of a received message, even if it is forwarded by other than the original sender. Assuming that the sender $p_i$ sends an authenticated message $m$ to a recipient $p_j$, the authenticated message $m$ is accompanied by an authentication token $\sigma (p_i)$ that allows  $p_j$ to verify that $p_i$ generated the message, e.g., {verify($m, \sigma (p_i)$)}. Authentication tokens are unforgeable:
\begin{itemize}[leftmargin=*]
  \item if $p_i$ is correct, then {verify($m, \sigma (p_i)$)} is true if and only if $p_i$ generated $m$.
  \item if $p_i$ is faulty, {verify($m, \sigma (p_i)$)} $\wedge$ {verify($m', \sigma (p_i)$)} $\Rightarrow$ $m = m'$. As such, a compromised $p_i$ cannot produce two valid different messages that can be verified with the same token $\sigma (p_i)$.
\end{itemize}
As an authentication token is transferable, it allows another recipient $p_k$ to evaluate {verify($m, \sigma (p_i)$)} in the same way even when $m$ and $\sigma (p_i)$ are forwarded from $p_j$.
}

\revcont{
Second, {\em non-equivocation} guarantees that a node cannot make conflicting statements to different nodes. Equivocation also manifests as network adversaries or replay attacks that send invalid messages or re-send valid but stale messages.
}

\revcont{The seminal paper~\cite{clement2012} proves that, given these two properties, a transformation from any CFT protocol to a BFT protocol is {\emph {always}} possible without increasing the number of participating nodes; e.g., a reliable broadcast can be implemented to tolerate up to $f$ Byzantine failures in an asynchronous system with $2f+1$ replicas, rather than the conventional $3f+1$.}
% An authentication token provides transferable authentication if the correct processes $p_j$ and $p_k$ always evaluate \texttt{verify($m, \sigma (p_i)$)} in the same way even when $p_k$ receives message $m$ and authentication token $\sigma (p_i)$ from $p_j$.

% To sum up, providing these two properties at the network level, we can {\em always} and {\em correctly transform} (any) CFT distributed system to operate in the BFT model~\cite{clement2012, byzantine-pratical}. 

\if 0
\noindent{\bf{Security properties for BFT.}} We seek to offer BFT while reducing its programmability and performance overheads. As such, we materialize the {\em minimum} security properties required to build trustworthy systems under the BFT model~\cite{clement2012}: 
\begin{itemize}[leftmargin=*]
    \item {\bf Transferable authentication} refers to a machine's capability to verify the original sender of a received message, even if it is forwarded by other than the original sender. %Authentication is transferable if the original sender can also be verified for forwarded messages. 
    \item {\bf Non-equivocation} guarantees that a node cannot make conflicting statements to different nodes. Equivocation also manifests as network adversaries or replay attacks that send invalid messages or re-send valid but stale messages.
\end{itemize}
\fi

\subsection{High-Performance Distributed Systems} \label{subsec::tees}
%The security properties discussed above suffice for building distributed systems that operate {\em correctly} under the BFT model. 
The aforementioned two security properties are sufficient to {\em correctly transform} (any) CFT distributed system to operate in the BFT model~\cite{clement2012, byzantine-pratical}. 
However, a fundamental design trade-off exists between efficiency and robustness for practical deployments in the cloud. Our work aims to resolve this tension.

\myparagraph{Trusted hardware for BFT} System designers established trusted hardware, TEEs, as the most effective way to eliminate a system's Byzantine counterparts~\cite{avocado, minBFT, hybster, 10.1145/3492321.3519568}. While TEEs can be used to offer BFT, prior research illustrated significant performance and architectural limitations in the context of networked systems~\cite{avocado, 10.1145/3492321.3519568, hybster, minBFT}. Based on performance and security studies~\cite{9460547, 9935045}, TEEs' overheads in the heterogeneous cloud, in addition to their heterogeneity in programmability and security guarantees, are incapable of offering high-performant trusted networking under the BFT model.

\myparagraph{SmartNICs for high-performance and BFT} We leverage the state-of-the-art hardware-level networking accelerators, i.e., SmartNICs~\cite{liquidIO_smartnics, u280_smartnics, bluefield_smartnics, broadcom_smartnics, netronome_smartnics, alibaba_smartnics, nitro_smartnics, msr_smartnics}, to address the trade-off between performance and security, overcoming the limitations of TEEs. Our design choice of leveraging SmartNICs is not hypothetical; SmartNIC devices have already been launched by major cloud providers~\cite{alibaba_smartnics, nitro_smartnics, msr_smartnics}, presenting great opportunities for performance thanks to their integrated fully programmable hardware (e.g., ARM cores~\cite{bluefield_smartnics, alibaba_smartnics, broadcom_smartnics, liquidIO_smartnics}, FPGAs~\cite{u280_smartnics, alveo_sn1000, msr_smartnics}). Precisely, we rely on two promising directions: {\em(1)} security and network processing offloading at the NIC-level hardware and {\em(2)} an efficient transformation for BFT. 

\section{Overview}

\subsection{System Overview}
%We advocate that distributed systems must be fast and trustworthy  in the Byzantine heterogeneous cloud infrastructure. %We further show that while TEEs could help in this direction, they cannot meet the requirements of such systems in terms of variations in performance, programmability overheads and complicated security analysis.

\begin{figure}[t!]
    \centering
    \includegraphics[width=0.7\linewidth]{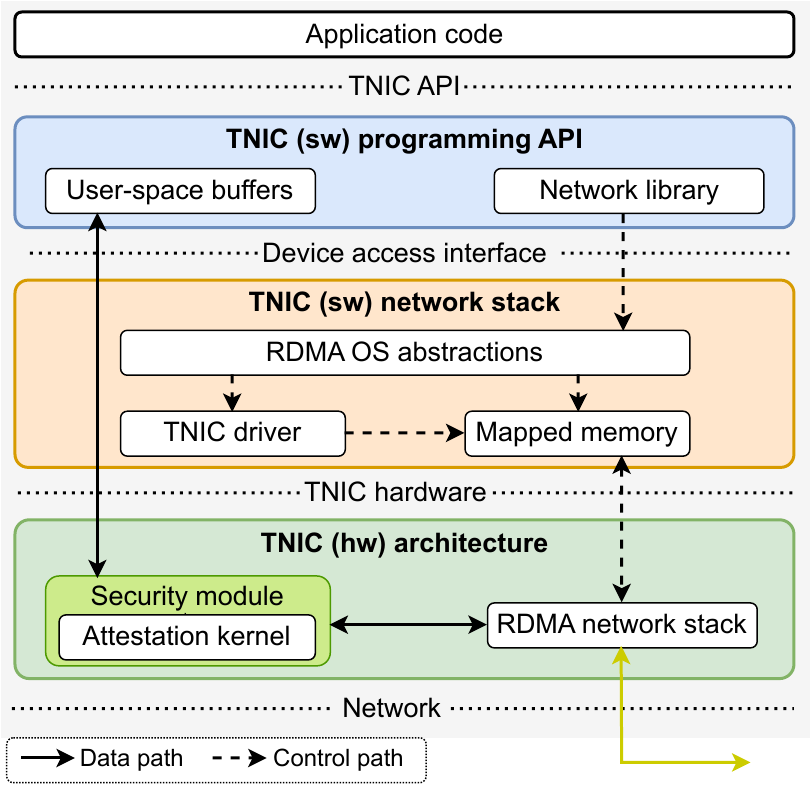}
    \caption{\projecttitle{} system overview.}
    \label{fig:overview}
\end{figure}

We propose \projecttitle{}, a trusted NIC architecture for high-performance, trustworthy distributed systems, formally guaranteeing their secure and correct execution in the heterogeneous Byzantine cloud infrastructure. 
% To this end, we propose \projecttitle{}, a trusted NIC architecture that offers a network abstraction for high-performance, trustworthy distributed systems under BFT that meets the performance requirements of modern systems while it formally guarantees their secure and correct execution in the heterogeneous (Byzantine) cloud infrastructure. 
\projecttitle{} is comprised of three layers (shown in Figure~\ref{fig:overview}): {\bf (1)  the \projecttitle{} hardware architecture} (green box) that implements trusted network operations on top of SmartNIC devices ($\S$~\ref{sec:t-nic-hardware}), \rev{(a)}{{\bf (2) the \projecttitle{} network stack}} (yellow box) that intermediates between the application layer and the \projecttitle{} hardware ($\S$~\ref{sec:t-nic-network}), and \rev{(a)}{{\bf (3)  the \projecttitle{} network library}} (blue box) that exposes \projecttitle{}'s programming APIs ($\S$~\ref{sec:t-nic-software}). 

Our \projecttitle{} hardware architecture implements the networking IB/RDMA protocol~\cite{rdma_specification} on FPGA-based SmartNICs~\cite{u280_smartnics}. 
It extends the conventional protocol implementation with a minimal hardware module, the attestation kernel, that materializes the security properties of the non-equivocation and transferable authentication. The \projecttitle{} network stack configures the \projecttitle{} device on the control path while it offers the data path as kernel-bypass device access for low-latency operations. Lastly, the  \projecttitle{} network library exposes \rev{(a)}{programming APIs} built on top of (reliable) one-sided RDMA primitives. 
% Our \projecttitle{} hardware architecture implements and extends the networking IB/RDMA protocol~\cite{rdma_specification} on top of the FPGA-based SmartNICs, Alveo U280~\cite{u280_smartnics}. Critically, it extends the conventional protocol implementation with a minimal hardware security module, the attestation kernel, that materializes the security properties of the non-equivocation and the transferable authentication. The \projecttitle{} trusted network stack runs in user space. It configures the \projecttitle{} device (MAC address, IP, etc.) on the control path, while the data path is offered as kernel-bypass device access for low-latency operations. Lastly, The  \projecttitle{} trusted network library exposes a {\em trusted} API that is built on top of one-sided RDMA (reliable) operations. 

%\section{Design}
%\dimitra{
%\begin{itemize}
%    \item Intro
%    \item Background + Motivation
%    \item Overview (including sys/data model etc.)
%    \item Design + Implementation (TNIC: hw architecture, TNIC libraries: sw abstraction, Applications: use-cases)
%\end{itemize}
%}

%\subsection{System Model}

%\myparagraph{Model sketch}
%We model the distributed system as a set of {\tt N} nodes each of which is attached to a single \projecttitle{} instance that is loaded into an FPGA-based SmartNIC as Alveo U280~\cite{u280_smartnics}. The nodes communicate by exchanging messages through bi-directional network links that connect their FPGAs. The system is managed and owned by the third-party cloud infrastructure which is untrusted.

\subsection{Threat Model} 

%\antonis{It is a bit weird we have Fault model in section 2 and Thread model in section 3}

We inherit the fault and threat model from the classical BFT~\cite{Castro:2002} and trusted computing domains~\cite{intel-sgx}. The cloud infrastructure (machines, network, etc.) can exhibit Byzantine behavior and also being subject to attackers that can control over the host CPU (e.g., the OS, VMM, etc.) and the SmartNICs (post-manufacturing). The adversary can attempt to re-program the SmartNIC, but they cannot compromise the cryptographic primitives~\cite{levin2009trinc, minBFT, Castro:2002}. The physical package, supply chain, and manufacturer of the SmartNICs are trusted~\cite{10.1145/3503222.3507733, 10.1145/2168836.2168866}. The \projecttitle{} implementation (bitstream) is synthesized by a trusted IP vendor with a trusted tool flow for covert channels resilience. %(in a trusted environment) The system designers source the \projecttitle{} from trusted IP vendors.

\rev{(a), B3, B4, C2}{
Since \projecttitle{} does not rely on CPU-based TEEs and its network stack and library run on the unprotected CPU, both software can be compromised by a potentially Byzantine actor on the machine. As such, \projecttitle{} does not distinguish between different types of untrusted software components. Whether the network library, the network stack, or the application code is compromised, the node is considered faulty (Byzantine) and must conform to the BFT application system model, which should specify its tolerance to Byzantine failures.
}

%We do not consider denial-of-service (DoS) attacks; the cloud provider has physical control of the hardware and can simply unpower it. Nevertheless such attacks affect availability and not the correctness.

\if 0
\begin{figure}[t!]
    \centering
    \includegraphics[width=.45\textwidth]{figures/trusted-nic-attestation_kernel.drawio.pdf}
    \caption{\trustedfpga{} attestation kernel overview (transmission path).}
    \label{fig:attestation_kernel}
\end{figure}    
\fi

\subsection{Design Challenges and Key Ideas} \label{subsec:challenges} While designing \projecttitle{}, we overcome the following challenges:

\myparagraph{\#1: Heterogeneous hardware} 
CPU-based TEEs in the cloud infrastructure are heterogeneous with different programmability~\cite{Baumann2014, scone, 10.1145/3079856.3080208, 10.1145/3460120.3485341, tsai2017graphene, Rkt-io} and security properties~\cite{10.1145/3600160.3600169, 7807249, 10.1007/978-3-031-16092-9_7} that complicate their adoption and the system's correctness~\cite{10.1145/3460120.3485341}. 
% For example, Intel SGX~\cite{cryptoeprint:2016:086} offers code confidentiality and attestation while ARM TrustZone~\cite{arm-realm} does not. Likewise, AMD-sev~\cite{amd-sev} and Intel TDX~\cite{intelTDX} offer OS-based programming interfaces, whereas Keystone~\cite{riscv-multizone} and SGX require specific SDKs~\cite{KeystoneSDK}.
Prior systems~\cite{hybster, 10.1145/3492321.3519568, minBFT, DBLP:journals/corr/LiuLKA16a} {\em could} not address this heterogeneity challenge as they require {\em homogeneous} {\tt x86} machines with SGX extensions of a specific version. This is rather unrealistic in modern heterogeneous distributed systems where system designers are compelled to {\em stitch heterogeneous TEEs together}. TEE's heterogeneity in programmability and security semantics hampers their adoption and adds complexity to ensuring the system's overall correctness. 

\myparagraph{Key idea: A host CPU-agnostic unified security architecture based on trustworthy network-level isolation} 
Our \projecttitle{} offers a unified and host-agnostic network-interface level isolation that guarantees the specific yet well-defined security properties of the non-equivocation and transferable authentication. 
\rev{B1}{\projecttitle{} shifts the security properties from CPU-hosted TEEs to NIC hardware, thereby addressing the heterogeneity and programmability issues associated with CPU-based TEEs.}
% \projecttitle{} is built upon SmartNIC hardware, which is highly favorable in the heterogeneous Byzantine cloud infrastructure. 
%\projecttitle{} is built upon SmartNIC hardware, which is highly favorable in the heterogeneous Byzantine cloud infrastructure. 
\projecttitle{} also offers generic programming APIs ($\S$~\ref{sec:net-lib}) that are used to {\em correctly} transform a wide variety of distributed systems for Byzantine settings. 
We demonstrate the power of \projecttitle{} with a generic transformation {\em recipe} ($\S$~\ref{subsec:transformation}) and its usage to transform prominent distributed systems ($\S$~\ref{sec:use_cases}).
% We build our \projecttitle{} system on SmartNIC hardware to achieve those goals, offering network-level isolation for our offered security properties. 
% Importantly, our host-agnostic \projecttitle{} network interface is highly favorable in the heterogeneous Byzantine cloud infrastructure. In contrast, the \projecttitle{}'s security properties have been proven to be sufficient for {\em correctly} transforming a wide variety of distributed systems for Byzantine settings. 

% Our \projecttitle{} offers a unified and host-agnostic network-interface level isolation that guarantees the specific and well-defined security properties of the non-equivocation and the transferable authentication. At the same time, we resolve the programmability burden through generic programming APIs. We build our \projecttitle{} system on SmartNIC hardware to achieve those goals, offering network-level isolation for our offered security properties. Importantly, our host-agnostic \projecttitle{} network interface is highly favorable in the heterogeneous Byzantine cloud infrastructure. In contrast, \projecttitle{}'s security properties of the non-equivocation and the transferable authentication have been proven to be sufficient for {\em correctly} transforming a wide variety of distributed systems for Byzantine settings. In fact, we show the power of \projecttitle{} with a generic transformation {\em recipe} ($\S$~\ref{subsec:transformation}) as well as its application to transform four widely adopted distributed systems ($\S$~\ref{sec:use_cases}).

\myparagraph{\#2: Large TCB in the TEE-based silicon root-of-trust} 
TEEs based on a {\em silicon root of trust} are promising for building trustworthy systems~\cite{avocado, minBFT, hybster, 10.1145/3492321.3519568}. Unfortunately, the state-of-the-art TEEs integrate a {\em large} TCB; for example, we calculate the TCB size of the state-of-the-art Intel TDX~\cite{intelTDX}. The TEE ports within the trusted hardware the entire Linux kernel (specifically, v5.19~\cite{linuxlifecircle}) and ``hardens'' at least 2000K lines of usable code, leading to a final TCB of 19MB. Such large TCBs have been accused of increasing the area of faults and attacks~\cite{10.1145/3379469, 10.5555/1756748.1756832} of commercial TEEs that are already under fire for their {security vulnerabilities}~\cite{intel_sgx_vulnerabilities1, intel_sgx_vulnerabilities2, intel_sgx_vulnerabilities3, intel_sgx_vulnerabilities4, intel_sgx_vulnerabilities5}. Importantly, TEE's large TCBs complicate their {security analysis and verification}, rendering their security properties {\em incoherent}. 
% Prior works~\cite{avocado, minBFT, hybster, 10.1145/3492321.3519568} have established the {\em silicon root of trust}, e.g., TEEs, to be a promising direction to build trustworthy systems. Unfortunately, even the state-of-the-art TEEs integrate a {\em large} TCB. For example, we calculated the TCB size of the state-of-the-art Intel TDX~\cite{intelTDX}. The TEE ports within the trusted hardware the entire Linux kernel (specifically, v5.19~\cite{linuxlifecircle}) and ``hardens'' at least 2K lines of usable code, leading to a final TCB of 19MB. Such large TCBs have been accused of increasing the area of faults and attacks~\cite{10.1145/3379469, 10.5555/1756748.1756832} of commercial TEEs that are already under fire for their {major security vulnerabilities}~\cite{intel_sgx_vulnerabilities1, intel_sgx_vulnerabilities2, intel_sgx_vulnerabilities3, intel_sgx_vulnerabilities4, intel_sgx_vulnerabilities5}. As such, modern TEEs suffer from security flaws, whereas their large TCBs complicate their {security analysis and formal verification}, rendering the derived ``trustworthy'' distributed system {\em incoherent}. 

\myparagraph{Key idea: A minimal and formally verifiable silicon root-of-trust with low TCB} 
In our work, we advocate that a {\em minimalistic silicon root of trust} (TCB) at the NIC level hardware is the foundation for building verifiable, trustworthy distributed systems. 
In fact, \projecttitle{} builds a minimalistic and verifiable attestation kernel ($\S$~\ref{subsec:nic_attest_kernel}) that guarantees the \projecttitle{} security properties at the SmartNIC hardware. 
% that guarantees the non-equivocation and transferable authentication properties for network messages. 
% In fact, \projecttitle{} builds a minimalistic and verifiable attestation kernel at the SmartNIC hardware that guarantees the non-equivocation and transferable authentication properties for network messages. 
Moreover, we have formally verified the \projecttitle{} secure hardware protocols ($\S$~\ref{subsec::formal_verification_remote_attestation}).
% since we rely on a minimalistic interface for trusted computing, 

% In our work, we advocate that a {\em minimalistic silicon root of trust} (TCB) at the NIC level hardware is the foundation for building verifiable, trustworthy distributed systems. In fact, \projecttitle{} builds a minimalistic and verifiable attestation kernel (TCB) at the SmartNIC hardware that guarantees the non-equivocation and the transferable authentication properties for the network messages. Moreover, since we rely on a minimalistic interface for trusted computing, we have formally verified the \projecttitle{} secure hardware protocols in the Tamarin theorem prover ($\S$~\ref{subsec::formal_verification_remote_attestation}).

\myparagraph{\#3: Performance} 
TEE's overheads are significant in the context of networked systems~\cite{avocado, treaty, minBFT,10.1145/3492321.3519568}. Prior research~\cite{avocado} reported 4$\times$---8$\times$ performance degradation with even a sophisticated network stack. Others~\cite{10.1145/3492321.3519568, hybster, minBFT} limit performance due to the communication costs between their untrusted and TEE-based counterparts~\cite{10.1145/2168836.2168866}. The actual performance overheads in heterogeneous distributed systems are expected to be more exacerbated~\cite{9460547, 9935045}. As such, TEEs cannot {\em naturally} offer high-performant, trusted networking. 
% Distributed systems in the third-party cloud infrastructure must be fast and trustworthy. Their overheads are significantly exacerbated in the context of networked systems~\cite{avocado, treaty, minBFT,10.1145/3492321.3519568}--- the foundational building block in the core of any distributed system. Prior research~\cite{avocado} have reported an average of 4$\times$---8$\times$ performance degradation for networking between two host-sided TEEs instances (Intel SGX) even with using a sophisticated network stack implementation that had carefully been optimized for this specific TEE version. Other systems on top of TEEs~\cite{10.1145/3492321.3519568, hybster, minBFT} also can limit performance due to the communication costs between the untrusted and TEE-based counterparts of the system~\cite{10.1145/2168836.2168866}. Based on performance analysis of heterogeneous TEEs~\cite{9460547, 9935045}, the actual performance overheads of a distributed system in the heterogeneous cloud can be even more exacerbated. As such host-sided trusted hardware cannot {\em naturally} offer high-performant trusted networking. 

\myparagraph{Key idea: Hardware-accelerated trustworthy network stack} 
Our \projecttitle{} bridges the gap between performance and security with two design insights. 
First, \projecttitle{} attestation kernel offers the foundations to transform CFT distributed systems to BFT systems without affecting the number of participating nodes, significantly improving scalability. 
% which significantly improves scalability. 
% First, \projecttitle{} minimalistic TCB, the attestation kernel, offers the foundations to transform CFT distributed systems to BFT systems without affecting the number of participating nodes, which significantly improves scalability. 
Second, \projecttitle{} user-space network stack ($\S$~\ref{sec:t-nic-network}) bypasses the OS and offloads security and network processing to the NIC-level hardware. 

\section{Trusted NIC Hardware}
\label{sec:t-nic-hardware}
Figure~\ref{fig:hardware-design} shows our \projecttitle{} hardware architecture that implements trusted network operations on a SmartNIC device. \projecttitle{} introduces two key components: \emph{(i)} the attestation kernel that guarantees the non-equivocation and transferable authentication properties over the untrusted network ($\S$~\ref{subsec:nic_attest_kernel}) and \emph{(ii)} the RoCE protocol kernel that implements the RDMA protocol including transport and network layers ($\S$~\ref{subsec:roce_protocol_kernel}). We also introduce a bootstrapping and a remote attestation protocol for \projecttitle{} ($\S$~\ref{subsec:nic_controller}) and formally verify them ($\S$~\ref{subsec::formal_verification_remote_attestation}).
% Lastly, we design a remote attestation protocol for our \projecttitle{} based on the \projecttitle{} Controller module that we discuss $\S$~\ref{subsec:nic_controller}. Importantly, we formally verify \projecttitle{}'s remote attestation and initialization protocol ($\S$~\ref{subsec::formal_verification_remote_attestation}).

\subsection{NIC Attestation Kernel}
\label{subsec:nic_attest_kernel}
\begin{figure}[t!]
    \centering
    %\includegraphics[width=.27\textwidth]{figures/trusted-nic-single-node-overview.drawio-1.pdf}
    %\includegraphics[width=0.45\textwidth]{figures/hw-implementation.drawio.pdf}
    %\caption{\projecttitle{} hardware architecture.}
    % \includegraphics[width=0.95\linewidth]{figures/trusted-nic-hardware-impl-improved.drawio.pdf}
     \includegraphics[width=0.75\linewidth]{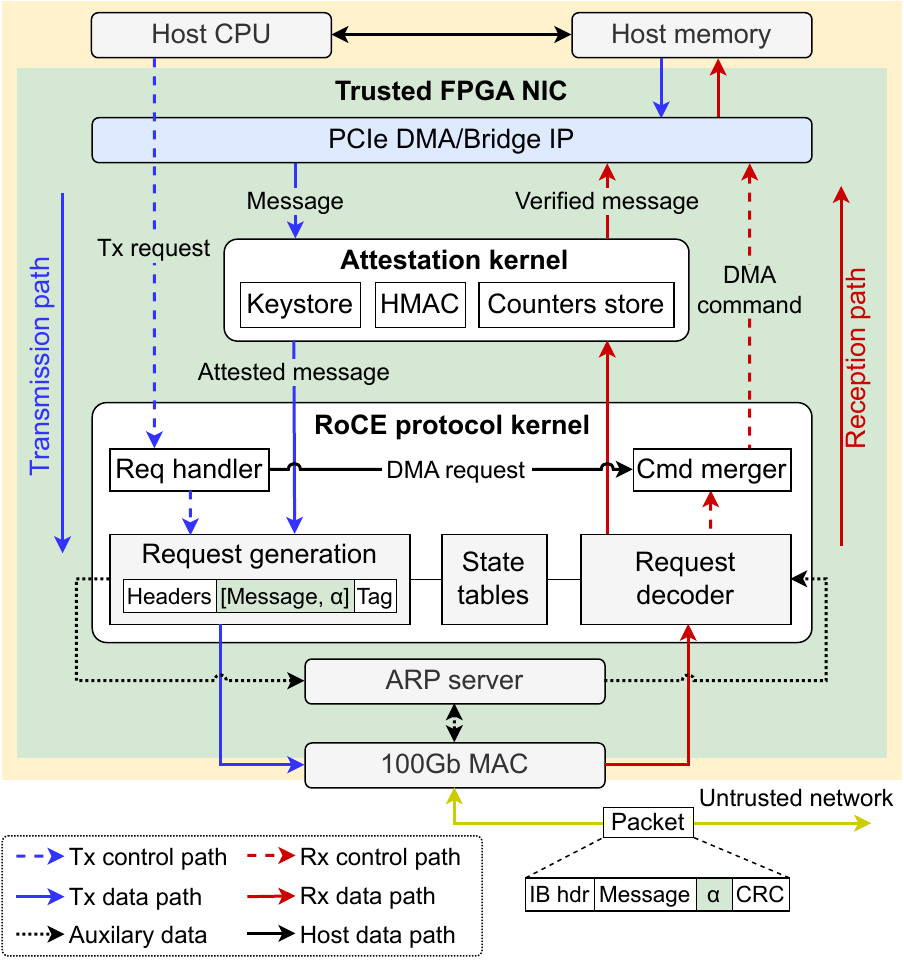}
    \caption{\projecttitle{} hardware architecture.}
     \label{fig:hardware-design}
\end{figure}

%\myparagraph{Overview} 
The attestation kernel {\em shields} network messages and materializes the properties of non-equivocation and transferable authentication by generating {\em attestations} for transmitted messages. As shown in Figure~\ref{fig:hardware-design}, the attestation kernel resides in the data pipeline between the RoCE protocol kernel that transmits/receives network messages and the PCIe DMA that transfers data from/to the host memory.
The kernel processes the messages as they {\em flow} from the memory to the network and vice versa to optimize throughput. 
% The module resides in the data pipeline between the RoCE protocol kernel that transmits/receives network messages and the PCIe DMA which fetches and pushes data {\em asynchronously} from the host memory to the FPGA memory.

\myparagraph{Hardware design} The attestation kernel is comprised of three components that represent its state and functionality: the HMAC component that generates the message authentication codes (MAC), the Keystore that stores the keys used by the HMAC module, and the Counters store that keeps the message's latest sent and received timestamp. 

The system designer initializes each \projecttitle{} device during bootstrapping with a unique identifier (ID) and a shared secret key---ideally, one shared key for each session---stored in static memory (Keystore). The keys are shared and, hence, unknown to the untrusted parties. 
% Each \projecttitle{} device is initialized by the system designer during bootstrapping ($\S$~\ref{subsec:nic_controller}) with a unique identifier (ID) and a shared secret key---ideally, one shared key for each session---which is stored in static memory (Keystore). The keys are shared and, hence, unknown to the untrusted parties. 

\projecttitle{} holds two counters per session in the Counters store: \texttt{send\_cnts}, which holds sending messages, and \texttt{recv\_cnts}, which holds the latest seen counter value for each session. The counters represent the messages' timestamp and are increased monotonically and deterministically after every send and receive operation to ensure that unique messages are assigned to unique counters for non-equivocation. Consequently, no messages can be lost, re-ordered, or doubly executed.
% \projecttitle{} holds two counters per session in the Counters store: {\tt send\_cnts} to be used for sending messages, and \texttt{recv\_cnts} that holds the latest seen counter value for each session. The counters represent the messages' timestamp and are increased monotonically and deterministically after every send and receive operation to ensure that unique messages are assigned to unique counters for non-equivocation. Consequently, no messages can be lost, re-ordered, or doubly executed.

\myparagraph{Algorithm} 
Algorithm~\ref{algo:primitives} shows the functionality of the attestation kernel. The module implements two core functions: {\tt Attest()}, which generates a unique and verifiable attestation for a message, and {\tt Verify()}, which verifies the attestation of a received message. The message transmission invokes {\tt Attest()}, and likewise, the reception invokes {\tt Verify()}. 
% When an application transmits a message, the message is attested by {\tt Attest()}. Likewise, upon a message reception, the message is verified with {\tt Verify()} before it is copied to the application's (host) memory. 
% Algorithm~\ref{algo:primitives} shows the functionality of the attestation kernel. The module implements two core functions: the {\tt Attest} function which generates a unique and verifiable attestation for a message and the {\tt Verify} function which verifies the attestation of a received message. When the application transmits a message, the data flow passes through the {\tt Attest} function, and upon a message reception, the message is verified (with {\tt Verify} function) before it is copied to the application's (host) memory. 

Upon transmission, as shown in Figure~\ref{fig:hardware-design}, the message is first forwarded to the attestation kernel. 
% , and then, the output is forwarded to the RoCE protocol kernel. 
The attestation kernel executes \texttt{Attest()} and generates an {\em attested} message comprised of the message data and its attestation certificate~$\alpha$. The function calculates $\alpha$ as the HMAC of the message concatenated with the counter {\tt send\_cnt} and the device ID (for transferable authentication) with the {\tt key} for that connection (Algo~\ref{algo:primitives}:~L4). It also increases the next available counter for subsequent future messages (Algo~\ref{algo:primitives}:~L2). The function forwards the message with its $\alpha$ to the RoCE protocol kernel (Algo~\ref{algo:primitives}:~L4).
% As shown in Figure~\ref{fig:hardware-design}, a request passes from the application layer to the \projecttitle{} through two separate paths, i.e., the data and control paths. The metadata and the request opcode pass through the control path from the host to the device and vice versa. On the other hand, the message data is first forwarded to the attestation kernel, and then, the output is forwarded to the RoCE protocol kernel. The attestation kernel upon transmission executes the \texttt{Attest} function and generates an {\em attested} message comprised of the message data and its attestation certificate, or simply attestation $\alpha$. The function calculates $\alpha$ as the HMAC of the message concatenated with the counter {\tt send\_cnt} using the {\tt key} for that connection (Algorithm~\ref{algo:primitives}:~L3). It also increases the next available counter for subsequent future messages (Algorithm~\ref{algo:primitives}:~L2). The function forwards to the RoCE protocol the original message along with $\alpha$ for transmission (Algorithm~\ref{algo:primitives}:~L4).

Upon reception, the received message passes through the attestation kernel for verification before it is delivered to the application. Specifically, \texttt{Verify()} checks the authenticity and the integrity of the received message by re-calculating the {\em expected} attestation $\alpha$' based on the message payload and comparing it with the received attestation $\alpha$ of the message (Algo~\ref{algo:primitives}:~L7---8). The verification also ensures that the received counter matches the expected counter for that connection to ensure \emph{continuity} (Algo~\ref{algo:primitives}:~L8).

\subsection{RoCE Protocol Kernel}
\label{subsec:roce_protocol_kernel}

%\myparagraph{Overview} 
The RoCE protocol kernel implements a reliable transport service on top of the IB Transport Protocol with UDP/IPv4 layers (RoCE v2)~\cite{infiniband} (transport and network layers). As shown in Figure~\ref{fig:hardware-design}, to transmit data, the {\tt Req handler} module in the RoCE kernel receives the request opcode ({\tt metadata}) issued by the host. The message is fetched through the PCIe DMA engine and passes through the attestation kernel. The request opcode and the attested message are forwarded to the {\tt Request generation} module that constructs a network packet. 
% The RoCE protocol kernel implements a reliable transport service on top of the IB Transport Protocol with UDP/IPv4 layers (RoCE v2)~\cite{infiniband} (transport and network layers), and it is connected to the attestation kernel as discussed in $\S$~\ref{subsec:nic_attest_kernel}. As shown in Figure~\ref{fig:hardware-design}, to send data at the transmission path, the {\tt Req handler} module in the RoCE kernel receives the request opcode (metadata) issued by the host. The message to be sent is then fetched through the PCie DMA engine and passes through the attestation kernel for processing. The request opcode and the attested message are forwarded to the {\tt Request generation} module that constructs the network packet. 

Upon receiving a message from the network, the RoCE kernel parses the packet header and updates the protocol state (stored in the State tables). The attested message is then forwarded to the attestation kernel. The message is delivered to the application’s (host) memory upon successful verification.
% Upon successful verification, the message is delivered to the application's (host) memory. 

\myparagraph{Hardware design} The RoCE protocol kernel is also connected to a 100Gb MAC IP and an ARP server IP. 

\noindent{\underline{100Gb MAC.}} The 100Gb MAC kernel implements the link layer connecting \projecttitle{} to the network fabric over a 100G Ethernet Subsystem~\cite{license}. The kernel also exposes two interfaces for transmitting (Tx) and receiving (Rx) network packets. 

\noindent{\underline{ARP server.}} 
The ARP server has a lookup table containing MAC and IP address correspondences. Right before the transmission, the RDMA packets at the {\tt Request generation} first pass through a MAC and IP encoding phase, where the {\tt Request generation} module extracts the remote MAC address from the lookup table in the ARP server.
% The ARP server contains a lookup table with the correspondences between MAC and IP addresses. Right before the transmission, the RDMA packets at the {\tt Request generation} first pass through a MAC and IP encoding phase. At this encoding phase, the {\tt Request generation} module extracts from the lookup (ARP table) in the ARP server the remote MAC address.

\begin{algorithm}[t]
\SetAlgoLined
\footnotesize
\textbf{function} \texttt{Attest(c\_id, msg)} \{ \\
\Indp
{\tt cnt} $\leftarrow$ {\tt send\_cnts[c\_id]++};\\
$\alpha$ $\leftarrow$ {\tt hmac(keys[c\_id], msg||ID||cnt)}; \\
\textbf{return} $\alpha${\tt ||msg||ID||cnt};\\
\Indm
\} \\

%\vspace{0.3cm}

\textbf{function} \texttt{Verify(c\_id, $\alpha$||msg||ID||cnt)} \{ \\
\Indp
    $\alpha$' $\leftarrow$ {\tt hmac(keys[c\_id], msg||ID||cnt)};\\
    \textbf{if} {\tt (}$\alpha$' $=$ $\alpha$ {\tt \&\&} {\tt cnt} $=$ {\tt recv\_cnts[c\_id]++} {\tt )} \\
    \Indp
        \textbf{return} ($\alpha${\tt ||msg||cnt)}; \\
    \Indm
    \textbf{assert(False)}; \\
\Indm
\} \\
%\vspace{0.1cm}
\vspace{-1pt}
\caption{\texttt{Attest()} and \texttt{Verify()} functions.}
\label{algo:primitives}
\end{algorithm}

\noindent{\underline{IB protocol.}} 
The RoCE protocol kernel implements the reliable version of the IB protocol. Similarly to its original specification~\cite{rdma_specification}, the kernel implements State tables to store protocol queues (e.g., receive/send/completion queues) as well as important metadata, i.e.,  packet sequence numbers (PSNs), message sequence numbers (MSNs), and a Retransmission Timer. % to detect packet losses.
% The RoCE protocol kernel implements the reliable version of the IB protocol. Similarly to its original specification~\cite{rdma_specification}, the kernel implements the State tables to store the protocol queues QPs (i.e., receive queue, send queue, completion queue, etc.). The State tables also store important metadata such as (1) the packet sequence numbers (PSNs) to distinguish valid, invalid, and duplicate PSN regions, (2) the message sequence numbers (MSNs) to reconstruct fragmented network messages, and (3) a Retransmission Timer to detect packet losses.

\myparagraph{Dataflow}
The transmission path is shown with the blue-colored axes in Figure~\ref{fig:hardware-design}. The {\tt Req handler} receives requests issued by the host and initializes a DMA command to fetch the payload data from the host memory to the attestation kernel. Once the attestation kernel forwards the attested message to the {\tt Req handler}, the module passes the message from several states to append the appropriate headers {\tt IB hdr} along with metadata (e.g., RDMA op-code, PSN, QP number). The last part of the processing generates and appends UDP/IP headers (e.g., IP address, UDP port, and packet length). The message is then forwarded to the 100Gb MAC module. 
 
In the reception path (red-colored axes in Figure~\ref{fig:hardware-design}), the {\tt Request decoder} extracts the headers, metadata, and attested message. The message is forwarded to the attestation kernel for verification and finally copied to the host memory.

% \myparagraph{{Message format:}} The IB protocol in \projecttitle{} processes the following headers: IP, UDP, BTH (Base Transport Header), RETH (RDMA Extended Transport Header), and AETH (ACK Extended Transport Header) as defined in the RDMA Protocol Specification~\cite{rdma_specification, storm}.  The RoCE kernel adds the Ethernet header (L2 header), the IPv4/UDP headers (L3 headers), and the IB headers (L4 headers). It also adds a 32b end-to-end CRC (ICRC) that covers all invariant fields of the packet and offers protection beyond the coverage of the Ethernet Frame Checksum (FCS). 
The message format in \projecttitle{} follows the original RDMA specification~\cite{rdma_specification}; only the difference between our \projecttitle{} and the original RDMA messages is that the attestation kernel {\em extends} the payload by appending a 64B attestation $\alpha$ and the metadata. The metadata includes a 4B id for the session id of the sender, a 4B ID for the device id (unique per device), and the appropriate {\tt send\_cnt}. This payload extension is negligible and does not harm the network throughput.  
% \atsushi{Can we say something positive: e.g., the RoCE protocol kernel does not need to be aware of this difference, or its extension doesn't badly affect the performance, etc.}\dimitra{Both are true, feel free to write in-place}

\subsection{\projecttitle{} Attestation Protocol} 
\label{subsec:nic_controller}
We design a remote attestation protocol to ensure that the \projecttitle{} device is genuine and the \projecttitle{} bitstream and secrets are flashed securely in the device. % along with the designer's configuration data (IPs, secrets, etc.).

\if 0
\myparagraph{Design assumptions} We base our design on a NIC Controller hardware component that drives the device initialization with no access to confidential information as in~\cite{10.1145/3503222.3507733}. 
The Controller can be implemented as a soft CPU~\cite{microblaze, nios, 10.1145/3503222.3507733} while after \projecttitle{}'s initialization, it monitors JTAG/ICAP interfaces to prevent physical attacks~\cite{secMon}. 
% The Controller can be implemented as a soft CPU~\cite{microblaze, nios, 10.1145/3503222.3507733} while after \projecttitle{}'s initialization, it monitors JTAG/ICAP interfaces to prevent physical attacks as commercial IPs~\cite{secMon} \atsushi{what do you mean 'as commercial IPs'?}. %ensure that the bitstream is not modified before use and 

\fi

\myparagraph{Boostrapping} 
% \rev{D1}{We assume that the FPGA cards have access to an AES key and the hash of a public encrypted key (ECDSA for Intel, RSA for Xilinx), which are embedded in secure, on-chip, non-volatile storage. The manufacturer, system designer, and IP vendor are trusted entities among each other.}
\rev{D1}{The \projecttitle{} hardware is securely bootstrapped in an untrusted third-party cloud by the Manufacturer, System designer, and IP vendor, who trust each other. At the device construction, the Manufacturer burns \texttt{HW$_{key}$}, a secret key unique to the device. It is possible with commercial FPGA cards that have access to an AES key and the hash of a public encrypted key embedded in secure, on-chip, non-volatile storage (Intel~\cite{intel-secure-dev}, AMD~\cite{amd-bootgen}).} The System designer shares the configuration with the IP vendor and instructs the cloud provider to load the (encrypted) FPGA firmware which is then decrypted with the \texttt{HW$_{key}$}. 
The firmware loads the controller binary \texttt{Ctrl$_{bin}$}, generates a key pair \texttt{Ctrl$_{pub, priv}$} for the specific device and binary, and signs the measurement of the \texttt{Ctrl$_{bin}$} and the \texttt{Ctrl$_{pub}$} with the  \texttt{HW$_{key}$} (\texttt{Ctrl$_{bin}$cert}). %Ctrl_bin_cert = sign{Ctrl_bin, pk(Ctrl_priv)}HW_key_priv
%The cert should include the freshly generated public key of the ctrl keypair %The Protocol designer shares the configurations with the IP vendor over a TLS/SSL connection. %participating machines and any configuration data. 

%\myparagraph{Boostrapping} At the manufacturing construction, the Manufacturer burns a secret, unique to the device~\cite{secure_FPGAs}, key \texttt{HW$_{key}$}. The System designer instructs the cloud provider to load the (encrypted) FPGA firmware from the storage medium using the embedded code in the BootROM device. The firmware is then decrypted using the \texttt{HW$_{key}$}. The firmware loads the controller binary \texttt{Ctrl$_{bin}$}, generates a key pair \texttt{Ctrl$_{pub, priv}$} that is bound to the specific device and binary, and lastly, generates and signs the measurement of the \texttt{Ctrl$_{bin}$} producing \texttt{Ctrl$_{bin}$cert}. The Protocol designer then establishes a TLS/SSL connection with the trusted IP vendor and sends the list of the participating machines and any configuration data. 

\begin{figure}[t!]
    \centering
    \includegraphics[width=0.75\linewidth]{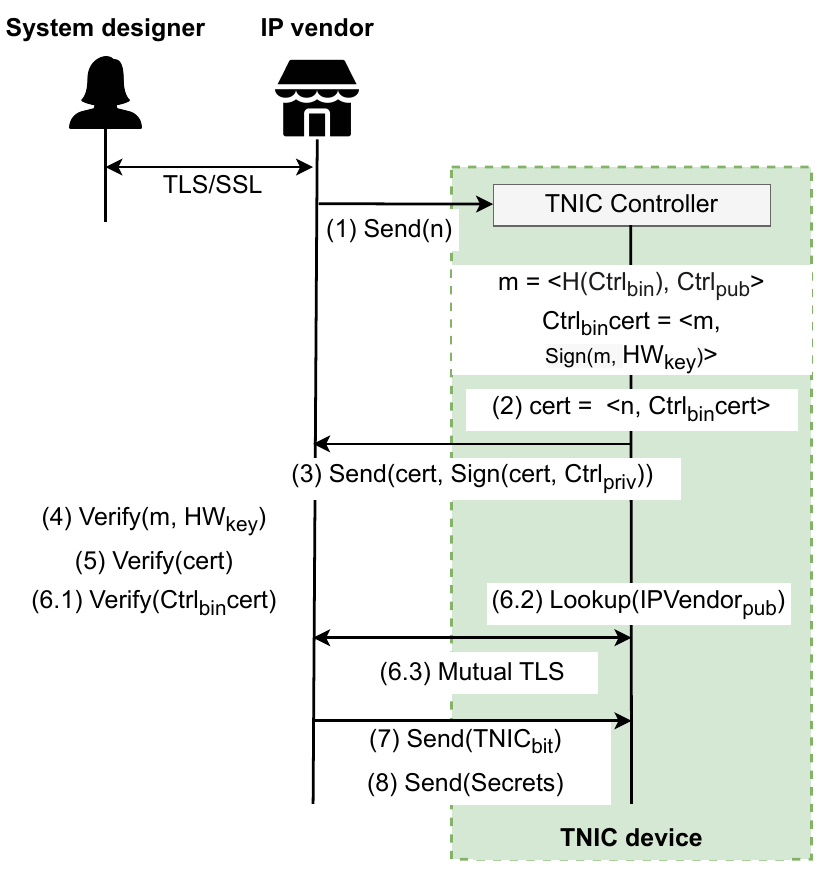}
    \caption{\projecttitle{} remote attestation protocol.}
    \label{fig:remote_attestation}
    % \vspace{-3mm}
\end{figure}

\myparagraph{Remote attestation} 
Figure~\ref{fig:remote_attestation} shows \projecttitle{} remote attestation. The IP vendor sends a random nonce \texttt{n} for freshness to the Controller. The IP vendors public key \texttt{IPVendor$_{pub}$} is embedded into the \texttt{Ctrl$_{bin}$}. The Controller generates a certificate \texttt{cert} over the \texttt{Ctrl$_{bin}$cert} and \texttt{n} (2) which signs with \texttt{Ctrl$_{pub}$} and sends it to the IP vendor (3).

The IP vendor verifies the authenticity of the \texttt{cert} (4)---(5) and establishes a TLS connection with the Controller. First, the vendor verifies the authenticity of \texttt{m}  with the \texttt{HW$_{key}$}, ensuring that a genuine \texttt{Ctrl$_{bin}$} and a genuine device has signed \texttt{m} (4). As such, the vendor ensures that the \texttt{Ctrl$_{bin}$}  runs in a genuine \projecttitle{} device by verifying that the measurement of the \texttt{Ctrl$_{bin}$} has been signed with an appropriate device key installed by the Manufacturer. Lastly, the vendor verifies the nonce \texttt{n} and \texttt{cert} to ensure freshness (with the \texttt{Ctrl$_{pub}$} included in \texttt{m}). 

Now, a mutually authenticated TLS connection is established; the IP vendor verifies authenticity by checking for the desired \texttt{Ctrl$_{pub}$} and the Controller checks for it's embedded \texttt{IPVendor$_{pub}$} (6.1)---(6.3).
Once the TLS connection is established the IP Vendor sends the Controller the secrets and the \projecttitle{} bitstream, \texttt{TNIC$_{bit}$}.

\subsection{Formal Verification of \projecttitle{} Protocols}
\label{subsec::formal_verification_remote_attestation}

\begingroup
\setlength{\abovedisplayskip}{0.2em}
\setlength{\belowdisplayskip}{0.1em}
%\atsushi{@Julian: can you update the section to address A1, A3? We plan to add a summary of our appendix (Tamarin's verification proofs) here to make the paper self-sustaining. }

% \pramod{Please tie to the high-level properties (tranferrable authentication and non-equivocation) of TNIC and then map it down to low-level TNIC protocols.}

\rev{A1, A3}{
We formally verify the safety and security properties of \projecttitle{} hardware using Tamarin~\cite{tamarin-prover}.
The proof details are covered in Appendix \S~\ref{sec:formal-verification-details}.
% , a security protocol verification tool that analyzes symbolic models of protocols specified as multiset rewriting systems. 
% We verify \projecttitle{}'s bootstrapping and the remote attestation protocol that provides a formal model to argue about non-equivocation and transferable authentication. 
% 
Our verification consists of a model for bootstrapping, remote attestation, message transmission, and reception, according to Figure~\ref{fig:remote_attestation}.
% Our verification approach consists of a model for bootstrapping, remote attestation, message transmission, and reception, according to Figure~\ref{fig:remote_attestation}.
This model is augmented with custom \textit{action facts}, which mark the occurrence of defined events in the execution trace. These include:
    \begin{enumerate}[leftmargin=*]
        %\item \(\text{D}_e(x)\), which marks the end of the attestation phase for device \(e\), i.e., a \projecttitle{} device or the IP vendor, with associated connection information \(x\). 
        \item \(\text{D}_e(x)\), which marks the end of the attestation phase for endpoint \(e\), with associated connection information \(x\). 
        \item \(\text{S}_e(\textit{m})\) and \(\text{A}_e(\textit{m})\) marking the sending and accepting of a message \(\textit{m}\), following Algorithm~\ref{algo:primitives}, respectively.
    \end{enumerate}
    %The intended temporal relationship of these action facts is expressed using lemmas. Tamarin is then used to apply automated deduction and equational reasoning to determine whether these lemmas can be falsified for our model. The relation \(\textit{a} ~@~ t_i\) expresses that action fact \(\textit{a}\) occurred at time \(t_i\) in the trace. % We use this relation to express our desired security properties. 
    The intended temporal relationship of these action facts is expressed using lemmas, which Tamarin then proves using automated deduction and equational reasoning. The relation \(\textit{a} ~@~ t_i\) expresses that action fact \(\textit{a}\) occurred at time \(t_i\).
    %in the trace. 
    Using this relation, we can express our desired security properties as follows:
}

\begingroup
% NOTE: temporary fix for the spacing, may need to be adjusted in case it breaks
\setlength{\belowdisplayskip}{-1.2em} 

\myparagraph{Remote attestation}  
\if 1
We model the bootstrapping and remote attestation protocols based on Figure~\ref{fig:remote_attestation}. We define lemmas to ensure the secrecy of the private information involved in the protocol and the main attestation lemma, which holds if and only if: once the last step of the remote attestation protocol is completed, the \projecttitle{} device is in a valid, expected state. 
\fi
\revcont{
We define the main attestation lemma for any \projecttitle{} device \(\textit{tnic}\) and associated IP Vendor \(\textit{ipv}\). The lemma holds if, after the last step of the remote attestation protocol, the \projecttitle{} device is in a valid, expected state:
\begin{equation}
    \begin{split}
        \forall~\textit{ipv}, \textit{tnic}, \textit{c}, t_i.~\text{D}_\textit{ipv}(\textit{c}) ~@~ t_i
        \implies \exists~t_j.~ t_j < t_i \land \text{D}_\textit{tnic}(\textit{c}) ~@~ t_j
    \end{split}
\end{equation}
}

% $
%     \forall~\textit{ipv}, \textit{tnic}, \textit{c}, t_i.~\text{D}_\textit{ipv}(\textit{c}) ~@~ t_i \implies 
%     \exists~t_j.~ t_j < t_i \land \text{D}_\textit{tnic}(\textit{c}) ~@~ t_j
% $.
% }

\myparagraph{Transferable authentication} 
% We extend the analysis of the bootstrapping and attestation protocol in Tamarin with models for the network operations that upon the message transmission and reception execute the {\tt Attest} and {\tt Verify} functions (Algorithm~\ref{algo:primitives}), respectively. In simple words we prove that  \projecttitle{} attestation kernel prevents equivocation by ensuring that two attestations of different messages will never be assigned identical message sequencers ({\tt send\_cnts}) and secondly, the transferable authentication is met because each generated attestation is uniquely verified by the secret keys jointly with the unique sender's \projecttitle{} device identifier.
\if 0
We extend the model with rules for the network operations that execute \texttt{Attest()} and \texttt{Verify()} (Algorithm~\ref{algo:primitives}) upon the message transmission and reception, respectively. The model extension allows us to reason about transferable authentication by defining an additional lemma: any accepted message was sent by an authentic \projecttitle{} device in a valid configuration.
\fi
\revcont{
%The \(\text{S}_e(\textit{m})\) action fact only marks send operations of authentic \projecttitle{} devices in a valid configuration because it is only used after attestation. Therefore, we define the lemma, which states that any accepted message was sent by an authentic \projecttitle{} device in a valid configuration:
We define the lemma, which states that any accepted message was sent by an authentic \projecttitle{} device in a valid configuration:
% $
%     \forall~e1, \textit{m}, t_i.~\text{A}_{e1}(\textit{m}) ~@~ t_i \implies 
%     \exists~e2, t_j.~ t_j < t_i \land \text{S}_{e2}(\textit{m}) ~@~ t_j
% $.
\begin{equation}
    \begin{split}
        \forall~e1, \textit{m}, t_i.~\text{A}_{e1}(\textit{m}) ~@~ t_i
        \implies \exists~e2, t_j.~ t_j < t_i \land \text{S}_{e2}(\textit{m}) ~@~ t_j
    \end{split}
\end{equation}
}

\endgroup

% We extend the model with rules for the network operations that upon the message transmission and reception execute the {\tt Attest} and {\tt Verify} functions (Algorithm~\ref{algo:primitives}), respectively. This allows us to reason about transferable authentication by defining the following additional lemma: Any message that is accepted was sent by an authentic \projecttitle{} device in a valid configuration.

% In simple words we prove that  \projecttitle{} attestation kernel prevents equivocation by ensuring that two attestations of different messages will never be assigned identical message sequencers ({\tt send\_cnts}) and secondly, the transferable authentication is met because each generated attestation is uniquely verified by the secret keys jointly with the unique sender's \projecttitle{} device identifier.

% Formally, we define one additional lemma to help reason about the transferable authentication: \emph{(i)} this message was sent by an authentic \projecttitle{} device in a valid configuration \emph{(ii)} there is no message that was sent before, but not accepted \emph{(iii)} there is no message that was sent after, but accepted before this one \emph{(iii)} this message has not been accepted before.

\myparagraph{Non-equivocation} 
\if 0
We further extend the model by three lemmas that help to reason about non-equivocation as follows: for any message that is accepted, it holds that \emph{(i)} there is no message that was sent before but not accepted, \emph{(ii)} there is no message that was sent after, but accepted before this one, \emph{(iii)} this message has not been accepted before. 
% The model is further extended by three lemmas that help to reason about non-equivocation: for any message that is accepted it holds that \emph{(i)} there is no message that was sent before, but not accepted \emph{(ii)} there is no message that was sent after, but accepted before this one \emph{(iii)} this message has not been accepted before. 
\fi
\revcont{
    We further extend the model by three lemmas that help to reason about non-equivocation. For any message that is accepted, it holds that \emph{(i)} there is no message that was sent before but not accepted:
    % \\ $
    %     \forall~e1, e2, \textit{m}_j, t_i, t_j.~\text{A}_{e1}(\textit{m}_j) ~@~ t_i \land \text{S}_{e2}(\textit{m}_j) ~@~ t_j \implies 
    %     (\forall \textit{m}_k, t_k.~ t_k < t_j \land \text{S}_{e2}(\textit{m}_k) ~@~ t_k \implies \exists~t_l.~ t_l < t_i \land \text{A}_{e1}(\textit{m}_k) ~@~ t_l)
    % $ \\
    \begin{equation}
        \begin{split}
            \forall~e1, e2, \textit{m}_j, t_i, t_j.&~\text{A}_{e1}(\textit{m}_j) ~@~ t_i \land \text{S}_{e2}(\textit{m}_j) ~@~ t_j \\\implies (&\forall \textit{m}_k, t_k.~ t_k < t_j \land \text{S}_{e2}(\textit{m}_k) ~@~ t_k \\
                      &\implies \exists~t_l.~ t_l < t_i \land \text{A}_{e1}(\textit{m}_k) ~@~ t_l)
        \end{split}
    \end{equation}
    \emph{(ii)} there is no message that was sent after, but accepted before:
    % \\ $
    %     \forall~e1, \textit{m}_i, \textit{m}_j, t_i, t_j.~t_i < t_j \land \text{A}_{e1}(\textit{m}_i) ~@~ t_i \land \text{A}_{e1}(\textit{m}_j) ~@~ t_j \implies 
    %     \exists~e2, t_k, t_l.~ t_k < t_l \land \text{S}_{e2}(\textit{m}_k) ~@~ t_k \land \text{S}_{e2}(\textit{m}_l) ~@~ t_l
    % $ \\
    \begin{equation}
        \begin{split}     
        \forall~e1, \textit{m}_i, \textit{m}_j, t_i, t_j.~t_i < t_j \land \text{A}_{e1}(\textit{m}_i) ~@~ t_i \land \text{A}_{e1}(\textit{m}_j) ~@~ t_j \\
        \implies \exists~e2, t_k, t_l.~ t_k < t_l \land \text{S}_{e2}(\textit{m}_k) ~@~ t_k \land \text{S}_{e2}(\textit{m}_l) ~@~ t_l
        \end{split}
    \end{equation}
    \emph{(iii)} this message has not been accepted before:
    % \\ $
    %     \forall~e1, \textit{m}, t_i, t_j.~\text{A}_{e1}(\textit{m}) ~@~ t_i \land \text{A}_{e1}(\textit{m}) ~@~ t_j \implies t_i = t_j
    % $ \\
    \begin{equation}
        \begin{split}
               \forall~e1, \textit{m}, t_i, t_j.~\text{A}_{e1}(\textit{m}) ~@~ t_i \land \text{A}_{e1}(\textit{m}) ~@~ t_j \implies t_i = t_j
        \end{split}
    \end{equation}
    Our complete verification includes additional action facts and lemmas to verify properties like the secrecy of private information and the implications of out-of-band key compromises.
}

To sum up, Tamarin successfully shows that there is no sequence of transitions that leads to any state where our lemmas are violated. Thus, the attestation and transferable authentication lemmas hold for our model, and the counters behave as expected for non-equivocation. 

\endgroup 
\section{\rev{(a)}{\projecttitle{} Network Stack}}
% \section{Trusted Network Stack}
\label{sec:t-nic-network}

\begin{figure}[t]
    \centering
    \includegraphics[width=0.75\linewidth]{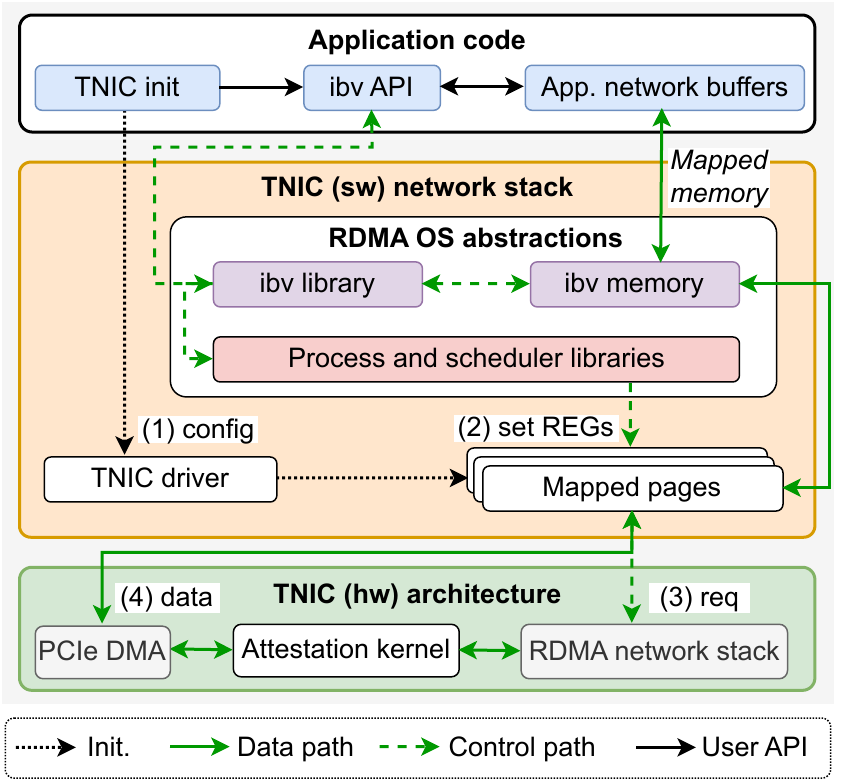}
    \caption{\projecttitle{} network system stack.}
    \label{fig:network_stack_design}
\end{figure}

We build a software \projecttitle{} system network stack that operates as the {\em middle layer} between the \projecttitle{} programming APIs (see $\S$~\ref{sec:net-lib}) and the hardware implementation of \projecttitle{}. 
Figure~\ref{fig:network_stack_design} shows an overview of the network stack design that is comprised of two core components: {\em (1)} the \projecttitle{} driver and mapped REGs pages that are responsible for initializing the device and configuring host---device communication and {\em (2)} the RDMA OS abstractions that execute networking operations.

\subsection{\projecttitle{} Driver and Mapped REGs Pages} 
The \projecttitle{} driver is invoked at the device initialization, before the remote attestation protocol ($\S$~\ref{subsec:nic_controller}), to configure the hardware with its static configuration (the device MAC address, the device QSFP port, and the network IP used by the application). 

The driver enables kernel-bypass networking---similar to the original (user-space) RDMA protocol---by mapping the \projecttitle{} device to a user-space addresses range, the Mapped REGs pages. \projecttitle{} reserves one page at the page granularity of our system for each connected device that is represented as pseudo-devices in {\tt /dev/fpga<ID>}. Read and write access to the pseudo-device is equal to accessing the control and status registers of the FPGA. Applications directly interact with the control path of the \projecttitle{} hardware bypassing the host OS.

\subsection{RDMA OS Abstractions} 
The RDMA OS abstractions are a user-space library that enables the networking operations in the \projecttitle{} hardware, bypassing the OS kernel for performance. 
As shown in Figure~\ref{fig:network_stack_design}, the RDMA OS library is comprised of two parts: \textit{the network (RDMA) library} (colored in purple) that implements the software part of the RDMA protocol and \textit{the OS library} (colored in red) that schedules the \projecttitle{} requests. 

% As shown in Figure~\ref{fig:network_stack_design}, invocations of the \projecttitle{} programming API calls into the RDMA OS library are comprised of two parts: \textit{the network (RDMA) library} (colored in purple) that implements the software part of the RDMA protocol and \textit{the OS library} (colored in red) that schedules the \projecttitle{} requests. 

\myparagraph{Network (RDMA) library}  The network (RDMA) library includes all the logic and data (e.g., Tx/Rx queues per connection, local and remote memory addresses, RDMA keys that denote memory access permissions) required to implement the RDMA protocol. It executes the application's networking operations by posting the requests to the hardware. More specifically, it creates an internal representation of the request and the associated data and metadata (i.e., request opcode, remote IP, source/destination addresses, data length, etc.) and writes them into specific offsets in the REGs pages to update the control registers of the \projecttitle{} hardware. 
% In case there are no empty transmission queues, the ibv library {\tt yields} until there is an empty slot in the queue.
% (i.e., request opcode, remote IP, local address, remote base address, offsets, data length, etc.), encodes all related parameters, and finally, 

As shown in Figure~\ref{fig:network_stack_design}, the transmission and reception of requests and responses mandate the allocation of application network buffers. 
We adopt memory management similar to that in widely used user-space networking libraries~\cite{erpc, dpdk, rdma}.
% We adopt similar memory management as in widely-used user-space networking libraries~\cite{erpc, dpdk, rdma}. 
Importantly, the network buffers need to be mapped to a specific \projecttitle{}-memory, called the ibv memory. The ibv memory area is allocated at the connection creation in the {\tt huge page} area by the application through the ibv library. It resides within the application's address space with full read/write permissions and is eligible for DMA transfers. 
% without involving the CPU. 
% It resides within the application's address space with full read/write permissions, it is eligible for DMA transfers, and it is registered to the \projecttitle{} for remote reads and writes without involving the CPU. 

\myparagraph{OS library} The \projecttitle{}-OS library is responsible for scheduling the requests and ensuring isolated access to the mapped REG pages. 
% The \projecttitle{}-OS library is a lightweight user-space library responsible for scheduling the requests and ensuring isolated access to the mapped REG pages.
For performance, the \projecttitle{} data path eliminates unnecessary data copies throughout the network stack; the data to be sent is directly fetched by the hardware through DMA transfers. The OS library creates a \projecttitle{}-process object to represent each \projecttitle{} device. This \projecttitle{}-process in \projecttitle{} is not a separate scheduling entity (i.e., a thread as in classical OSes). 
In contrast, it is an object handle, exposed to the ibv library but managed by the \projecttitle{}-OS library that acquires locks on the respective REG pages to ensure isolated access to the \projecttitle{} hardware.

% In fact, the data to be sent (shown in blue lines in Figure~\ref{fig:network_stack_design} is directly fetched from the hardware through DMA transfers. 
\section{\rev{(a)}{\projecttitle{} Network Library}}
% \section{Trusted Network Library}
\label{sec:t-nic-software}
We present \projecttitle{}'s programming APIs ($\S$~\ref{sec:net-lib}), and a generic recipe to transform existing distributed systems ($\S$~\ref{subsec:transformation}).

\begin{table}
\begin{center}
\footnotesize
\begin{tabular}{ |c|c| } 
 \hline
 \multicolumn{2}{|c|}{{\bf Initialization APIs}} \\ [0.5ex] \hline 
 {\tt ibv\_qp\_conn()} & Creates an ibv struct \\
 {\tt alloc\_mem()}    & Allocates host ibv memory \\
 {\tt init\_lqueue()}    & Registers local memory to \projecttitle{} \\
 {\tt ibv\_sync()} & Exchanges ibv memory addresses \\
 \hline
 \hline
 \multicolumn{2}{|c|}{{\bf Network APIs}} \\ [0.5ex] \hline 
 {\tt local\_send/verify()}    & Generates/verifies attested messages \\
 %{\tt local\_verify()}    & Verifies a message \\
 {\tt auth\_send()} &  Transmits an attested message \\
 {\tt poll()} &  Polls for incoming messages \\
 {\tt rem\_read/write()} &  Fetches/writes remote memory \\
 \hline
\end{tabular}
\end{center}
%\vspace{-10pt}
\caption{\projecttitle{} programming APIs.}
\label{tab:apis}
\vspace{-10pt}
\end{table}

\subsection{Programming APIs}
\label{sec:net-lib}
\projecttitle{} implements a programming API (Table~\ref{tab:apis}) that resembles the traditional RDMA programming API~\cite{erpc} used in modern distributed systems\cite{f04eb9b864204bab958e72055062748c, farm, hermes, rdma-design, rdma-scale, octopus, Mitchell2013}. We further extend the security semantics, offering the properties of non-equivocation and transferable authentication  ($\S$~\ref{subsec:trustworthy_ds}). %Our API is shown in Table~\ref{tab:apis}.
%as the lower bound to build trustworthy distributed systems
% Our API, shown in Table~\ref{tab:apis}, allows system designers to leverage modern RDMA-enabled networks as they have been proven to be the most efficient design in modern distributed systems\cite{f04eb9b864204bab958e72055062748c, farm, hermes, rdma-design, rdma-scale, octopus, Mitchell2013}.  
%In contrast to traditional RDMA, our \projecttitle{} extends the security semantics, offering the properties of non-equivocation and transferable authentication as the lower bound to build trustworthy distributed systems ($\S$~\ref{subsec:trustworthy_ds}).

\myparagraph{Initialization APIs} 
The \projecttitle{} application first needs to configure the \projecttitle{} system to establish peer-to-peer RDMA connections. % using the initialization APIs. 
% Prior to exchanging network messages, the \projecttitle{} system needs to be configured by the application to establish RDMA connections. 
The application creates one ibv struct for each connection with {\tt ibv\_qp\_conn()}, which sets up and stores the queue pair, the local and remote IP addresses, device UDP ports, and others. 
%The application also invokes {\tt alloc\_mem()} to allocate the ibv memory and then calls into {\tt init\_lqueue()} that initializes the local queue pairs and registers the ibv memory to the \projecttitle{} hardware. 
The application also invokes {\tt alloc\_mem()} to allocate the ibv memory and then register the ibv memory to the \projecttitle{} hardware. 
Lastly, the application synchronizes with the remote machine using {\tt ibv\_sync()} to exchange necessary data (e.g., ibv memory address, queue pair numbers).
% Lastly, the application synchronizes itself with the remote machine using {\tt ibv\_sync()} to exchange necessary data (ibv memory address, queue pair numbers, etc.). 
%which exchanges important information such as the ibv memory address, queue pair numbers, etc. 
% In addition, the application needs to allocate the ibv memory that will be registered to the \projecttitle{} hardware. The code calls into the {\tt init\_lqueue()} that initializes the local queue pairs and registers the ibv memory to the \projecttitle{}. 

% Once the application configures its \projecttitle{} appropriately, it synchronizes itself with the remote side using {\tt ibv\_sync()}. 
% This function communicates with the remote machine and exchanges important information such as its registered memory base address, queue pair numbers, etc. 

%After all these steps are complete, the application can use the \projecttitle{} network API to exchange data in a secure manner. 

\myparagraph{Network APIs}
\projecttitle{} executes trusted one-sided, reliable RDMA with the same reliability guarantees as the classical one-sided RDMA over Reliable Connection (RC), i.e., a FIFO ordering (per connection), similar to TCP/IP networking. % The networking stack imposes 

% We expose to the user the {\tt auth\_send()} operation which sends an attested message by implementing an RDMA reliable write of the message concatenated with its attestation. 
\projecttitle{} offers {\tt auth\_send()} to send an attested message with RDMA reliable writes.
% that concatenates the message with its attestation. 
We support classical RDMA operations for reads and writes: {\tt rem\_read()} and {\tt rem\_write()}. The remote side runs {\tt poll()} to fetch the number of completed operations; {\tt poll()} is updated only when the message verification succeeds at the \projecttitle{} hardware. We offer local operations for generating and verifying attested messages within a single-node setup: {\tt local\_send()} and {\tt local\_verify()}.
% In contrast to classical RDMA, we have not used background threads for polling. The function returns the number of the completed operations. 
% The poll function is only updated when the verification at the \projecttitle{} hardware has been successful. 

\projecttitle{} does not support a hardware-assisted multicast, but it can offer equivocation-free multicast uni-casting the same attested message generated by {\tt local\_send()} as in~\cite{levin2009trinc}.
 % Even in the standard RDMA libraries, hardware-based multicast is barely used in practice~\cite{f04eb9b864204bab958e72055062748c}. %, whereas it is only supported over unreliable connections.

\subsection{A Generic Transformation Recipe}
\label{subsec:transformation}
\myparagraph{Transformation properties} 
We show how to use \projecttitle{} APIs to transform an existing (CFT) distributed system into one that targets Byzantine settings. Our transformation is defined as wrapper functions on top of the send and receive operations~\cite{clement2012}. For safety, our transformation needs to meet the following properties to provide the same guarantees expected by the original CFT system~\cite{clement2012, making_distributed_app_rob, 268272}:
% We demonstrate how to use \projecttitle{} programming API to construct a generic transformation of a distributed system operating under the CFT model to a system that targets Byzantine settings.  Our transformation is defined as a set of wrapper functions on top of the send and receive operations~\cite{clement2012}. For safety, our transformation further needs to meet the following properties in order to provide the same guarantees that are expected by the original CFT system~\cite{clement2012, making_distributed_app_rob, 268272}.

\noindent\underline{{\bf{Safety.}}} If a correct receiver receives a message $m$ from a correct sender $S$, then $S$ must have sent with $m$.

\noindent\underline{{\bf{Integrity.}}} If a correct receiver receives and delivers a message $m$, then $m$ is a {\em valid} message.
%\begin{itemize}[leftmargin=*]
%    \item {\bf{Safety.}} If a correct receiver receives a message $m$ from a correct sender $S$, then $S$ must have sent with $m$.
%    \item {\bf{Integrity.}} If a correct receiver receives and delivers a message $m$, then $m$ is a {\em valid} message.
%\end{itemize}

\if 0
{\small
\begin{lstlisting}[frame=h,style=customc,
                    label={lst:transformation},
                    caption= Generic application send and recv wrapper functions for transformation using \projecttitle{}: blue sections (native) and orange sections (\projecttitle additions).]
// generic send wrapper function
void send(P_id, char[] msg) {
    // current state
@\Hilight@    state = hash(my_state);
@\Hilight@    tx_msg = msg || state || prev_receiver_msg;
@\HilightYlinewidth@    auth_send(follower, tx_msg);
}
// upon receiving a send(follower, msg) message
void recv(recv_msg) {
@\HilightYlinewidth@ // message authenticity and integrity checked in TNIC hw
auto [attestation, msg || state || prev_receiver_msg] = deliver();
@\HilightYlinewidth@    [msg, cnt] = verify_msg(msg);
@\Hilight@    verify_sender_state(state);
@\HilightYlinewidth@    local_verify(prev_receiver_msg);
@\Hilight@    verify_system_view(prev_receiver_msg);
@\Hilight@    apply(msg);
}
\end{lstlisting}
}
\pramod{should we it to the appendix and summarize with a forward pointer?}
\fi

{\small
\begin{lstlisting}[frame=h,style=customc,
                    label={lst:transformation},
                    caption= Generic send and recv wrapper functions using \projecttitle{}. \projecttitle additions are highlighted in orange.]
void send(P_id, char[] msg) {
    state = hash(my_state);  
    tx_msg = msg || state || receiver_state;
@\HilightYlinewidth@    auth_send(follower, tx_msg);
}
void recv(recv_msg) {
    auto [att, msg || state || receiver_state] = deliver();
@\HilightYlinewidth@    [msg, cnt] = verify_msg(msg);
    verify_sender_state(state);
@\HilightYlinewidth@    local_verify(receiver_state);
    verify_system_view(receiver_state); apply(msg);
}
\end{lstlisting}
}
\vspace{-1mm}

Listing~\ref{lst:transformation} shows our proposed {\tt send} (L1-5) and {\tt recv} (L7-13) operations, providing a general method for transforming a CFT system into a BFT system. 
We assume a two-node scenario where the first node (sender) receives client requests and forwards them to the second node (receiver). For transmission, the sender sends the client message {\tt msg}, its current state (e.g., the sender's action to the \texttt{msg}), and the receiver's previous state (known to the sender). 
% To avoid replaying the entire message history in order to reconstruct the system's state (as done in \cite{clement2012}), we require that nodes simulate the sender's state to verify that the sender’s response to the request is as expected
The receiver's state is optional depending on the consistency guarantees of the derived system and can be used to ensure that both nodes have the same system view. 

% Listing~\ref{lst:transformation} shows our proposed generic {\tt send} (L2-7) and {\tt recv} (L10-18) operations that we applied in a scenario between two nodes where the first node (sender) receives client requests and forwards the requests to the second node (receiver) who is following the sender. For each transmitted message ({\tt tx\_msg}), the sender sends the (client) msg, its current state (e.g., sender's output as a response to the current client msg), and the follower's previous state known to the sender. The last piece of data is optional depending on the consistency guarantees of the derived system and can be used to ensure that both the sender and the receiver have the same view of the system (L4-6). 

Upon receiving a valid message (L8-9), the receiver {\em simulates} the sender's state to verify that the sender's action to the request is as expected (L10). \rev{A5, (b)}{The way to simulate the states depends on the applications, e.g., in our BFT protocol implementation ($\S$~\ref{sec:use_cases}), each replica maintains copies of counters that represent the expected counter values for all other participating nodes. The simulation allows nodes to avoid replaying the entire message history in order to reconstruct the system's state, as done in \cite{clement2012}.} The receiver also ensures that it does not lag, and both nodes have the same ``view'' of the system inputs by verifying that the sender has {\em seen} the receiver's latest state (e.g., message) (L11-12).
% \rev{A5}{The state of the sender/receiver depends on the applications, e.g., in our BFT protocol implementation ($\S$~\ref{sec:use_cases}), each replica maintains copies of counters that represent the expected counter values for all other participating nodes. }
% The receiver, on the other hand, {\em simulates} the sender's state to verify its actions. Importantly, upon a valid message reception (L12-13), the receiver verifies that the sender's action as a response to the client's request is the expected (L14). In addition, the receiver ensures that it is not lagging behind and it has the same ``view'' of the system inputs as the sender by verifying that the sender has {\em seen} the receiver's latest sent message (L15-16).

% \myparagraph{Correctness arguments for the transformation} 
Our generic transformation satisfies the requirements listed above. First, \projecttitle{}'s transferable authentication property directly satisfies the safety requirement. A faulty node cannot impersonate correct nodes; if \projecttitle{} validates a message $m$ from a sender, the sender must have sent $m$. \projecttitle{}'s reliable network operations ensure liveness properties between correct nodes. % are , where the \projecttitle{} hardware transparently retries failed sends,
% Next we explain how our generic transformation recipe satisfies the listed above requirements. First, the safety requirement is directly satisfied by \projecttitle{}'s transferable authentication property. A faulty node cannot impersonate a correct node; if \projecttitle{} validates and delivers a message $m$ from a sender, then the sender must have sent $m$. \projecttitle{} reliable network operations, where the \projecttitle{} hardware transparently retries failed sends ensures liveness properties between correct nodes. 
% 
Second, our transformation satisfies the integrity property. \rev{A7, (b)}{The integrity property is ensured by validating that the sender’s response to the client’s request follows the protocol specification. The transferable authentication mechanism allows correct receivers to prove the integrity flow by simulating the sender's output and state, e.g., by maintaining a copy of the sender’s state. }
\myparagraph{Consistency property for replication} 
Our transformation further needs to meet the consistency property~\cite{clement2012}. If correct receivers $R_1$ and $R_2$ receive valid messages $m_i$ and $m_j$  respectively from sender $S$, then either (a) $Bpg_{i}$ is a prefix of $Bpg_{j}$ , (b) $Bpg_{j}$ is a prefix of $Bpg_{i}$, or (c) $Bpg_{i}$ = $Bpg_{j}$ (where $Bpg_{x}$ is the process behavior that supports the validity of message $m_x$).

%for the specific case of the CFT replication protocols
% to transform distributed systems for Byzantine settings:
% \begin{itemize}[leftmargin=*]
% \item \myparagraph{Consistency} If correct receivers $R_1$ and $R_2$ receive valid messages $m_i$ and $m_j$  respectively from sender $S$, then either (a) $Bpg_{,i}$ is a prefix of $Bpg_{,j}$ , (b) $Bpg_{,j}$ is a prefix of $Bpg_{,i}$, or (c) $Bpg_{,i}$ = $Bpg_{,j}$ (where $Bpg_{,x}$ is the process behavior that supports the validity of message $m_x$).
% \end{itemize}

The consistency requirement is enforced through the \projecttitle{}'s non-equivocation primitive that assigns a (unique) monotonic sequence number to each outgoing message, enforcing a total order on the sender's outgoing messages. 
\rev{D2, (b)}{Along with the integrity requirement, the total order can prevent equivocation and suffice for consistency. } 
% \atsushi{@Dimitra: can we detail more why the total order can suffice for consistency? }
Importantly, \projecttitle{} ensures that correct receivers cannot miss any past messages. Following this, two followers that receive from the same sender (using the equivocation-free multicast discussed in $\S$~\ref{subsec:net_lib}) follow the same transition (execution) path. 
\rev{D2, (b)}{\projecttitle{} cannot transform systems with non-deterministic specifications.}
%\rev{D2, (b)}{\projecttitle{} cannot transform systems that might have non-deterministic outputs for a given input.} %\atsushi{@Dimitra: D2: Can you check the text? I think they should be more detailed}
% Our suggested transformation based on the sender's state simulation (leader) can be effectively applied to state machine replication.

% The consistency requirement is enforced through the \projecttitle{}'s non-equivocation primitive that assigns a (unique) monotonic sequence number to each outgoing message, enforcing a total order on the sender's outgoing messages. Along with the integrity requirement, the total order suffices for consistency. 
% Importantly, \projecttitle{} ensures that correct receivers cannot miss any past messages. Following this, two followers that receive from the same sender (using the equivocation-free multicast discussed in $\S$~\ref{subsec:net_lib}) follow the same transition (execution) path. 

\section{Trusted Distributed Systems}
\label{sec:use_cases}

%\pramod{@Dimitra - please make a carefull pass over the text, please cite all related seminal papers, and also explain how TNIC helps in improving these systems explicitly. }

Using \projecttitle{}, we transform the following four distributed systems into BFT systems (see Appendix $\S$~\ref{sec:use_cases-appendix} for details).  
% in Appendix~\ref{sec:use_cases-appendix}.

% \antonis{Below we use \projecttitle{}-A2M, but further down, we do not! (i.e., no \projecttitle{}-BFT, \projecttitle{}-CR..)}

\myparagraph{Attested Append-Only Memory (A2M)} We design an Attested Append-Only Memory (A2M)~\cite{A2M} leveraging \projecttitle{}, which can be used to shield and optimize various systems~\cite{AbdElMalek2005FaultscalableBF, Castro:2002, Li2004, 10.5555/1298455.1298473}. The original A2M, and hence our implementation over \projecttitle{}, builds append-only (trusted) logs, associating each entry with a monotonically increasing sequence number to combat equivocation. While A2M has a large TCB and ports the log within the TEE, our implementation has only a minimal TCB in hardware and it can robustly store the log in the untrusted host memory, improving memory efficiency~\cite{levin2009trinc}.

As in the original A2M, we build the \texttt{append} and \texttt{lookup} operations. The \texttt{append} calls into  \projecttitle{} to generate an attestation for the log entry while associating it with a monotonically increased sequence number ({\tt sent\_cnt}). The sequence number denotes the entry's position in the log. The \texttt{lookup} operation retrieves entries locally without verification.
%, and the \texttt{truncate} operation deletes entries based on a provided sequence number. The append operation involves adding entries with a sequence number, context, and an authenticator field. %The lookup operation retrieves log entries without verification, while the truncate operation removes entries based on sequence numbers and utilizes a \textsc{manifest} log to keep track of state changes.

\myparagraph{Byzantine Fault Tolerance (BFT)} 
We design a Byzantine Fault-Tolerant protocol (BFT) using \projecttitle{}. The protocol builds a replicated counter as a foundational service for various systems~\cite{rafthyperledger, Kafka, boki, 10.1145/3286685.3286686, scalog}. Our system model considers a network of replicas with at most $f$ Byzantine replicas out of $N=2f+1$ total replicas. One replica serves as the leader, and the others act as followers. 
The system prevents equivocation through \projecttitle{}, which enforces and validates the ordering of messages. This reduces the number of replicas required and the message complexity compared to the classical BFT ($3f+1$).
% , thus cutting deployment costs and message complexity.

%The protocol operates in a partial synchrony model for liveness and assumes deterministic protocol specifications.

Clients send increment counter requests to the leader, who performs the requests and broadcasts the change along with a {\em proof of execution} (PoE) message to followers. The proof of execution is a \projecttitle{}-attested message with the original client's request, the leader's counter value, and metadata. The followers leverage their local state machine to detect a faulty leader (or follower)~\cite{268272}. Subsequently,
if and only if a follower has not applied the message before, it applies the incremented counter value to its state machine before forwarding its own PoE message to all other replicas and replying to the client.
% , who will also validate the followers' outputs. 
A quorum of at least $f+1$ identical messages from different replicas guarantees a correctly committed result for the client. %Overall, \projecttitle{} optimizes the replication factor and message complexity of BFT.

\myparagraph{Chain Replication (CR)} 
We design a Byzantine CR system~\cite{10.1007/978-3-642-35476-2_24} using \projecttitle{} as the replication layer of a Key-Value store. As in the CFT version of CR, the replicas, e.g., head, middle, and tail, are connected in a chained fashion. 
%We assume a centralized configuration service for error detection and reconfiguration, which always provides clients with a correct configuration.
%The system model assumes Byzantine fault tolerance with a centralized configuration service for error detection and reconfiguration. The head triggers reconfiguration if it intentionally fails to forward messages.

Clients execute requests by forwarding them to the head. The head orders and executes the request, creating his own {\em proof of execution message} (PoE), which is sent along the chain. The PoE consists of the original request and the head's output that \projecttitle{} attests. Each node in the chain verifies the previous node's PoE, executes the request, and creates its own PoE, which consists of the last PoE and the node's output. 

%Unlike the CFT CR assuming that cryptographic operations on the CPU are not compromised, local operations in the tail cannot be trusted in the BFT model.
\rev{D5}{Unlike the CFT CR, local operations in the tail (e.g., reads) are untrusted in the BFT model. Therefore, all operations must traverse the entire chain. Replicas reply to clients with their output after forwarding their PoE message, and clients wait for identical replies from all chained nodes. We discuss the performance-security trade-offs of an alternative TEE-based design of porting the entire CR protocol into the TEE (that would allow clients to read only from the tail) in $\S$~\ref{subsec:use_cases_eval}.} %While such a system , it targets a weaker threat model compared to \projecttitle{}.}
%Such a system would adhere to the protocol specification, with clients only needing to communicate with the tail. However, this design would target a weaker threat model compared to \projecttitle{}.
% Unlike the CFT CR, all operations must traverse the chain as local operations in the tail cannot be trusted. Replicas reply to clients with their output after forwarding their PoE message. Clients wait for identical replies from all chained nodes.
% % For \texttt{get} requests, clients traverse the chain or consult the majority, broadcasting the request to $f+1$ replicas, including the tail.

% \rev{D5}{We base our protocol implementation on~\cite{10.1007/978-3-642-35476-2_24}, where operations must traverse the entire chain, similar to Chain Replication for the Crash Fault Model. While \cite{10.1007/978-3-642-35476-2_24} assumes that cryptographic operations on the CPU are not compromised, we have implemented the system practically using \projecttitle{}. A hypothetical TEE-based design would involve porting the entire Chain Replication protocol into the TEE. Such a system would adhere to the protocol specification, with clients only needing to communicate with the tail, as in CFT Chain Replication. However, this hypothetical design would target a weaker threat model compared to \projecttitle{}.}

\myparagraph{Accountability (PeerReview)}
Lastly, we design an accountability system with \projecttitle{} based on the PeerReview system~\cite{peer-review} to {\em detect} malicious actions in large deployments~\cite{nfs, 10.1145/1218063.1217950}.  We detect faults impacting the system's network messages logged into the participant's tamper-evident log. We frame the protocol within an overlay multicast protocol for streaming systems where the nodes are organized in a tree topology. Witnesses assigned to each node audit the node's log to detect faults or non-responsive nodes. The witnesses replay the log entries, comparing them with a reference deterministic implementation to identify inconsistencies. 
Our \projecttitle{} prevents equivocation at NIC hardware efficiently, which eliminates the expensive all-to-all communication of the original PeerReview that does not use trusted hardware~\cite{levin2009trinc}.
% Moreover, \projecttitle{} optimizes accountability by efficiently handling equivocation.~\cite{trinc}

\begin{figure*}[t!]
%\begin{center}
\minipage{0.33\textwidth}
%\begin{figure}
    \centering
     \vspace{-3mm}
    \includegraphics[width=\linewidth]{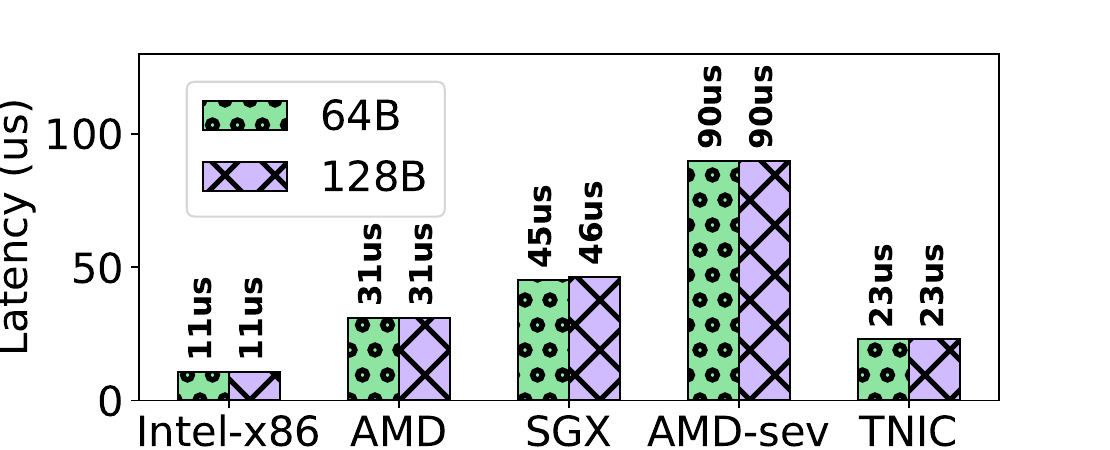} 
  \caption{{\tt Attest} function latency.}
  \label{fig:attest_kernel}
%\end{figure}
\endminipage
\minipage{0.33\textwidth}
%\begin{figure}
    \centering
     \vspace{-3mm}
  \includegraphics[width=\linewidth]{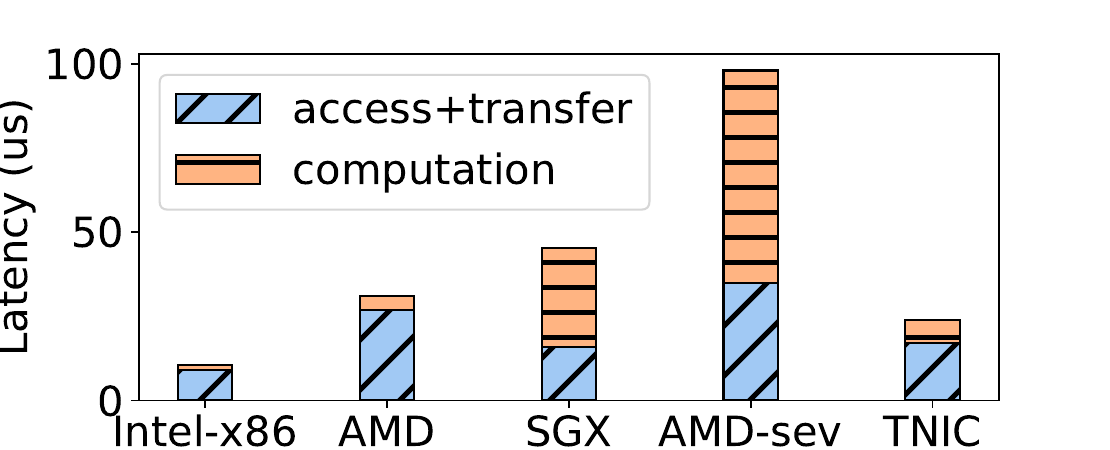}
  \caption{{\tt Attest} latency breakdown.}\label{fig:latency_breakdown}
%\end{figure}
\endminipage
\minipage{0.33\textwidth}
%\begin{figure}
    \centering
  \includegraphics[width=\linewidth]{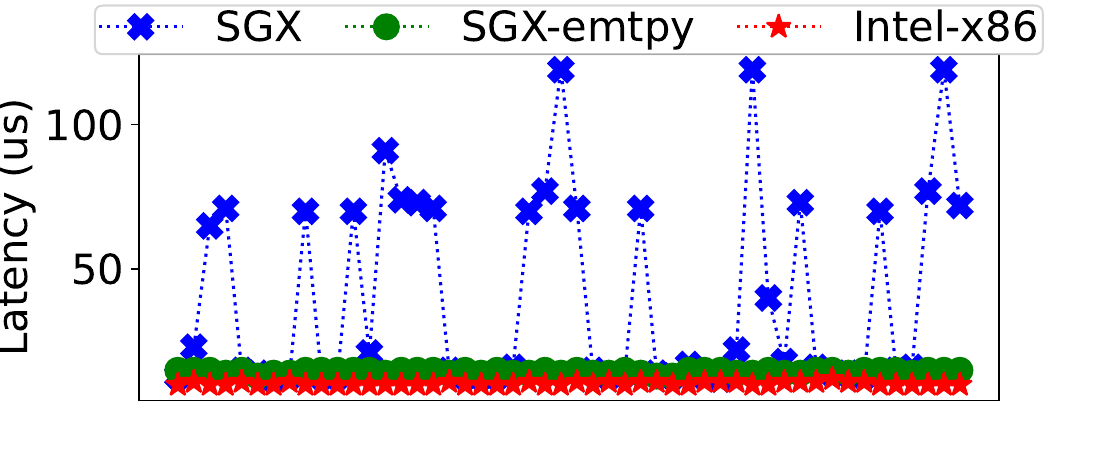}
    \vspace{-7mm}
  \caption{Latency over time (SGX).}\label{fig:latency_distribution}
%\end{figure}
\endminipage
\vspace{-3mm}
%\end{center}
\end{figure*}

\begin{table}
\begin{center}
\footnotesize
\begin{tabular}{ |c|c|c| } 
 \hline
 System &  (host) TEE-free & Tamper-proof \\ [0.5ex] \hline \hline
 SSL-lib & Yes & No\\
 SSL-server/Intel-x86*/AMD  &  Yes & No\\
 SGX/AMD-sev &  No & Yes\\
% TEE-P [Hybster, Damysus, Trinc] & Yes & Yes & Yes & Assumes trusted local persistent state\\
 %TEE-DS & Yes & Yes & Yes & Builds a DS of $f+2u+1$ TEEs\footnote{$f$ is the compromised TEEs and $u$ is the number of unresponsive TEEs}\\
 \projecttitle{} & Yes & Yes\\
 \hline
\end{tabular}
\end{center}
%\vspace{-10pt}
\caption{Host-sided baselines and \projecttitle{}. (*) We use the term SSL-server for this run unless stated otherwise.}
% \caption{(Trusted) Host-sided hardware baselines and \projecttitle{}. (*) We also use the term SSL-server for this run unless stated otherwise.}
\label{tab:hw_options}
\vspace{-8pt}
\end{table}

\section{Evaluation}
\label{sec:eval}

We evaluate \projecttitle{} across three dimensions: {\em (i)} 
hardware (\S~\ref{subsec:hw_eval}), {\em (ii)} network stack (\S~\ref{subsec:net_lib}) and {\em (iii)} distributed systems (\S~\ref{subsec:use_cases_eval}).

\myparagraph{Evaluation setup}
We perform our experiments on a real hardware testbed using two clusters: AMD-FPGA Cluster and Intel Cluster. AMD-FPGA Cluster consists of two machines powered by AMD EPYC 7413 (24 cores, 1.5 GHz) and 251.74 GiB memory. Each machine also has two Alveo U280 cards~\cite{u280_smartnics} that are connected through 100 Gbps QSFP28 ports. Intel Cluster consists of three machines powered by Intel(R) Core(TM) i9-9900K (8 cores, 3.2 GHz) with 64 GiB memory and three Intel Corporation Ethernet Controllers (XL710).

\subsection{Hardware Evaluation: \trustedfpga{}}
\label{subsec:hw_eval}

\myparagraph{Baselines} 
We evaluate the performance of {\tt Attest()} of the \projecttitle{}'s attestation kernel  ($\S$~\ref{subsec:nic_attest_kernel}) compared with four host-sided systems shown in Table~\ref{tab:hw_options}. For these host-sided versions, we build OpenSSL \rev{C3}{v3.1} servers that run natively or within a TEE \rev{C3}{with the same BIOS configuration (AES-NI enabled)}. The servers attest and forward network messages to the host application. We use the terms Intel-x86 and AMD for a native run of the server process (SSL-server) and SGX and AMD-sev for their tamper-proof versions within a TEE. 
\rev{B5}{The TEE baselines follow the same system model as in state-of-the-art hybrid systems~\cite{10.1145/3492321.3519568, minBFT, 10.1145/2168836.2168866, levin2009trinc}, where the host BFT application runs on the untrusted CPU and communicates with a separate TEE-based process to generate and verify message attestations.}
\rev{D6}{\projecttitle{} implements similar abstractions for counter and message attestation. Thus, \projecttitle{} does not introduce additional protocol alterations compared to them.}%as those used in the hybrid systems.
% (SGX, AMD-sev)
The server and host process run in the same machine to eliminate network latency and communicate through TCP sockets. We implement SGX using the {\sc scone} framework~\cite{scone} while AMD-sev runs in a QEMU VM using the official VM image~\cite{AMDSEV}. In addition, we build (non-temper-proof) SSL-lib, which integrates the {\tt Attest} function as a library. 

\myparagraph{Methodology and experiments}
We use Vitis XRT v2022.2 and the shell \texttt{xilinx\_u280\_gen3x16\_xdma\_base\_1} for the stand-alone evaluation of the \projecttitle{} attestation kernel: synchronous data transfers between the host and device (U280). We measure and report the average latency and communication costs by executing an empty function body of \texttt{Attest()}.
%To isolate the latencies between data transfers and computation we further execute the same experiment without computing the HMAC (empty kernel). 

% We use Vitis XRT v2022.2 for the stand-alone evaluation of \projecttitle{} attestation kernel. We load the shell \texttt{xilinx\_u280\_gen3x16\_xdma\_base\_1} and use Vitis XRT synchronous data transfers from host to device and vice versa. We measure and report the {\tt Attest} function average latency as well as the communication costs executing an empty funcation body.

%We use Vitis XRT v2022.2 for the stand-alone evaluation of \projecttitle{} attestation kernel on top of which we build the host and the \trustedfpga{} processes (\texttt{xilinx\_u280\_gen3x16\_xdma\_base\_1}). We use Vitis XRT synchronous data transfers from host to device and vice versa. We measure and report the HMAC average latency on these competitive systems. To isolate the latencies between data transfers and computation we further execute the same experiment without computing the HMAC (empty kernel). \atsushi{I wanna simplify the texts here... work on it later}

\myparagraph{Results}
Figure~\ref{fig:attest_kernel} shows the average latency of {\tt Attest()} based on the HMAC algorithm for 64B and 128B data inputs. The latency of {\tt Verify()} is similar, and as such, it is omitted. Our \projecttitle{} achieves performance in the microseconds range (23 us) and outperforms its equivalent TEE-based competitors at least by a factor of 2. Importantly, \projecttitle{} is approximately 1.2$\times$ faster than AMD, which is not tamper-proof. 
% although it is approximately 2$\times$ slower than Intel-x86. Recall that neither AMD nor Intel-x86 are tamper-proof.

Figure~\ref{fig:latency_breakdown} shows the latency breakdown of {\tt Attest()}. Accessing the \projecttitle{} device and TEEs can be expensive, ranging from 30\% to 90\% of the total operation latency among the systems. 
Regarding \projecttitle{}, the transfer time (16us) accounts for 70\% of the execution time. We expect that \projecttitle{} effectively eliminates this cost by enabling asynchronous (user-space) DMA data transfers. 
% Specifically, for \projecttitle{}, the transfer time takes about 16 us, which accounts for 70\% of the execution time. This is not a concern in \projecttitle{} as it effectively eliminates this cost by enabling asynchronous (user-space) DMA data transfers. 
% Regarding the native runs, i.e., Intel-x86 and AMD, the communication costs account for $\sim$90\% of the latency.  
% Figure~\ref{fig:latency_breakdown} shows the latency breakdown of {\tt Attest()}. Accessing the \projecttitle{} device and the TEEs can be expensive ranging from 30\% to 90\% of the total operation latency among the systems. Specifically, for \projecttitle{}, the transfer time takes about 16 us, which accounts for 70\% of the total execution time. This is not a concern in \projecttitle{} as it effectively eliminates this cost by enabling asynchronous (user-space) DMA data transfers. Regarding the native runs, i.e., Intel-x86 and AMD, the communication costs including the syscalls execution and data transfers between the two processes account for the $~$90\% of the latency.  
% 
Regarding the TEE-based systems (SGX, AMD-sev), the communication and system call execution costs account for up to 40\% of the total execution. To our surprise, this implies that the HMAC computation within any of the two TEEs experiences more than 30$\times$ overheads compared to its native run. To analyze TEEs' behavior, we instrument the client's code to measure the operations' individual latency at various periods of time during the experiment accurately. 
% Our evaluation shows that in the TEE-based systems, e.g., SGX and AMD-sev, the communication and system call execution costs account for up to 40\% of the total execution (on average). To our surprise, this implies that the HMAC computation itself within any of the two TEEs experiences more than 30$\times$ overheads compared to its native run. We further analyzed TEEs' behavior instrumenting the client's code to accurately measure the operations' individual latency at various periods of time during the experiment. 

Figure~\ref{fig:latency_distribution} illustrates the individual operation latency, where SGX-empty indicates SGX without HMAC computation. 
% for three systems: SGX, SGX-empty (SGX without HMAC computation), and Intel-x86. 
% SGX (the SSL-server runs in the SGX enclave), SGX-empty (the SGX SSL-server without HMAC computation) and Intel-x86 (the SSL-server runs natively). 
As shown in Figure~\ref{fig:latency_distribution}, the HMAC execution within the TEE often experiences huge latency spikes. 
% quite often experiences huge spikes in latency. 
% While these spikes are very frequent, they are not guaranteed. 
% We calculate an average mean of 45us and a geometric mean of 30 us. 
\rev{A6}{We attribute this behavior to the scheduling effects and asynchronous exitless system calls inherent in our SGX framework, {\sc scone}~\cite{scone}. We observe similar latency variations during executions on AMD systems, spiking up to 200--500 us. }

\begin{figure}
    \centering
   \includegraphics[width=0.75\linewidth]{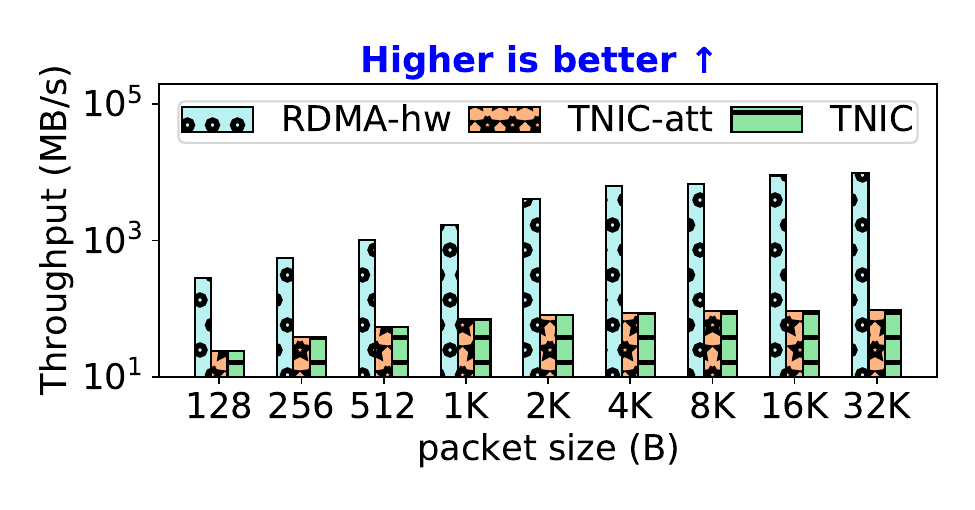}
   \vspace{-10pt}
    \caption{Throughput of send operations across the three selected network stacks.}
    % \caption{Throughput evaluation of send operations for various packet sizes across five competitive network stacks with various security properties.}
    \vspace{-4pt}
  \label{fig:net_throughput}
\end{figure}

\begin{figure*}
    \centering
   \includegraphics[width=0.80\linewidth]{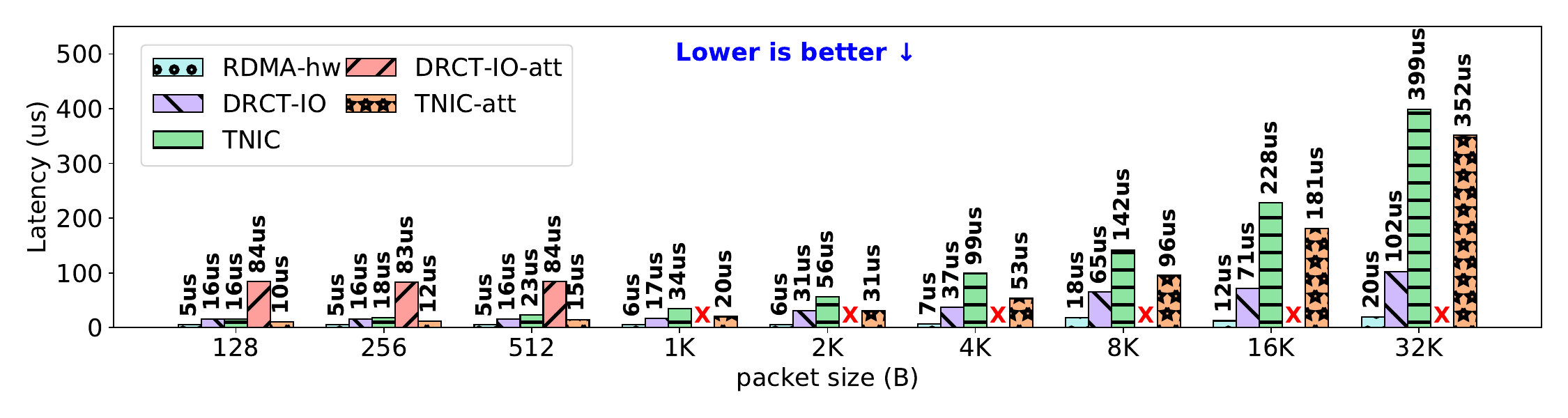} 
   \vspace{-12pt}
  \caption{Latency of send operations across five competitive network stacks with various security properties.}
  % \caption{Latency evaluation of send operations for various packet sizes across five competitive network stacks with various security properties.}
  \label{fig:net_latencies}
   \vspace{-10pt}
\end{figure*}

%%%%%%%%

\vspace{-4pt}
\subsection{Software Evaluation: \projecttitle{} Network Stack}\label{subsec:net_lib}
\vspace{-2pt}

\myparagraph{Baselines} 
To evaluate the \projecttitle{} performance, we discuss (1) the effectiveness of offloading the network stack to the \projecttitle{} hardware and (2) the overheads incurred by the CFT systems transformation for the BFT model. We compare \projecttitle{} across four other software/hardware network stacks with different security properties as follows: 
% We evaluate the \projecttitle{} performance to show {\em{(i)}} the effectiveness of offloading the network stack from host CPUs to accelerators and {\em{(ii)}} the overheads that our system incurs due to materializing the requirements for CFT systems transformation (discussed in $\S~\ref{sec:requirements-ds}$). As such, we evaluate and compare \projecttitle{} across four other network stacks implemented on software or hardware with different security properties. 
%Specifically, we use the acronym D-I/O to refer to a Direct I/O network stack that bypasses the kernel stack (for performance). The acronyms D-I/O w/ A. and \projecttitle{} w/ A. means that the network stack generates and sends attested messages without verifying them at the receiving side. 
%\atsushi{What do you think new labels like this: RDMA-hw, DRCT-IO, DRCT-IO-acc, TNIC, TNIC-acc}
% 
% Specifically, we evaluate five different network stacks
{\em (i)} RDMA-hw, an untrusted RoCE protocol on FPGAs, {\em (ii)} DRCT-IO (direct I/O), untrusted software-based kernel-bypass stack, {\em (iii)} DRCT-IO-att, altered DRCT-IO that offers trust by sending attested messages but does not verify them, and {\em (iv)} \projecttitle{}-att, altered \projecttitle{} that similarly sends attested messages without verification. We build {\em (i)} RDMA-hw on top of Coyote~\cite{coyote} network stack similarly to \projecttitle{}. For {\em (ii) (iii)} DRCT-IOs, we base our design on eRPC~\cite{erpc} with DPDK~\cite{dpdk} that offers similar reliability guarantees with RDMA-hw. The hardware network stacks run on AMD-FPGA Cluster, whereas the rest run on Intel Cluster.

\myparagraph{Methodology and experiments} 
Our experiments measure the latency and throughput for respective network stacks, which run through a single-threaded client-server implementation.
For the latency measurement, the client sends one operation at a time, whereas for the throughput measurement, one client can have multiple outstanding operations.

\myparagraph{Results} 
Figure~\ref{fig:net_latencies} and~\ref{fig:net_throughput} show the latency and throughput of the send operation with various packet sizes. First, regarding (1) the effectiveness of network stack offloading, RDMA-hw is 3$\times$---5$\times$ faster than DRCT-IO, which indicates that the network offloading boosts performance. Although DRCT-IO offers minimal latency (16-16.6us) for small packet sizes up to 1~KiB due to its zero-copy transmission/reception optimizations~\cite{erpc}, they are only effective for up to 1460B (MTU is 1500B, but 40B are reserved for metadata), and RDMA-hw still achieves 3$\times$ lower latency (5-5.5us). For bigger data transfers, the RDMA-hw latency increases steadily up to 19~us, whereas DRCT-IO does not scale well with latencies up to 100us.

Second, regarding (2) the \projecttitle{} performance overhead, \projecttitle{} offers trusted networking with 3$\times$---20$\times$ higher latencies than the untrusted RDMA-hw. 
% The latencies of both \projecttitle{} and \projecttitle{}-att increase linearly with the packet size. Whereas, the (untrusted) RDMA-hw latencies remain stable (5---7us) for packet sizes up to 4KiB and are tripled thereafter (18---20us). 
The latency increase stems from the HMAC calculation of the \projecttitle{} hardware. As this algorithm fundamentally cannot be parallelized, the higher the message size, the higher the latency our \projecttitle{} incurs. More specifically, for packet sizes less than 1~KiB, doubling the packet size in \projecttitle{} results in a 13\%---20\% increment in the overall latency. For packet sizes bigger or equal to 1~KiB, doubling the packet size increases the latency by 30\%---40\%. 
Compared to DRCT-IO-att (82us), \projecttitle{} is up to 5.6$\times$ faster. Importantly, DRCT-IO-att reports extreme latencies (2000us or more) for packet sizes larger than 521B, which are omitted to avoid plot distortion. We attribute these latencies to the scheduling effects of {\sc scone}~\cite{scone}.

\subsection{Distributed Systems Evaluation}
\label{subsec:use_cases_eval}

\if 0

\fi
%We implement the four systems of $\S~\ref{sec:use_cases}$ with \projecttitle{} in a three-node setup ($N=3$) except for the single-node A2M system.
% the systems' properties where $N$ refers to the number of machines used for the protocol, and $f$ is the number of failures the system can tolerate.

We next evaluate four distributed systems described in $\S~\ref{sec:use_cases}$.% based on \projecttitle{}.

\myparagraph{Methodology and experiments} 
We execute all four of our codebases on Intel Cluster in three servers (as the minimum required setup capable of withstanding a single failure, $N=2f+1$, where $f=1$). \rev{E4}{We only use a single port of the U280 for network communication because of a limitation introduced in our system by the Coyote codebase~\cite{coyote}, on top of which we base \projecttitle{} implementation.} Due to our limited resources, we cannot install Alveo U280 cards on all these servers.  Instead, we build our codebase using the DRCT-IO stack (detailed in $\S$~\ref{subsec:net_lib}) and inject busy waits to emulate the delays incurred by \projecttitle{} for generating and verifying attested messages.
% in the \projecttitle{} system. %Our code uses busy waiting to accurately emulate latency rather than sleep functions. 
% We execute all four of our codebases on Intel Cluster, utilizing all its three servers (as the minimum required setup capable of withstanding a single failure, $N=2f+1$, where $f=1$). Furthermore, due to our limited resources, with only two U280 cards available and the physical separation of Intel Cluster from AMD-FPGA Cluster, we cannot access U280 cards for these specific experiments. Instead, we compiled our code using the DRCT-IO network stack as previously detailed in $\S$~\ref{subsec:net_lib} and do busy waiting to accurately replicate the \projecttitle{} delays within the CPU for generating and verifying attested messages.% in the \projecttitle{} system. %Our code uses busy waiting to accurately emulate latency rather than sleep functions. 

We evaluate each codebase using five systems that generate and verify the attestations: {\em (i)} SSL-lib (no tamper-proof), {\em (ii)} SSL-server (no tamper-proof), {\em (iii)} SGX, {\em (iv)} AMD-sev, and {\em (v)} \projecttitle{}. To perform a fair comparison, we integrate into our codebases a library that accurately emulates all latencies (measured in $\S$~\ref{subsec:hw_eval}) within the CPU. For the AMD latency, we use 30us, representing the lower bound of the latencies measured in $\S$~\ref{subsec:hw_eval}. We do not emulate the SSL-lib latency. 
% We evaluate each protocol using five systems that generate and verify the attestations: {\em (i)} SSL-lib, an SSL library that is integrated into the codebase (no tamper-proof), {\em (ii)} SSL-server (no tamper-proof), {\em (iii)} SGX, {\em (iv)} AMD-sev, and {\em (v)} \projecttitle{}. To perform a fair comparison, we integrate into our protocols' codebases an library that accurately emulate all latencies (measured in $\S$~\ref{subsec:hw_eval}) within the CPU. We do not emulate the SSL-lib latency and for the AMD latency we use 30us which represents the lower bound of the latencies measured in $\S$~\ref{subsec:hw_eval}.

 %which are all an integrated library to the codebase and further adds a configurable delay to represent the operation's latency in each system. The added delay for each system is the respective latency we measured in

Given that DRCT-IO, which is used for the emulation, is at least 3$\times$ slower than the hardware RDMA network stack (RDMA-hw), the latencies outlined in this section are anticipated to reflect the upper limit for all four systems with \projecttitle{}.

\rev{(c)}{We additionally evaluate two TEEs-hosted CFT replication protocols (TEEs-Raft and TEEs-CR) where the entire protocol codebase (Raft~\cite{raft} and Chain replication~\cite{chain-replication} respectively) resides within the TEE. We compare the TEEs-hosted systems with \projecttitle{} and discuss the trade-offs between their threat model, TCB, and performance.}
% Given that DRCT-IO for small messages operates roughly three times the speed of the hardware-based RDMA network stack implementation, the latencies outlined in this section are anticipated to reflect the upper limit for all four systems with \projecttitle{}.

\begin{table}[t!]
\begin{center}
\small
\minipage{0.22\textwidth}
  \centering
\begin{tabular}{lrr}
\hline
& \multicolumn{2}{c}{Throughput (Op/s)} \\
System          & append    & lookup  \\
\hline
SSL-lib         & 790K      & 256M      \\
SGX-lib             & 380K      & 3.8M       \\
AMD-sev         & 30K       & 263M      \\
\projecttitle{} & 158K      & 257M      \\
\hline
\end{tabular}
\endminipage
\hfill
\minipage{0.18\textwidth}
\centering
\begin{tabular}{rrr}
\hline
\multicolumn{2}{c}{Latency (us)} \\
 append     & lookup  \\
\hline
 1.26       & 0.0039      \\
 2.6        & 0.26       \\
 32.37      & 0.0038      \\
 6.34       & 0.0039      \\
 \hline
\end{tabular}
\endminipage
\end{center}
\caption{Throughput and latency of A2M.}%\dimitra{@experiments: maybe you try run A2M into an SGX server to also show the communication costs}}
\label{fig:a2m_eval}
\vspace{-4pt}
\end{table}

\myparagraph{A2M} We first evaluate our \projecttitle{}-A2M system. 
% We evaluate our prototype of the A2M system with \projecttitle{}. 
% For SGX, we port the entire log within the TEE, labeled as SGX-lib. All other versions place the attested log in the untrusted host memory, using the trusted systems to generate attestations as in~\cite{levin2009trinc}.
\rev{B6}{
We evaluate two TEE baselines: SGX-lib, which places the entire log within the TEE, and AMD-sev, which places the attested log outside the TEE as in the implementation of TrInc~\cite{levin2009trinc} and has been shown to be effective. 
% adapting the implementation of TrInc~\cite{levin2009trinc}
% We evaluate another TEE baseline using SGX, labeled SGX-lib, which places the entire log within the TEE. We also evaluate AMD-sev adapting the implementation of TrInc~\cite{levin2009trinc}, which places the attested log outside the TEE and has been shown to be effective. 
}
%, using the trusted systems to generate attestations as in~\cite{levin2009trinc}. 
% In this experiment, we first construct a log of size 9GB with 100M entries and then sequentially we lookup for them individually. Each log entry is comprised of 64B of appended data (context) and an extra 36B for the metadata.
In this experiment, we construct a 9.3GiB log with 100 million entries and then lookup them sequentially/individually. %Each log entry is 100B, which contains 64B appended context and 36B metadata.

\noindent{\underline{Results.}} 
Table~\ref{fig:a2m_eval} shows the throughput and mean latency of the append/lookup operations. The native execution (SSL-lib) achieves the highest throughput as it incurs no communication costs. 
% Specifically, its latency is 1.26us, which is dominated by the HMAC computation.
Compared to SSL-lib, SGX-lib experiences only a 2$\times$ slowdown because we avoid the costly communication w.r.t. an SGX-based server implementation. On the other hand, AMD-sev, which runs the SSL server, incurs a 15$\times$ slowdown. Lastly, \projecttitle{} incurs 5$\times$ and 2.4$\times$ slowdown compared to SSL-lib and SGX-lib, respectively, due to the HMAC calculation.

% Table~\ref{fig:a2m_eval} shows the throughput (operations/s) and the mean latency of our A2M system using various systems. The append operation throughput in the SSL-lib case illustrates the throughput upper bound as it incurs no communication costs. Specifically, the A2M with SSL-lib running natively in the CPU reports a latency of 1.26us, which is dominated by the HMAC computation latency.
% Placing the log within the SGX (SSL-lib), we avoid the costly communication w.r.t. to an SGX-based server implementation, and as such, the system only experiences a 2$\times$ slowdown compared to the native case. The communications costs are reflected in the AMD-sev case; that runs the SSL-server. Compared to when porting to SGX, AMD-sev incurs 15$\times$. Lastly for \projecttitle{}, we observe approximately 5$\times$ and 2.4$\times$ slowdown compared to the SSL-lib and SGX-lib execution, respectively which is due to the HMAC calculation.

Regarding the lookup operation, SSL-lib, AMD-sev, and \projecttitle{} report similar throughput and latency because they lookup untrusted host memory for the requested entry. However, SGX-lib reports a 66$\times$ slowdown due to its trusted memory size constraints and expensive paging mechanism~\cite{treaty} \rev{C3}{because we have to support a log of 9GB within the SGX enclave that only provides 94MB of memory. In contrast, AMD-sev is faster as it only accesses the untrusted host memory. Similar findings have also been demonstrated in~\cite{levin2009trinc}}. As a result, while \projecttitle{} increases append latencies, it greatly optimizes lookup latencies due to its minimal TCB.
% As a result, while \projecttitle{} offers slower append operations than porting the entire log into the TEE, it greatly optimizes lookup latencies due to its minimal TCB.

\begin{figure*}
\centering
\minipage{0.33\textwidth}
\centering
    \includegraphics[width=\linewidth]{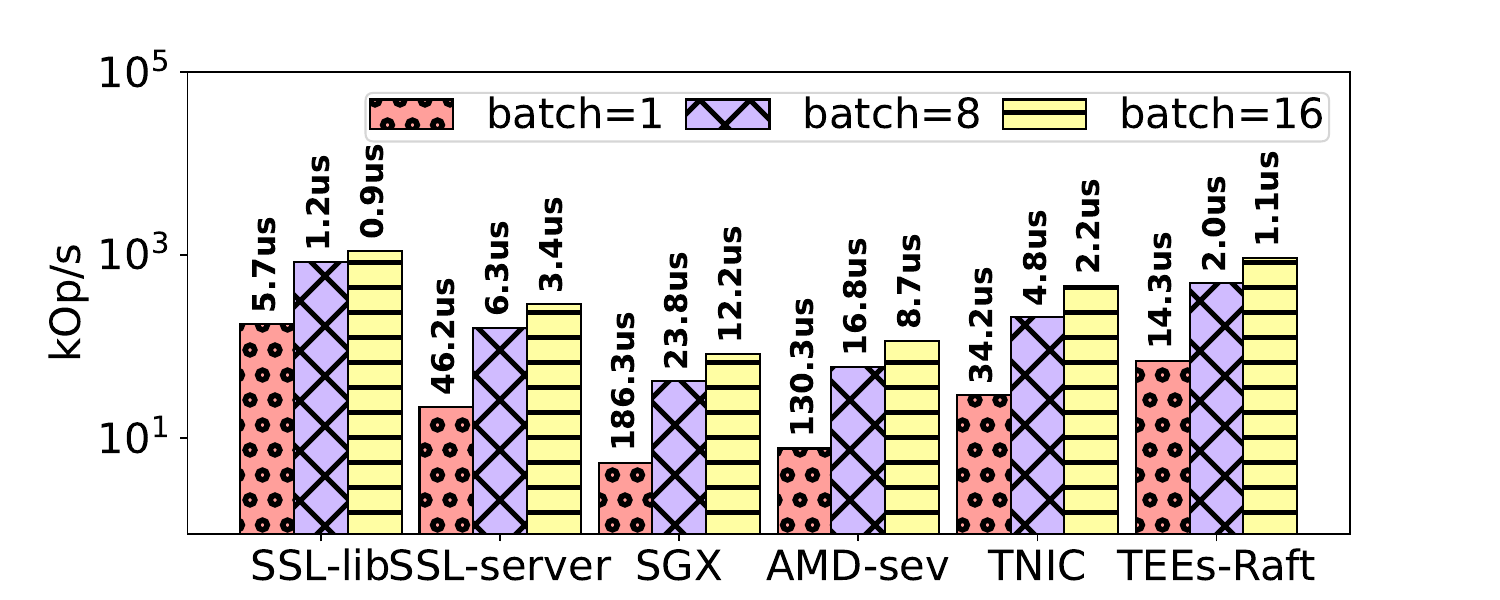} 
    \vspace{-4mm}
    \caption{Throughput (and latency numbers) of BFT.} \label{fig:byz_smr_throuthput}
\endminipage%
\minipage{0.33\textwidth}
  \centering
  \includegraphics[width=\linewidth]{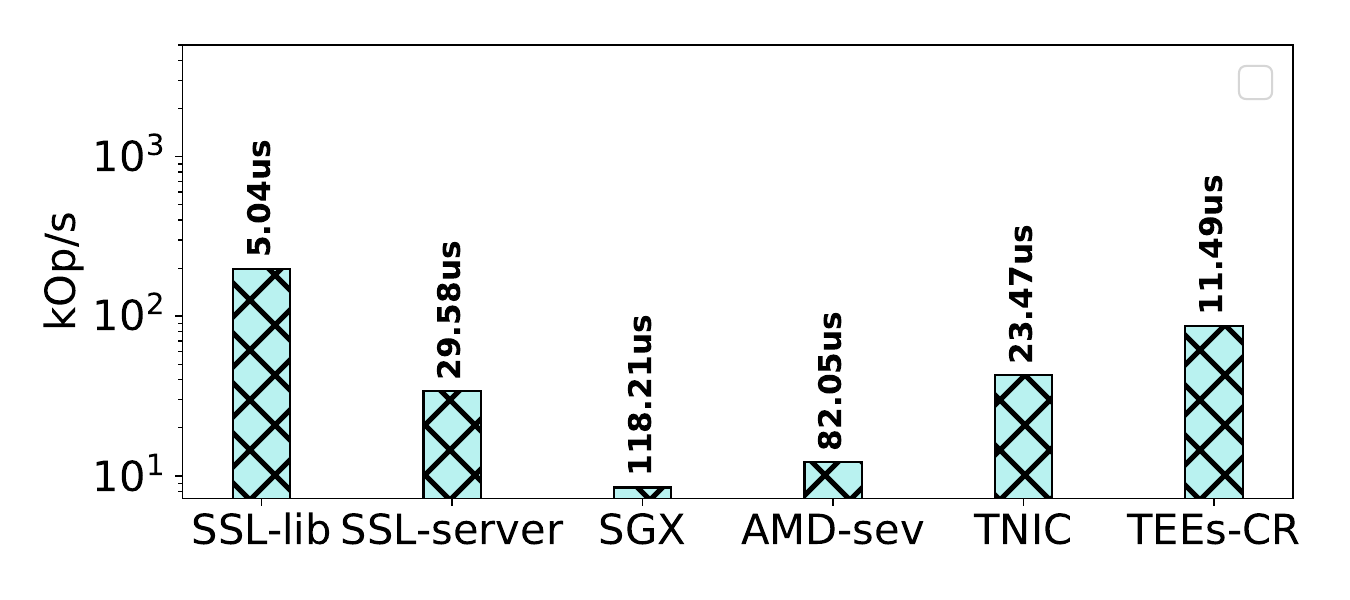} 
    \vspace{-6mm}
    \caption{Throughput (and latency numbers) of Chain Replication.} \label{fig:byz_chain_replication}
\endminipage %
\minipage{0.33\textwidth}
     \includegraphics[width=\linewidth]{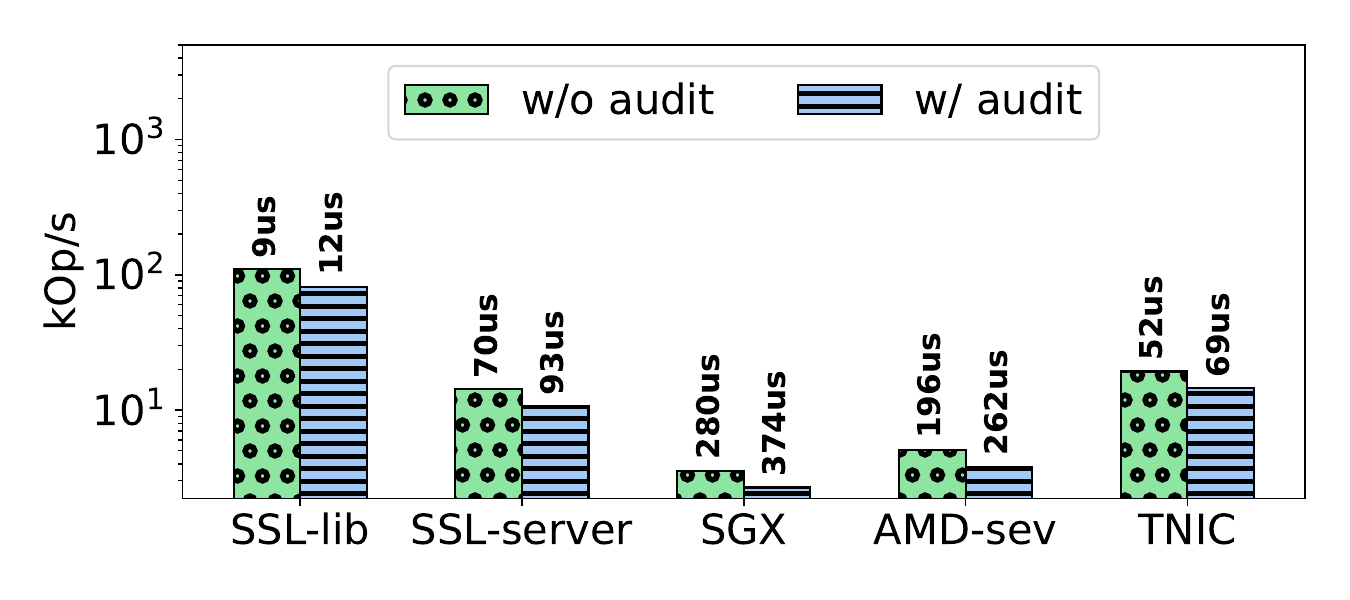} 
    \vspace{-6mm}
    \caption{Throughput (and latency numbers) of PeerReview.} \label{fig:accountability_protocol}
\endminipage
\vspace{-2mm}
\end{figure*}

\myparagraph{BFT} We evaluate the performance of our BFT protocol with various network batching factors. We implement network batching as part of the application's message format. % In this experiment, we allocate one {\em message} structure for each client's request, which contains the initial command, the results of the command's execution in a node, the incremented counter values, etc. % and the signed hashes of the replicas' states known to each node.

% In this experiment, we allocate one {\em message} structure for each client's request. The {\em message} contains command and output fields (16B each), which store the initial command and results of the command's execution in a node, respectively. In addition, the message contains metadata, which includes a sequencer and signed hashes of the replicas' states known to each node. The sequencer is used for serialization, and it is equivalent to the counter value assigned to each message from \projecttitle{}. The signed hashes representing the replicas' states are obtained from the last messages received from each replica.

\noindent{\underline{Results.}} Figure~\ref{fig:byz_smr_throuthput} shows the throughput and latency of the protocol, which highlights that \projecttitle{} significantly outperforms TEE-based versions (SGX, AMD-sev), improving the throughput and latency 4---6$\times$. On the other hand, \projecttitle{} incurs 2.4$\times$ throughput overhead and up to 7$\times$ higher latency compared to SSL-lib. 
We recall that SSL-lib is not tamper-proof (Table~\ref{tab:hw_options}) and eliminates the communication overheads incurred by other tamper-proof solutions (SGX, AMD-sev).

We also observe that batching improves the throughput and latency proportionally to the number of batched messages. For all except SSL-lib, the batching factors equal to 8 and 16 achieve 7$\times$ and 15$\times$ higher throughput than without batching, respectively. For SSL-lib, they are moderately effective: approximately 4---6$\times$ faster. It is primarily because the native execution of the attestation function is fast enough to saturate the network bandwidth. 
As such, conventional techniques can drastically eliminate the overheads for BFT and improve \projecttitle{}'s adoption into practical systems.
% In addition, we observe that batching increases throughput (and decreases latency) proportionally to the number of batched messages. For all but the SSL-lib version, we report a 7$\times$ and 15$\times$ throughput boost for batching factor to be equal to 8 and 16 respectively compared with the experiment run with batching factor equal to 1. This is because our batching technique improves the network utilization and reduces the overall attestation calculations, e.g., one attestation per 8 and 16 messages. The technique is moderately effective when using the SSL-lib, approximately 4---6$\times$ faster than without batching. This is primarily because, in that case, the latency is dominated by the network stack latency as the attestation function running natively generates attestations fast. As such, we show that conventional techniques can drastically eliminate the overheads for BFT and improve \projecttitle{}'s adoption into practical systems.

%\begin{figure}[t!]
%    \centering
%    \includegraphics[width=\linewidth]{atc-submission-plots/bft_pb.pdf} 
%    \caption{Throughput (and latency numbers) of BFT using various trusted components.} \label{fig:byz_smr_throuthput}
%\end{figure}

%\begin{figure}[t!]
%    \centering
%    \includegraphics[width=0.7\linewidth]{atc-submission-plots/bft_pb_lat.pdf} 
%    \caption{Average latency of the BFT SMR using various trusted components.} \label{fig:byz_smr_lat}
%\end{figure}

\myparagraph{CR} 
In this experiment, we evaluate the performance of our CR. 
% We evaluate the performance of our Chain Replication. In this experiment, 
We allocate one message structure per client request comprising 60B context, 4B operation type, and a 32B signature.

% (that includes metadata, e.g., source/destination nodes, message ID), 
%We implemented the replication protocol without any underlying Key-Value store data structure.
%of the message allocates 60B comprised of an 8B key and a 32B value as well as 16B for metadata (e.g., source and destination nodes, message idx, etc.) and
\noindent{\underline{Results.}} Figure~\ref{fig:byz_chain_replication} shows the throughput and latency of our Chain Replication. We highlight that our \projecttitle{} is 5$\times$ and 3.4$\times$ faster than SGX and AMD-sev, respectively. While \projecttitle{} incurs 4.6$\times$ overheads compared to SSL-lib, it is 30\% faster than SSL-server, which is not tamper-proof. The performance benefit stems primarily from hardware acceleration by the \projecttitle{}'s attestation kernel on the transmission/reception data path.

%\begin{figure}
%    \centering
%  \includegraphics[width=\linewidth]{atc-submission-plots/bftcr_lat_throughput.pdf} 
%    \caption{Throughput (and latency numbers) of Chain Replication using various trusted components.} \label{fig:byz_chain_replication}
%\end{figure}

%\begin{figure}
%  \includegraphics[width=\linewidth]{atc-submission-plots/bftpr_lat_throughput.pdf} 
%    \caption{Throughput (and latency numbers) of PeerReview using various trusted components.} \label{fig:accountability_protocol}
%\end{figure}

\if 0
\begin{figure*}[t!]
\begin{center}
\minipage{0.5\textwidth}
  \centering
  \includegraphics[width=0.8\linewidth]{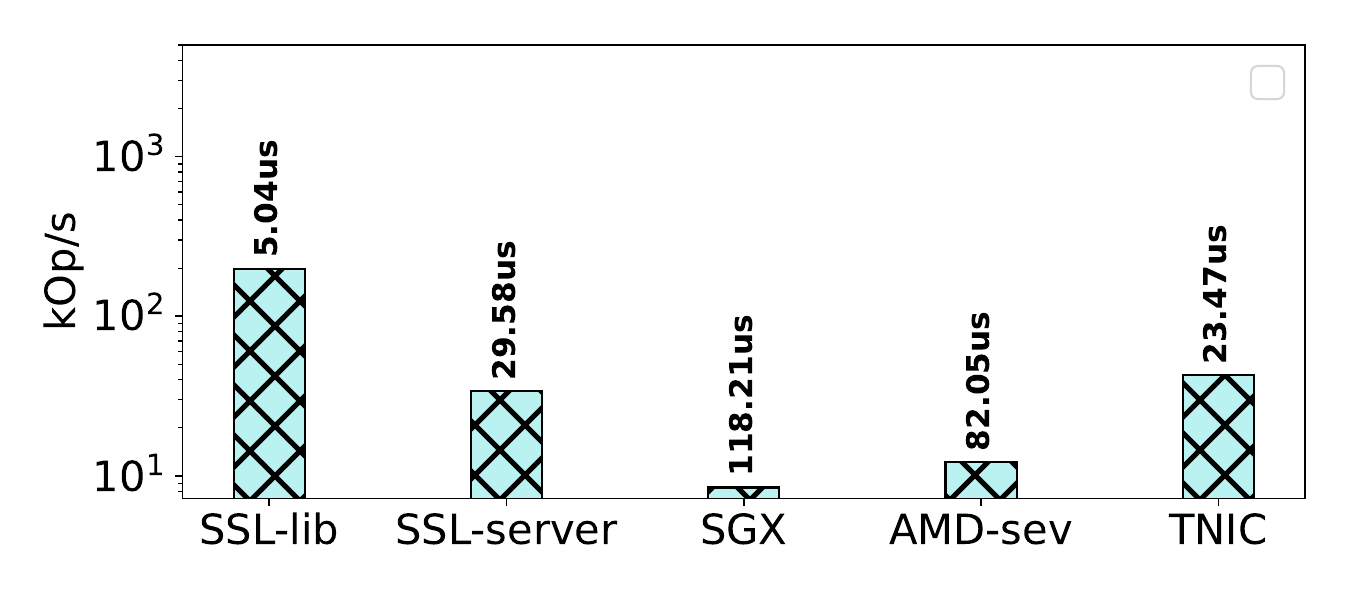} 
    \caption{Throughput-latency evaluation of a Byzantine Chain Replication using various trusted components.} \label{fig:byz_chain_replication}
\endminipage
\minipage{0.5\textwidth}
  \centering
  \includegraphics[width=0.8\linewidth]{atc-submission-plots/bftpr_lat_throughput.pdf} 
    \caption{Throughput-latency evaluation of the accountability protocol using various trusted components.} \label{fig:accountability_protocol}
\endminipage
\end{center}
\end{figure*}
\fi

\if 0
\begin{figure}[t!]
    \centering
    \includegraphics[width=\linewidth]{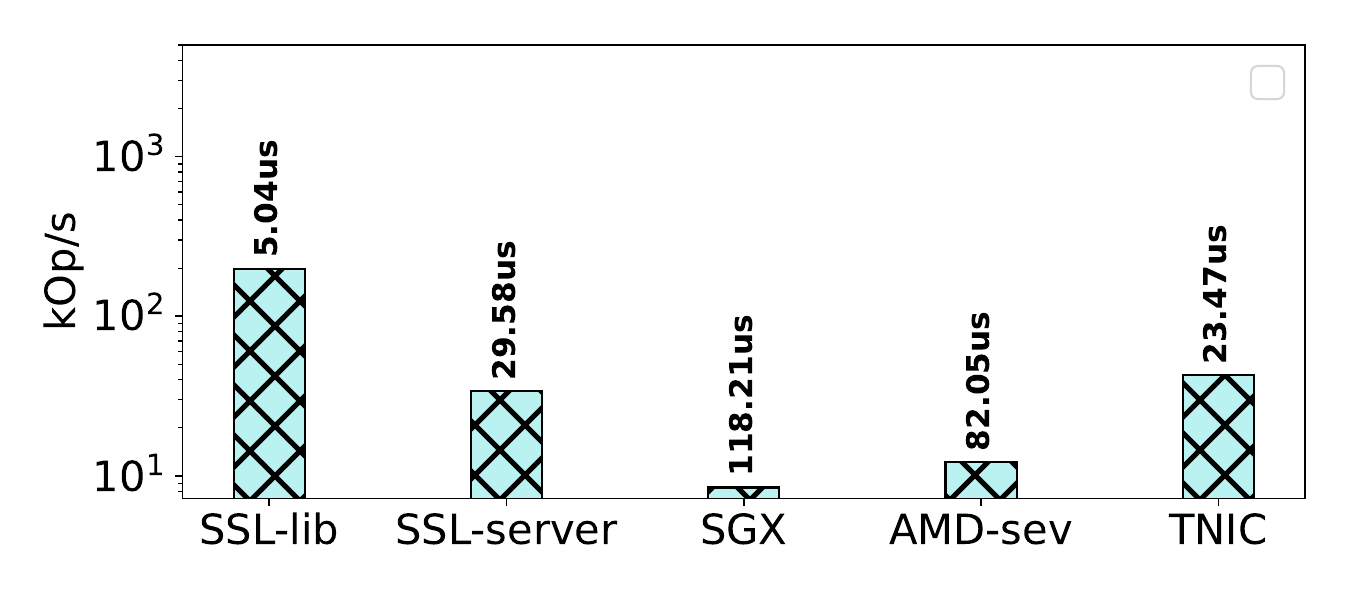} 
    \caption{Throughput-latency evaluation of a BFT version of CR using various trusted components.} \label{fig:lat_throughput_kernel}
\end{figure}
\fi

\myparagraph{PeerReview} 
We evaluate our PeerReview system's performance by both activating and deactivating the audit protocol. The system uses one witness for the source node that {\em periodically} audits its log. 
% The witness process is co-located in the same node, and as such, reading the log implies reading a shared memory. We decide in favor of that implementation to carefully isolate the overheads for inspection and replay of the log. 
In our experiments, the witness audits the log after every send operation in the source node until both clients acknowledge the receipt of all source messages. % Each message allocates about 200B, comprising a context, its hash, the cumulative digest of the hashes, etc. 

\noindent{\underline{Results.}} Figure~\ref{fig:accountability_protocol} shows the throughput and latency of our PeerReview system with and without enabling the audit protocol.
Without the audit protocol, the TEE-based systems (SGX, AMD-sev) result in up to 30$\times$ slower throughput than SSL-lib, whereas our \projecttitle{} mitigates the overheads: 3---5$\times$ better throughput compared to AMD-sev and SGX.

Similarly, \projecttitle{} outperforms AMD-sev and SGX by 3.7---5$\times$ with the audit protocol. Importantly, when using \projecttitle{}, the audit protocol itself consumes about 25\% (17us) of the overall latency, leading to 1.33$\times$ performance slowdown. % compared to when being disabled. 
% The audit protocol itself consumes approximately 25\% (17us) of the overall execution latency. 
However, even with the audit protocol, \projecttitle{} offers 3.7---5.42$\times$ lower latency compared to its TEE-based competitors.
% With the audit protocol, we observe similar system behavior; \projecttitle{} outperforms both AMD-sev and SGX by 3.7---5$\times$. Importantly, when using \projecttitle{}, the audit protocol leads to a performance slowdown of 1.33$\times$ compared to when being disabled. The audit protocol itself consumes approximately 25\% (17us) of the overall execution latency. Even with the audit protocol, \projecttitle{} offers 3.7---5.42$\times$ lower latency compared to its TEEs-based competitors.

%\atsushi{a bit difficult to read because 'log' is used as both noun and verb. Can we somehow use another word?} When the auditing protocol is de-activated, the source node sends the (streaming) data to the client nodes. The clients execute and append the message to their log, and then reply to the source. The source appends its own message to the log along with the client's reply. This is required to make sure that any participant cannot lie to a witness about the data it sends to the clients. \atsushi{The sentences until here explain the application workflow, but we should show the results first and then start the discussion with an extra explanation.} We see that using a TEE to generate attested messages increases the throughput slowdown up to 30$\times$ compared to the native embedded library (Nat-lib). Using our \projecttitle{} we improve the overheads---specifically, we incur 5.8$\times$ slowdown, that is 3---5$\times$ better throughput compared to when using AMD-sev and SGX as trusted messages' singers.

\begin{table}[t]
    \centering
    \small
    \setlength\tabcolsep{4.2pt}
    \begin{tabular}{ccrrrrr}
        \hline
        & &\multicolumn{4}{c}{TCB size (LoC)} \\ \cmidrule{3-6}
        System          &Threat model &OS & Att. kernel & App & Total \\ \midrule
        TEEs-Raft       &CFT &2,307K & 1,268 & 856 & 2,309K \\
        TEEs-CR         &CFT &2,307K & 1,268 & 992 & 2,309K \\ 
        \projecttitle{} &BFT & -     & 2,114 & -   & 2,114  \\ \hline
    \end{tabular}
    % \caption{\rev{(c)}{\projecttitle{} compared with TEE-hosted applications.}}
    \caption{\rev{(c)}{\projecttitle{} compared with TEE-hosted applications.}}
    \label{table:qualitative_comparison}
    \vspace{-2mm}
\end{table}

%~ configured with virtio with batch factor equal to one
\myparagraph{\rev{(c)}{TEEs-hosted baselines}} 
\revcont{We compare \projecttitle{} with TEEs-hosted systems implementing two prototypes based on the failure-free execution of Raft (TEEs-Raft) and CR (TEEs-CR). The code runs within three AMD-sev machines. Prior works~\cite{avocado, nimble} suggested this setup for performance---however, at the cost of (1) significantly increased TCB size and (2) a weaker threat model from the application perspective. 
Table~\ref{table:qualitative_comparison} summarizes the security costs. Regarding (1), the TCB of TEEs-hosted systems includes the entire OS~\cite{gramine_tdx}, OpenSSL libraries for messages authentication~\cite{openssl_hmac} (labeled as Att. kernel), and the application codebase, which is over $2$M LoCs in total. In contrast, \projecttitle{}'s TCB only includes our hardware attestation kernel, which is 2,114 LoC of HLS/HDL code. It is only 0.09\% of TEE-hosted systems. 
% In contrast, our \projecttitle{}'s attestation kernel is minimal; only 2,114 LoC of HLS/HDL code (0.09\% of TEE-hosted systems). 
Regarding (2), the TEE-hosted application can only fail by crashing; it can be thought to remain protected from a potentially Byzantine cloud environment, whereas \projecttitle{} targets BFT settings, handling up to $f$ arbitrary failures. 
}

\revcont{We compare TEE-Raft with our \projecttitle{}-based BFT (Figure~\ref{fig:byz_smr_throuthput}) as both are broadcast-based protocols, and TEEs-CR with our \projecttitle{}-based CR (Figure~\ref{fig:byz_chain_replication}) as both require all messages to traverse the entire chain of nodes. TEE-Raft achieves approximately $2.5\times$ higher throughput than \projecttitle{}-based BFT. The performance difference is primarily due to Raft's one-phase commitment compared to our \projecttitle{}-based BFT. Similarly, TEE-CR achieves $2\times$ higher throughput than the \projecttitle{}-based CR. While both versions of CR involve the same number of network Round-Trip Times (RTTs), 
\projecttitle{} involves a higher number of the attestation kernel invocations to verify all the chained messages in the PoE.
}

% \begin{table}[t]
% \centering
% \small
% \begin{tabular}{ c c  c  }
% \hline
%  System & Threat model &  TCB size \\ 
%  \hline
%  \projecttitle{} & BFT  & $<2.5$K LoC  \\  
%  TEEs-app & CFT  & OS  $>2000$K LoC\\
%  \hline
% \end{tabular}
% \caption{\projecttitle{} compared with TEE-hosted applications.}\label{table:qualitative_comparison}
% \vspace{-3mm}
% \end{table}

\subsection{\revcont{FPGA Resource Usage}}
\rev{(d), E2}{Lastly, we perform a resource utilization analysis to show \projecttitle{}'s scalability capabilities. We measure the resource consumption of \projecttitle{}'s primary hardware components~\cite{easynet} and estimate maximum connections on the latest FPGA. }
% , e.g., how many instances of the attestation kernel as well as the RoCE can fit on the same FPGA.

\revcont{
Table~\ref{tab:resource-utilization} shows the resource consumption details. The overall \projecttitle{} design consumes 16.6\% of LUTs, 16.3\% of Flip-Flops (FF), and 16.6\% of RAMB36 (3.46~\% of the entire on-chip memory) on the U280 FPGA. Note that \projecttitle{} only requires commodity FPGA NIC designs to add the attestation kernel, whose utilization is comparable with the other modules, XDMA and RoCE. 
% , which occupies 15.7\% of LUTs, 13.4\% of FFs, and 24.1\% of RAMB36 compared to the entire design.
}

\begin{figure}
    \centering
    \includegraphics[width=0.9\linewidth]{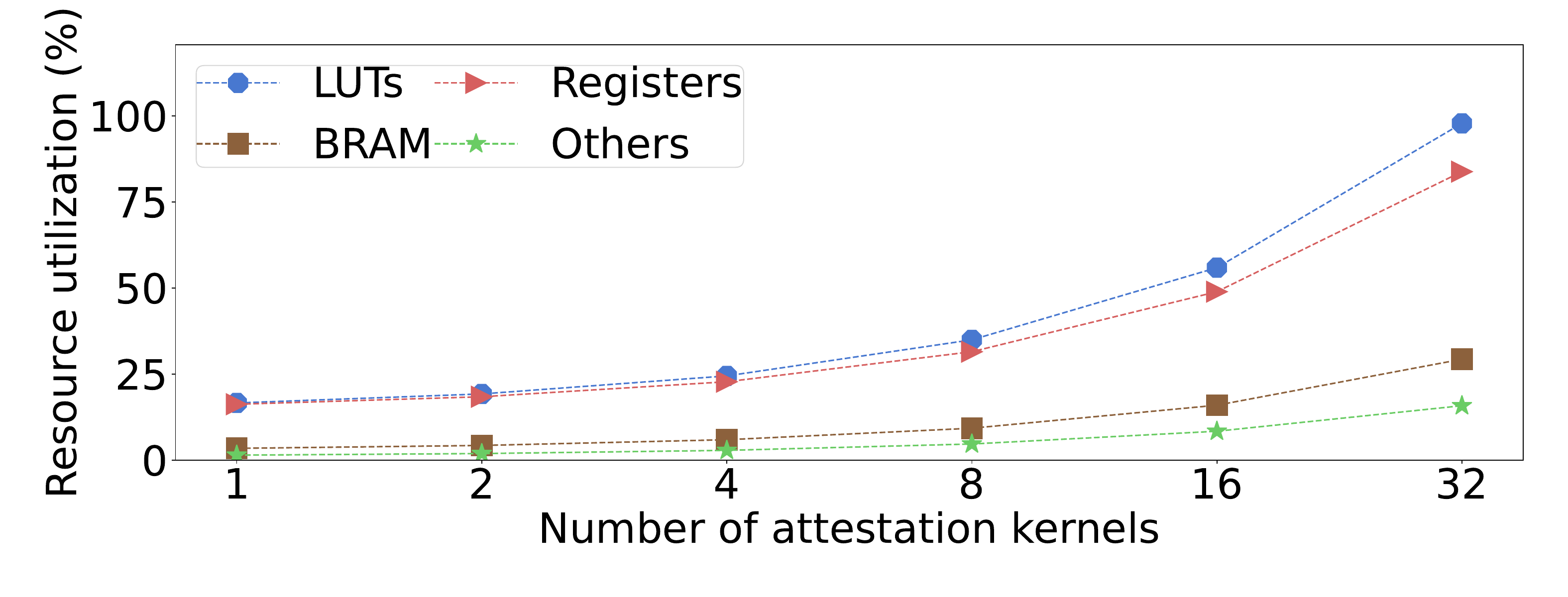}
    \vspace{-10pt}
  \caption{\rev{(d)}{The scalability analysis of \projecttitle{} hardware. The resource usage is normalized to the U280 FPGA capacity.} }
  % \caption{Latency evaluation of send operations for various packet sizes across five competitive network stacks with various security properties.}
    % \felix{Figure: BRAM = RAMB36 (including RAMB18) + URAM. "CARRY8" is the 4th most used resource of the attestation kernel and is used for "Others" as an upper bound for all other resources.}
    % More resource utilization details: https://github.com/dgiantsidi/replication-protos/blob/main/plots/asplos_submission/attestation_kernel_util.md
    \label{fig:scalability}
   %\vspace{-2pt}
\end{figure}

\revcont{
Figure~\ref{fig:scalability} shows the scaling capabilities of \projecttitle{} hardware. 
% The XDMA module, CMAC module, and RoCE kernel are single for each in the design because the XDMA and CMAC modules are independent of the number of connections and the RoCE kernel is configured to hold 500 queue pairs to establish the same number of network connections~\cite{storm}. 
As the number of network connections increases, we only need to replicate the attestation kernel because the XDMA and CMAC modules are independent of the number of connections, and the RoCE kernel is configured to hold up to 500 connections~\cite{storm}. The result demonstrates that \projecttitle{} can support up to 32 concurrent connections on a single U280 FPGA. 
%  to sustain the throughput per connection
% 500 queue pairs to establish the same number of
}
% \rev{E2}{Our implementation is based on~\cite{coyote}, a fork of which has been used in prior works~\cite{storm} showing that 500 queue pairs (QPs) occupy 9\% of the on-chip memory, while the logic resource usage remains below 1\% when scaling from 500 to 16,000 QPs. Our evaluation and modern deployments use more powerful FPGAs, suggesting that even a larger number of connections could be supported compared to the work in~\cite{storm}. }

\newcolumntype{C}[1]{>{\centering\arraybackslash}p{#1}}
\newcolumntype{L}[1]{>{\raggedright\arraybackslash}p{#1}}
\newcolumntype{R}[1]{>{\raggedleft\arraybackslash}p{#1}}

\begin{table}[t]
    \small
    \centering
    \setlength\tabcolsep{4.2pt}
    \begin{tabular}{C{14mm}R{10mm}R{7mm}R{10mm}R{7mm}R{6mm}R{7mm}}
    \hline
        Name & \multicolumn{2}{c}{LUT (\%)} & \multicolumn{2}{c}{FF (\%)} & \multicolumn{2}{c}{RAMB36 (\%)} \\ \cmidrule(lr){1-1} \cmidrule(lr){2-3} \cmidrule(lr){4-5} \cmidrule(lr){6-7}
        U280               & 1303680 &  (100) & 2607360 &  (100) & 2016 &  (100) \\ \cmidrule(lr){1-1} \cmidrule(lr){2-3} \cmidrule(lr){4-5} \cmidrule(lr){6-7}
        TNIC               &  216905 & (16.6) &  423891 & (16.3) &  335 & (16.6) \\               
        XDMA               &   48258 &  (3.7) &   50701 &  (1.9) &   64 &  (3.1) \\ 
        Att. kernel        &   34138 &  (2.6) &   56914 &  (2.2) &   81 &  (4.0) \\ 
        RoCE               &   30379 &  (2.3) &   75804 &  (2.9) &   46 &  (2.3) \\ 
        CMAC               &    1484 &  (0.1) &    3433 &  (0.1) &    0 &  (0.0) \\ \hline
    \end{tabular}
    \caption{\rev{(d)}{\projecttitle{}'s resource usage. The relative (\%) compares with the U280 FPGA capacity. \projecttitle{} means the entire design.}} 
    % available resources on the 
    \label{tab:resource-utilization}
    \vspace{-2mm}
\end{table}

% Nevertheless, our \projecttitle{} does not assume a specific FPGA board. Therefore, the findings from previous works on other boards are still relevant.

\if 0
\begin{figure}[t!]
    \centering
    \includegraphics[width=\linewidth]{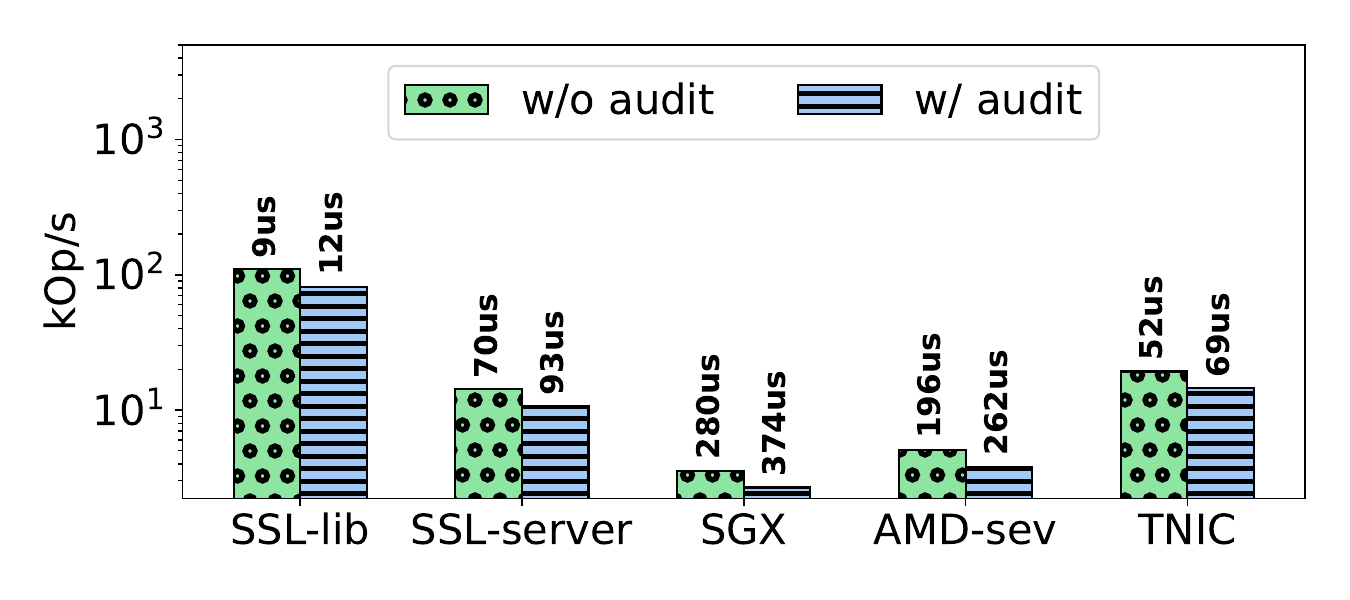} 
    \caption{Throughput-latency evaluation of a BFT version of PR using various trusted components.} \label{fig:lat_throughput_kernel}
\end{figure}
\fi
\subsection{\revcont{Discussion}}
\myparagraph{\rev{B2}{\projecttitle{}'s applicability}}
\revcont{As FPGA-based SmartNICs are widely adopted by major cloud providers for hardware acceleration~\cite{211249}, we believe that \projecttitle{} has the potential for broader industry application. In addition,  ASIC-based NICs can also provide the same functionalities by integrating \projecttitle{}'s hardware modules into an optimized System-on-Chip (SoC).}
%\revcont{As FPGA-based SmartNICs are widely adopted by major cloud providers for hardware acceleration~\cite{211249}, we believe that \projecttitle{} has the potential for broader industry application. In addition, \projecttitle{}’s APIs are generic and CPU-agnostic, making them more suitable for cloud environments than programming heterogeneous TEEs.}

%\revcont{\projecttitle{} demonstrates its minimalistic root of trust, which suffices for building BFT distributed systems. 
%ASIC-based NICs can also provide the same functionalities by integrating \projecttitle{}'s hardware modules into an optimized System-on-Chip (SoC). }
%which would be more optimized than the FPGA-based approach

\myparagraph{\rev{C1}{Use cases}}
\revcont{The paper deliberately focuses on distributed cloud applications as \projecttitle{}'s primary use cases. Trust in shared third-party clouds is a more critical concern than in other environments, posing unique challenges in trust, performance, and scalability. While the current scope is specific, the underlying principles could extend to other use cases, such as HPC or on-premise computing.}
%However,  \projecttitle{} addresses these using SmartNICs, which are already widely deployed in modern clouds~\cite{211249}. 

\myparagraph{\rev{D3}{Message drops}}
\revcont{
\projecttitle{} guarantees packet retransmission between two correct nodes until their successful reception  extending a RoCE implementation that supports reliable operations. The application need not re-send the message as it receives a different sequence number, which is not accepted (or verified) by the remote \projecttitle{} until all preceding messages have been received. %Instead, \projecttitle{} extends a RoCE implementation that supports reliable operations where the remote side acknowledges the receipt of a message using specific ACK messages. 
% \projecttitle{} guarantees packet retransmission between two correct nodes until their successful reception. The application need not re-send the message as it would receive a different sequence number, which would not be accepted (or verified) by the remote \projecttitle{} until all preceding messages have been received. Instead, \projecttitle{} extends a RoCE implementation that supports reliable operations where the receipt of a message is acknowledged by the remote side using specific ACK messages.
}

\myparagraph{\rev{D4, E3}{View-change and recovery}}
\revcont{Detailing view-change and recovery in \projecttitle{} protocols are beyond the scope of our work. \projecttitle{} could adopt similar techniques as in TrInc~\cite{trinc} without disrupting these operations. In a new leader's election, replicas can establish new connections with new identifiers. As such, previous connections will not block execution, and state transfers, e.g., view-change, can be performed effectively. } 

%\revcont{Detailing view-change and recovery in \projecttitle{} protocols are beyond the scope of this paper, while \projecttitle{} could adopt techniques such as the ones used by TrInc~\cite{trinc}. Importantly, \projecttitle{} does not interfere with these operations. During a new leader's election, replicas can establish new connections with new connection identifiers. As a result, previous connections will not block execution, and state transfers (e.g., view-change) can be carried out effectively. } 
% \rev{D4, E3}{We do not implement view-change or recovery in \projecttitle{} protocols, while \projecttitle{} does not interfere with these operations. During a new leader's election, replicas can establish new connections with new connection identifiers. As a result, previous connections will not block execution, and state transfers (e.g., view-change) can be carried out effectively. For recovery, \projecttitle{} could adopt techniques such as the ones used by TrInc~\cite{trinc}. } 

% \myparagraph{\rev{E1}{Impact of RDMA on security}}
\myparagraph{\rev{E1}{PCIe transaction encryption}}
\revcont{\projecttitle{} encrypts PCIe transactions for CPU-to-device communication, allowing attackers to modify the PCIe transactions. This vulnerability is not unique to \projecttitle{}; it applies to any network stack, including the OS-based ones, since the \emph{untrusted} OS drives PCIe transactions. }

\section{Related work}
%\pramod{the related work bucket should not start with individual papers. It should first summarize the field/area/bucket area, and then give example systems, and lastly, it should end with an overall comparison of the the research field with our solution. In this way, you avoid direct comparison with individual work. Can you please re-write it?}

\myparagraph{Trustworthy distributed systems} Classical BFT systems~\cite{Castro:2002, Suri_Payer_2021, DBLP:journals/corr/abs-1803-05069, Chan2018PaLaAS, DBLP:journals/corr/abs-1807-04938, Chan2018PiLiAE, bft-smart, 6681599} provide BFT guarantees at the cost of high complexity, performance, and scalability overheads. \projecttitle{} bridges the gap between BFT and prior limitations, designing a {\em silicon root-of-trust} with generic trusted networking abstractions that materialize the BFT security properties.

\myparagraph{Trusted hardware for distributed systems} Trustworthy systems~\cite{10.1145/3492321.3519568, minBFT, 10.1145/3552326.3587455, 10.1145/3492321.3519568, treaty, avocado, ccf} leverage trusted hardware to optimize the performance of classical BFT at the cost of generalization and easy adoption. The systems suffer from high latencies (50us---105ms)~\cite{levin2009trinc, 10.1145/2168836.2168866}, build large TCBs~\cite{treaty, avocado}, and rely on specific TEEs~\cite{minBFT, hybster}. In contrast, our \projecttitle{} aims to offer performance and generality, while our minimalistic TCB is verifiable and unified in the heterogeneous cloud. 

\myparagraph{SmartNIC-assisted systems} Networked systems offer fast network operations with emerging (programmable) SmartNIC devices~\cite{liquidIO_smartnics,u280_smartnics,bluefield_smartnics,broadcom_smartnics,netronome_smartnics,alibaba_smartnics,nitro_smartnics,msr_smartnics}. Some of them offload the network functions to the hardware and reduce the host processing and energy overheads~\cite{246498,211249,10.1145/3387514.3405895,10.1145/3365609.3365851,10.1145/3127479.3132252,258971,246486,179716,227655,10.1145/3286062.3286068,shan2022supernic,10.1145/3390251.3390257} or re-design generic networking protocols, from RDMA/RoCE to TCP/IP network stacks, on top of FPGA-based SmartNICs for performance~\cite{coyote,corundum,storm,8891991,280712,9114811,opennic_project}. Others~\cite{10.1145/3341302.3342079,10.1145/2872362.2872367,234944,9220629,6853195,10292786,10.1145/3477132.3483555,280678,10.1145/3132747.3132756,honeycomb,288659,10329593} build generic execution frameworks to optimize various distributed systems. Our \projecttitle{} follows a similar approach by building a high-performant unified network stack with SmartNICs and extending its security semantics with the properties of non-equivocation and transferable authentication.% and offloading security to the NIC hardware.

%\myparagraph{SmartNIC-assisted network stacks} SmartNICs effectively provide high-performance network stacks. Another line of research~\cite{coyote, corundum, storm, 8891991, 280712, 9114811, opennic_project} re-designs generic networking protocols, from RDMA/RoCE to TCP/IP network stacks, on top of FPGA-based SmartNICs for performance. Our \projecttitle{} further extends its security semantics with the properties of non-equivocation and transferable authentication. 

\myparagraph{Programmable HW for network security} Programmable hardware, SmartNICs, and switches are used to shield networking. Recent systems~\cite{10.1145/3603269.3604874, 10.1145/3620678.3624786, 10.1145/3563647.3563654, 10.1145/3321408.3323087, 278292} leverage programmable switches and FPGAs to offload security processing and boost performance in the context of blockchain systems~\cite{10.1145/3603269.3604874} or security functions (e.g., access control, DNS traffic inspection)~\cite{10.1145/3620678.3624786, 10.1145/3563647.3563654, 10.1145/3321408.3323087}. Our \projecttitle{} similarly offloads security into the hardware, but it carefully uses SmartNICs to overcome the processing bottlenecks of the switches.

%\input{related_work2}
% \section{Supplementary Material}
% \label{sec:appendix}
% As part of the appendix, we present (1) formally verified proofs for the \projecttitle security protocols and  (2) the implementation details of four distributed
% systems using \projecttitle.

\subsection*{Acknowledgement}
This work was supported by an ERC Starting Grant (ID: 101077577), the Intel Trustworthy Data Center of the Future (TDCoF), and the Chips Joint Undertaking (JU), European Union (EU) HORIZON-JU-IA, under grant agreement No. 101140087 (SMARTY).
% This project has also received funding from the CHIPS Joint Undertaken as part of the European Union's Horizon Europe research and innovation programme, SMARTY Project, grant agreement No. 101140087. 
The authors acknowledge the financial support by the Federal Ministry of Education and Research of Germany in the programme of "Souverän. Digital. Vernetzt.". Joint project 6G-life, project identification number: 16KISK002. 

\appendix
\section{Artifact Appendix}

%%%%%%%%%%%%%%%%%%%%%%%%%%%%%%%%%%%%%%%%%%%%%%%%%%%%%%%%%%%%%%%%%%%%%
\subsection{Abstract}
Our artifacts include the \projecttitle{} codebase as well as the software artifact with the four \projecttitle{} applications, i.e., A2M, BFT, CR, and PeerReview. In addition, we provide the codebases of all the microbenchmarks we discuss in the paper including those of the TEE-based systems. Lastly, we attach the security proofs of \projecttitle{} system operations and attestation protocol based on Tamarin~\cite{tamarin-prover}. This appendix provides the necessary information to set up, build, and run the experiments we present in the paper.

\subsection{Artifact check-list (meta-information)}

%{\em Obligatory. Use just a few informal keywords in all fields %applicable to your artifacts
%and remove the rest. This information is needed to find appropriate reviewers and gradually 
%unify artifact meta information in Digital Libraries.}

{\small
\begin{itemize}
 % \item {\bf Algorithm: }
  \item {\bf Program:} \projecttitle{} hardware implementation codebase. \projecttitle{} software codebases that include the systems where \projecttitle{} has been applied (run in emulated hardware) and microbenchmarks (e.g., network benchmark). \projecttitle{}'s security proofs based on Tamarin~\cite{tamarin-prover}.
  \item {\bf Compilation:} Requires Vitis HLS~\cite{vitis-hls}, Vivado~\cite{vivado}, CMake, C++, Boost, eRPC~\cite{erpc}, DPDK~\cite{dpdk}, Tamarin~\cite{tamarin-prover}.
  %\item {\bf Transformations: }
  %\item {\bf Binary: }
  %\item {\bf Model: }
  %\item {\bf Data set: }
  \item {\bf Run-time environment:} Requires NixOS, 5.15.4, {\sc scone}~\cite{scone} (for SGX-based experiments).
  \item {\bf Hardware:} Requires Alveo U280 cards~\cite{u280_smartnics}, Intel(R) Core(TM) i9-9900K with Intel Corporation Ethernet Controllers (XL710) (or any other DPDK compatible NIC) and AMD EPYC 7413. 
  % \item {\bf Run-time state: }
  \item {\bf Execution:} The time of the experiments are configurable. Each of our experiments did not take more than 10 minutes. However, the compilation and synthesis phases of the \projecttitle{} hardware implementation might take up to 4 hours. 
  \item {\bf Metrics:} Throughput and latency
  %\item {\bf Output: }
  % \item {\bf Experiments:} 
  %\item {\bf How much disk space required (approximately)?: }
  %\item {\bf How much time is needed to prepare workflow (approximately)?: }
  %\item {\bf How much time is needed to complete experiments (approximately)?: }
  \item {\bf Publicly available:} Yes.
  \item {\bf Code licenses:} MIT License. \projecttitle{} doesn't use any external license.
  %\item {\bf Data licenses (if publicly available)?: }
  %\item {\bf Workflow framework used?: }
  \item {\bf Archived (DOI):} \url{10.5281/zenodo.14775354}
\end{itemize}
}

%%%%%%%%%%%%%%%%%%%%%%%%%%%%%%%%%%%%%%%%%%%%%%%%%%%%%%%%%%%%%%%%%%%%%
\subsection{Description}

\subsubsection{How to access}

The open-source version of the \projecttitle{} codebase can be found on GitHub at the following address:

\url{https://github.com/TUM-DSE/TNIC-main.git}

%{\em Obligatory}

\subsubsection{Hardware dependencies}
For AMD-SEV and \projecttitle{}-hardware setups, you need three machines with AMD EPYC 7413 CPU. Each machine is equipped with an Alveo U280 card~\cite{u280_smartnics} and one of every U280's QSFP28 ports connects to the 100Gbps network. For Intel SGX setups, you need machines with Intel(R) Core(TM) i9-9900K with Intel Corporation Ethernet Controllers (XL710) (or any other DPDK compatible NIC) for network connection. 

\subsubsection{Software dependencies}
The software build process involves building the low-level 
Linux kernel driver and the high-level user application layers.
All codebases run on top of NixOS, 5.15.4. We provide the appropriate \texttt{.nix} files to set up a \texttt{nix-shell} environment with all necessary system dependencies. 

The code has been built with \texttt{Makefile} and \texttt{cmake}. The applications, as well as the TEE-based code and application layer, are  written in C++17. We depend on Boost libraries and gflgas for
the parsing of the command line arguments. We rely on several other dependencies, which we explain in our README files, including; {\sc scone}~\cite{scone} for SGX-based experiments, Vivado~\cite{vivado} and Vitis HLS~\cite{vitis-hls} for building \projecttitle{} hardware, eRPC~\cite{erpc}, DPDK~\cite{dpdk}, and Tamarin~\cite{tamarin-prover}.

%\subsubsection{Data sets}

%\subsubsection{Models}

%%%%%%%%%%%%%%%%%%%%%%%%%%%%%%%%%%%%%%%%%%%%%%%%%%%%%%%%%%%%%%%%%%%%%
\subsection{Installation}
The artifact is linked to the repository as submodules. Each repository provides analytical instructions in their \texttt{README.md} files of how to build and run the binaries.

To build the \projecttitle{}'s hardware implementation, please follow the instructions provided in ~\cite{build-hardware-for-fpga}.

To build the software including the driver and the benchmarks, please follow the instructions in~\cite{build-software}.

To run the experiments for the \projecttitle{} hardware implementation, you need to first load the \projecttitle{}'s kernel module and then run the compiled binary. Detailed instructions are available in ~\cite{tnic-run}.  

Similar instructions have been documented for the applications~\cite{tnic-apps} and the security proofs~\cite{tnic-proofs}.

%{\em Obligatory}

%%%%%%%%%%%%%%%%%%%%%%%%%%%%%%%%%%%%%%%%%%%%%%%%%%%%%%%%%%%%%%%%%%%%%
%\subsection{Experiment workflow}

%%%%%%%%%%%%%%%%%%%%%%%%%%%%%%%%%%%%%%%%%%%%%%%%%%%%%%%%%%%%%%%%%%%%%
\subsection{Evaluation and expected results}
Each of the experiments will output information about its progress; this is a hint that the script is still
running and hasn’t halted. The output of the experiment reports important measurements about the execution. The results are expected not to vary significantly (less than 5$\%$) when compared to the results presented in the paper. However, as discussed, we observed quite a significant variance in some TEE-based systems (Intel SGX and AMD-SEV).
%{\em Obligatory}

%%%%%%%%%%%%%%%%%%%%%%%%%%%%%%%%%%%%%%%%%%%%%%%%%%%%%%%%%%%%%%%%%%%%%
%\subsection{Experiment customization}
%A wide variety of experiemnt customization can be available
%through different execution parameters. The users can cre-
%ate different versions of the system through combinations of
%vFPGAs, network and memory stacks.
%%%%%%%%%%%%%%%%%%%%%%%%%%%%%%%%%%%%%%%%%%%%%%%%%%%%%%%%%%%%%%%%%%%%%
%\subsection{Notes}

%%%%%%%%%%%%%%%%%%%%%%%%%%%%%%%%%%%%%%%%%%%%%%%%%%%%%%%%%%%%%%%%%%%%%
\subsection{Methodology}

Submission, reviewing, and badging methodology:

\begin{itemize}
  \item \url{https://www.acm.org/publications/policies/artifact-review-badging}
  \item \url{http://cTuning.org/ae/submission-20201122.html}
  \item \url{http://cTuning.org/ae/reviewing-20201122.html}
\end{itemize}

%%%%%%%%%%%%%%%%%%%%%%%%%%%%%%%%%%%%%%%%%%%%%%%%%%%%
% When adding this appendix to your paper, 
% please remove below part
%%%%%%%%%%%%%%%%%%%%%%%%%%%%%%%%%%%%%%%%%%%%%%%%%%%%

%\bibliographystyle{ACM-Reference-Format}
%\bibliography{sample}
%\end{document}

% \if 0
% \newpage

% \appendix

% \section*{Appendix}
\section{Formal verification Proofs}
\label{sec:formal-verification-details}
In this appendix, we also present the detailed security proofs for \projecttitle{} security protocols using
the Tamarin Prover~\cite{tamarin-prover}.

\myparagraph{Proof artifact} The complete proofs, including the detailed formal models used to generate them, can be found under the following link:\\{\color{blue}\url{https://github.com/TUM-DSE/TNIC-proofs.git}}

\myparagraph{Symbolic model} We prove the security properties of \projecttitle{}  in a symbolic model because Tamarin analyzes protocols in symbolic models and can prove properties by verifying user-defined lemmas. We leverage Tamarin's built-in primitives and automated and interactive analysis to verify the security protocols.

We impose a set of assumptions on our proofs motivated by Tamarin's symbolic model: \emph{(i)}~The symbolic model does not reason about bitstrings directly. Instead, it assumes a set of atomic terms and functions that operate on these terms. All messages that are part of the model are composed of such atomic terms and functions applied on these terms \emph{(ii)}~These cryptographic functions are assumed to be perfect with no side-effects, e.g., hash functions are irreversible, and hash collisions are impossible. This allows for proving lemmas without considering the probabilities of violating specific properties and thus significantly reducing the complexity. The computational model is an alternative to the symbolic model that considers such probabilities. \emph{(iii)}~Attackers can read and delete all messages that are sent on the network and modify them in accordance with the set of defined functions.

Tamarin works on symbolic models specified using multiset rewriting rules that operate on the system's state. Different states of the system are expressed as a set of facts with rules capturing the available transitions from one system state to another. Rules are used to model the actions of agents running the protocol and the adversary’s capabilities. In addition to the rules, Tamarin also makes use of restrictions. Restrictions further refine the sources of facts in the protocol to improve the efficiency of the proof generation.

 Our verification work relies on properties of the already analyzed TLS handshake~\cite{tamarin_tls_proof}. It provides a model and lemmas for the security properties of the protocols presented in this paper.

To prove the correctness of our lemmas, Tamarin computes possible executions for each rule. Tamarin employs constraint solving to refine its knowledge about the sequence of protocol transitions. To check the correctness of the protocol model we also employ sanity lemmas which ensure that there exists a sequence of transitions to reach a predefined valid state. These lemmas ensure that the protocol can be executed as intended. 

In the following paragraphs, we give an overview of the rules and lemmas used to model the \projecttitle{} protocols.

\myparagraph{Rules}
The bootstrapping rules, in accordance with the bootstrapping steps in Section~\ref{subsec:nic_controller}:
\begin{itemize}

\item \emph{bootstrapping\_1}: Models step (1), the generation and burning of the hardware key by the \projecttitle{} manufacturer.
\item \emph{bootstrapping\_2}: Models step (2), the loading and verification of firmware from the insecure storage medium.
\item \emph{bootstrapping\_3\_4\_5}: Models step (3-5), the loading, key and certificate generation of the controller.
\item \emph{publish\_firmware}: Models the hardware manufacturer publishing a new firmware version.
\item \emph{get\_tnic\_public\_key}: Models the retrieval of the public \projecttitle{} key for verification.
\item \emph{compromise\_tnic\_private\_key}: Models an attacker compromising a specific \projecttitle{} device and retrieving the private key.

\end{itemize}
The attestation rules, in accordance with the attestation steps in Section~\ref{subsec:nic_controller}:
\begin{itemize}

\item \emph{attestation\_6\_7}: Models step (6-7), the receiving of the configuration data from the protocol designer, and the start of the secure channel establishment with the \projecttitle{} device.
\item \emph{attestation\_8a}: Models step (8) on the \projecttitle{} side.
\item \emph{attestation\_8b\_9}: Models step (8) on the IP Vendor side and step (9), the sending of the encrypted configuration bitstream.
\item \emph{attestation\_10\_11\_12}: Models step (10-12), the report generation of the \projecttitle{} device. 
\item \emph{attestation\_13\_14\_15\_16}: Models step (13-16), the report retrieval and verification, as well as the sending of the bitstream encryption key.
\item \emph{attestation\_17}: Models step (17), the decryption and configuration of the bitstream, as well as the acknowledgment.
\item \emph{attestation\_18}: Models the final step of the IP Vendor after which the attestation protocol is completed.
\item \emph{add\_bitstream}: Models the addition of a new bitstream to the IP Vendor, which potentially contains sensitive information.

\end{itemize}
The communication rules, in accordance with the functions provided in Algorithm~\ref{algo:primitives}:
\begin{itemize}

\item \emph{init\_ctrs}: Models the initialization of the send and receive counters for each session. Is restricted to guarantee the uniqueness of the session counters.
\item \emph{send\_msg}: Models send an arbitrary message by attesting it before sending it over the secure channel. Is restricted to guarantee the session counters are increased.
\item \emph{recv\_msg}: Models receive an arbitrary message by only accepting it after a successful verification.

\end{itemize}

\myparagraph{Lemmas}
The sanity lemmas, which ensure the protocol can be executed as intended:
\begin{itemize}

\item \emph{sanity} (verified in 26 steps): Ensures that the protocol allows for successfully completing the bootstrapping \& attestation phase, such that the IP Vendor and uncompromised \projecttitle{} device are in an expected state.
\item \emph{send\_sanity} (verified in 23 steps):  Ensures that the protocol allows for successfully verifying a message sent during the communication phase after two \projecttitle{} devices are successfully initialized. 

\end{itemize}
The attestation lemmas, which ensure the bootstrapping \& attestation phase behaves as expected:
\begin{itemize}

\item \emph{HW\_key\_priv\_secret} (verified in 3 steps): Ensures that the private key of the \projecttitle{} device is not obtainable from any messages sent as part of the \projecttitle{} protocols of the model. 
\item \emph{S\_key\_secret} (verified in 97 steps): Ensures all symmetric keys established during initialization phases are secret. It also ensures that past symmetric keys stay secret even if the hardware key is compromised in the future after the session is completed.
\item \emph{bitstream\_secret} (verified in 83 steps): Ensures all bitstreams shared during initialization phases are secret. It also ensures that past shared bitstreams stay secret even if the hardware key is compromised in the future after the session is completed.
\item \emph{initialization\_attested} (verified in 5540 steps): Ensures that after the IP Vendor finished the attestation during the initialization phase, the \projecttitle{} device is in an expected state and loaded the correct configuration.

\end{itemize}
The transferable authentication lemma:
\begin{itemize}

\item \emph{verified\_msg\_is\_auth} (verified in 31795 steps): Ensuring that each message that is successfully accepted by a \projecttitle{} device is sent by a genuine \projecttitle{} device, assuming
the hardware of the \projecttitle{} devices was not compromised.

\end{itemize}
The non-equivocation lemmas:
\begin{itemize}

\item \emph{no\_lost\_messages} (verified in four steps): Ensures that for all messages that are successfully accepted by a genuine \projecttitle{} device, there are no messages that were sent before but not accepted by the same \projecttitle{} device.
\item \emph{no\_message\_reordering} (verified in 5447 steps): Ensures that for all messages that are successfully accepted by a genuine \projecttitle{} device, there are no messages that were sent after that message but accepted before.
\item \emph{no\_double\_messages} (verified in 10850 steps): Ensures that a genuine \projecttitle{} device does not accept the same message multiple times.

\end{itemize}

\section{Protocols Implementation}\label{sec:use_cases-appendix}
We next present the implementation details of four distributed systems shown in Table~\ref{tab:use_cases_options} using \projecttitle{}, presented in Section~\ref{sec:use_cases}.

\begin{table}
\begin{center}
\small
\begin{tabular}{ |c|c|c|c| } 
 \hline
 System & $N$ & $f$ ($N=3$) & Byzantine faults \\ [0.5ex] \hline \hline
 A2M    & 1 & 0 & Prevention\\
 BFT &  $2f+1$ & $f=1$ & Prevention\\
 Chain Replication &  $f+1$ & $f=2$ & Prevention\\
 PeerReview & $f+1$ & $f=2$ & Detection\\
 \hline
\end{tabular}
\end{center}
\caption{Properties of the four trustworthy distributed systems implemented with \projecttitle{}.}
\label{tab:use_cases_options}
\end{table}

\subsection{Clients} Clients in a \projecttitle{} distributed system execute requests by sending singed request messages to \projecttitle{} nodes through the network. \projecttitle{} assumes Byzantine (untrusted) clients; as such, its installed shared keys cannot be outsourced. We assume that at the initialization, the System Designer also loads to \projecttitle{} devices a (per-device) key pair $C_{pub, priv}$ where the $C_{pub}$ is distributed to clients. \projecttitle{} then replies to a client by verifying the (under transmission) attested message and signing it with $C_{priv}$. As such, \projecttitle{} is restricted to only sending valid attested messages to clients where clients can prove the transferable authentication and validity of the message. The only attack vector open to a Byzantine machine is to try to equivocate by sending a stale, valid, attested message that does not reflect the current execution round. However, clients can detect this by verifying that the original request is theirs.    % the such replies to clients are signed with a private key within the \projecttitle{} Furthermore, clients do not need to have access to a \projecttitle{}

%Similarly to classical BFT systems, \projecttitle{} clients require a {\em quorum certificate}, a set of identical messages collected from different participants~\cite{10.1145/800215.806583} to consider their request as committed. In contrast to the traditional BFT, where any $f$ out of the total 3$f$+1 nodes could equivocate, our strategy to prevent equivocation improves message complexity, allowing clients to wait for (at least) $f+1$ identical replies to consider their request committed.

\subsection{Attested Append-Only Memory (A2M)}\label{sec:use_cases::a2m}
We designed a single-node trusted log system based on the A2M system (Attested Append-Only Memory)~\cite{A2M} using \projecttitle{}. A2M has been proven to be an effective building block in improving the scalability and performance of various classical BFT systems~\cite{sundr, Castro:2002, AbdElMalek2005FaultscalableBF}. We show the {\em how} to use \projecttitle{} to build this foundational system while we also show that \projecttitle{} minimizes the system's TCB jointly with the performance improvements demonstrated in $\S$~\ref{sec:eval}.

\myparagraph{System model} Our \projecttitle{} version and the original A2M systems are single-node systems that target a similar goal; they both build a trusted append-only log as an effective mechanism to combat equivocation. The clients can only append entries to a log; each log entry is associated with a monotonically increasing sequence number. Each data item, e.g., a network message, is bound to a unique sequence number, a well-known approach for equivocation-free operations~\cite{clement2012, hybster}. 

A2M was originally built using CPU-side TEEs---specifically, Intel SGX--- whereas we build its \projecttitle{} derivative. While the original A2M system keeps its entire state and the log within the TEE, we use \projecttitle{} to keep the (trusted) log in the untrusted memory. As such, in contrast to the original A2M, \projecttitle{} effectively reduces the overall system's TCB. Our evaluation showed that naively porting the application within the TEE has adverse performance implications in lookup operations.

%the trusted component \atsushi{Here explains that TNIC-log brings better memory efficiency than A2M, which could also be written in the first paragraph to highlight the advantage of TNIC}.

\myparagraph{Execution} Similarly to A2M, we expose three core operations: the \texttt{append}, \texttt{lookup}, and \texttt{truncate} operations to add, retrieve, and delete items of the log, respectively. A2M stores the lowest and highest sequence numbers for each log. Upon appending an entry, A2M increases the highest sequence number and associates it with the newly appended entry. When truncating the log, the system advances the lowest sequence number accordingly. We next discuss how we designed the operations using \projecttitle{} APIs.

\begin{algorithm}
\SetAlgoLined
\small
%\vspace{0.02cm}
\textbf{function} \texttt{append(id, ctx)} \{ \\
\Indp
 [$\alpha$,\texttt{i},\texttt{ctx}] $\leftarrow$ \texttt{local\_send(id,ctx)};\\
 \texttt{log[id].append(log\_entry($\alpha$,\texttt{i},\texttt{ctx}))};\\
 {\bf return} \texttt{[$\alpha$,\texttt{i},\texttt{ctx}]};\\
\Indm
\} \\

\vspace{0.15cm}

\textbf{function} \texttt{lookup(id, i)} \{ \\
\Indp
    {\bf return} \texttt{log[id].get(i)};\\
\Indm
\} \\
\vspace{0.15cm}
\textbf{function} \texttt{truncate(id, head, z)} \{ \\
\Indp
    [$\alpha$,\texttt{tail},\texttt{ctx}] $\leftarrow$ \texttt{append(id,} \textsc{trnc}\texttt{||id||z||head)};\\
        
    %[$\alpha_{2}$,\texttt{idx},\texttt{ctx}$_{2}$]
    \texttt{e} $\leftarrow$ \texttt{append(}\textsc{manifest}\texttt{,[$\alpha$,\texttt{tail},\texttt{ctx}])};\\
    {\bf return} \texttt{e};\\
\Indm
\} \\

\vspace{0.15cm}
\textbf{function} \texttt{verify\_lookup(id, e, head, tail)} \{ \\
\Indp
    \textbf{assert}(\texttt{e.i}$>=$\texttt{tail)};\\
    \texttt{local\_verify(id, e)};\\
\Indm
\} \\
\vspace{-1pt}
\caption{Attested Append-Only Memory (A2M) using \projecttitle{}.}
\label{algo:tnic_log}
\end{algorithm}

\myparagraph{Append operation} The \texttt{append(id,ctx)} operation takes a data item, \texttt{ctx}, and appends it to the log with identifier \texttt{id}. A log entry at index \texttt{i} is comprised of three items: the sequence number of that entry (\texttt{i}), the context of the entry (\texttt{ctx}), and the {\em authenticator} field, namely the digest of the \texttt{ctx||i} as in~\cite{levin2009trinc}. In our implementation, we additionally support the original A2M {\em authenticator} format calculated as the cumulative digest \texttt{c\_digest[i]} for that entry which is calculated as \texttt{c\_digest[i]=hash(ctx||sq||c\_digest[i-1])} where \texttt{c\_digest[0]=0}. The sequence number \texttt{i} is then increased to distinguish any entry that will be appended in the future. %With the cumulative digest, we create a set of chains, and as such, our method does not cause any values to be forgotten.

\myparagraph{Lookup operation} The \texttt{lookup(id, i)} retrieves the log entry at index \texttt{i} of log with identifier \texttt{id}. Compared to A2M, where lookups are compelled to access the trusted hardware, \projecttitle{}-log only performs a local memory access. 
The function does not verify whether the entry is legitimate. Developers need to implement the \texttt{verify\_lookup(id, entry, head, tail)} to verify the attestation. The boundaries of the log (i.e., \texttt{head} and \texttt{tail}) can constantly be retrieved by replaying a specific log, which keeps the state changes, the \textsc{manifest}. We explain how \textsc{manifest} works in the next paragraph.

\myparagraph{Truncate operation} The \texttt{truncate(id, head, z)}, where \texttt{z} is a nonce provided by the client for freshness, ``forgets'' all log entries with sequence numbers lower than \texttt{head}. A non-Byzantine client can never successfully verify a forgotten log entry. To do that, \projecttitle{}-log uses an additional log \textsc{manifest}, which keeps the logs' state changes. First, the operation attests to the {\em tail} of the log by appending a specific entry, which includes the nonce for a correct client to be later able to verify the operation. Then, the algorithm will append the last attested message of the log to the \textsc{manifest} log and return the attested message for the second append. To retrieve the boundaries of a log, clients can always attest to the tail of the \textsc{manifest} and read backward until they find a \textsc{trnc} entry.

\noindent\fbox{\begin{minipage}{\columnwidth}
\myparagraph{System design takeaway} \projecttitle{} minimizes the required TCB in the A2M system while offering faster lookup operations than its original version.
\end{minipage}}
%\atsushi{Can we explain more which point is improved thanks to TNIC?}

%Since the logs reside in the untrusted host memory, their integrity can be compromised by malicious adversaries. However, these adversaries cannot impersonate \projecttitle{} and generate verifiable attestations in any way. As such we do not worry about entries that are not written yet (these will never be verifiable). 
%If an adversary compromises the \textsc{MANIFEST}, \projecttitle{}-log might become responsive. However, this affects availability but not safary and it is beyond the scope of this work similar to other systems [A2M, Trinc, Damysus]. 

\subsection{Byzantine Fault Tolerance (BFT)}\label{sec:use_cases::byz_smr}
%\atsushi{unclear to me what is the advantage of using TNIC to implement the Byzantine SMR. Can we somehow highlight this point?}
%\dimitra{@Atnoni: Can we afford a network partition? 2f+1 w/ f=2, Assume Byz. leader and Byz. follower that drive execution with one correct replica---the others are on purpose exluded by the faulty leader. The client will have a correct reply always because it will wait for f+1 (=3) identical replies. Although if the most up to date correct replica afterwards is partitioned out, then we just block; the remaining correct replicas will have lost one message and block until they get it ... }
%\pramod{ToDo: Pesudocode.}

As a second example of \projecttitle{} applications, we build a Byzantine Fault-Tolerant protocol (BFT) that implements a robust counter based on {\em state machine replication} (SMR). Clients send increment counter requests to the SMR and receive the updated value of the counter. Despite its simplicity, this particular system can represent an ordering service, which is a fundamental building block of various distributed applications ranging from event logging and database systems to serverless and blockchain~\cite{rafthyperledger, Kafka, boki, 10.1145/3286685.3286686, scalog}. Our BFT combats equivocation by leveraging the attestation kernel of \projecttitle{}. As such, via \projecttitle{}, it reduces \textit{(i)} the number of replicas and \textit{(ii)} the message complexity (and latency) required by classical BFT.

\myparagraph{System model} We consider a system of $N=2f+1$ replicas (or {\em nodes}) that communicate with each other over unreliable point-to-point network links. At most $f$ of these replicas can be Byzantine (aka {\em faulty}), i.e., can behave arbitrarily. The rest of the replicas are {\em correct}. Recall that classical BFT protocols require an extra set of $f$ replicas, in total $3f+1$, to handle $f$ Byzantine failures.  One of the replicas is the {\em leader} that drives the protocol, whereas the remaining replicas are (passive) followers. There is only a single active leader at a time.

For liveness, we assume a partial synchrony model~\cite{FLP, 10.1145/226643.226647}. We have only explored deterministic protocol specifications; the correct replicas begin in the same state, and receiving the same inputs in the same order will arrive at the same state, generating the same outputs. Lastly, as in classical BFT protocols, we cannot prevent Byzantine clients who otherwise follow the protocol from overwriting correct clients' data.

\myparagraph{Execution} We implement BFT with \projecttitle{} as a leader-based SMR protocol for a Byzantine model that stores and increases the counter's value. The leader receives clients' requests to increment the counter. The leader, in turn, executes the protocol and applies the changes to its state machine---in our case, the leader computes and stores the next available counter value. Subsequently, the leader broadcasts the request along with some metadata to the passive followers. The metadata includes, among others, the leader's calculated output in response to the client's command, namely, the increased counter value the leader has calculated.
% and the {\em state} of the followers known to the leader.
% In our implementation, the {\em state} is represented as the signed hash of the counter value of each follower. %(known to the leader).

The followers, in turn, execute and apply the incremented counter value to their state machines. However, they first attest to the leader's (and other followers') actions to detect misbehavior. Importantly, followers validate if the state (counter) of the replicas (including the leader and all other replicas) match the expected value.

%The followers, in turn, execute and apply the incremented counter value to their state machines. owever, they first attest to the leader's actions to detect misbehavior. o do so, they audit and validate its sent output through re-execution. recisely, the followers except for their state machine, {\em simulate} the leader's state machine. ach follower replica must add an extra counter representing the state the value counter is expected to have, leading to a $2\times$ extra space complexity.   follower will update the leader's value based on the commands received and then compare its calculated leader's value with the received one. n addition, the followers will validate the state of the replicas (including the leader and all other replicas). hey only have to check if their previous state equals the other replicas' state. 

After a follower applies the increments to its local counter, it replies to the client.
% and the leader with the result of the operation. 
In addition, it forwards the leader's request to every other replica to ensure that all correct replicas will eventually receive and apply the same command. Replicas that have already applied the request ignore it; otherwise, they validate it and apply it. The leader, upon successful validation, will also reply to the client. The client can trust the result if they receive identical replies from a majority quorum, i.e., at least $f+1$ identical messages from different replicas (including the leader). This guarantees that at least one correct replica has responded with the correct result.

\myparagraph{Failure handling} Our strategy to verify the replica's execution jointly with the primitives of non-equivocation and transferable authentication offered by \projecttitle{} shields the protocol against Byzantine behavior. The leader cannot equivocate; even if it attempts to send different requests for the same round to different followers, executing the {\tt local\_send()} will assign different counter values, which healthy followers will detect. As such, a leader in that case will be exposed. 

Likewise, the equivocation mechanism allows correct followers to discard stale message requests sent through replay attacks on the network. If a follower is Byzantine, a healthy leader or replica can detect it. For $f\geq2$, it is impossible for a faulty leader and, at most, $f-1$ remaining Byzantine followers to compromise the protocol. Either these faults will be detected by a healthy replica during the validation phase, or the protocol will be unavailable, i.e., if the leader in purpose only communicates with the Byzantine followers. This directly affects BFT correctness requirements; a client will never get at least $f+1$ matching replies. Even in the extreme case of a network partition or a faulty leader that purposely excludes some healthy replicas from its multicast group, when the network is restored, these replicas will not accept any future messages unless they receive all missed ones. Suppose the leader fails in the middle of the broadcast. In that case, the last step in the follower's protocol ensures that if a correct replica accepts the requests, all correct replicas will eventually apply the same request. Since the reliability aspect and FIFO ordering are implemented in hardware, healthy replicas will ultimately receive all past messages in the proper order. For protocols to progress in the case of a faulty leader, they must pass through a recovery protocol or view-change protocols similar to those described in previous works~\cite{minBFT, Castro:2002}. Recovering is beyond the scope of this work, and as such, we did not implement it.

\noindent\fbox{\begin{minipage}{\columnwidth}
\myparagraph{System design takeaway} \projecttitle{} optimizes the replication factor and the message rounds compared to classical BFT.
\end{minipage}}

\begin{algorithm}
\SetAlgoLined
\small
%\vspace{0.02cm}
\textbf{function} \texttt{leader(req)} \{ \\
\Indp
 {\tt output} $\leftarrow$ \texttt{execute(req)};\\
 {\tt msg} $\leftarrow$ \texttt{req||output};\\
 {\tt attested\_msg} $\leftarrow$ \texttt{local\_send(msg)};\\
 \texttt{rem\_write(}\textsc{followers[:]}{\tt, attested\_msg)};\\

{\bf upon reception of {\tt ack} from \textsc{followers}:}\\
    \Indp
        {\tt [{$\alpha$ || f\_attested\_msg || f\_output || f\_id}]} \\\hspace{22pt} $\leftarrow$ \texttt{upon\_delivery(ack)};\\
        {\bf assert(}\texttt{validate\_follower(f\_attested\_msg,\\\hspace{22pt} f\_output)}{\bf)};\\
        \texttt{incr\_req\_acks\_if\_not\_incr\_before(f\_id)};\\
    \Indm

 \texttt{auth\_send(}\textsc{client}{\tt,msg)};\\
\Indm
\} \\

\vspace{0.15cm}

% \textbf{function} \texttt{follower(}\textsc{sender}{\tt, attested\_msg)} \{ \\
\textbf{function} \texttt{follower()} \{ \\
\Indp
{\bf upon reception of {\tt attested\_msg}:}\\
    \Indp
        {\tt [{$\alpha$ || req || output}]} $\leftarrow$ \\\hspace{22pt}\texttt{upon\_delivery(attested\_msg)};\\

    {\bf assert(}\texttt{validate\_sender(req, output)}{\bf)};\\
    {\bf if }{\tt (in\_order\_not\_applied(req))}\\
    \Indp
    {\tt current\_output} $\leftarrow$ \texttt{execute(req)};\\
    {\tt f\_attested\_msg} $\leftarrow$ \texttt{local\_send(req||current\_output)};\\
    ack $\leftarrow$ {\tt f\_attested\_msg}\\
    % \texttt{auth\_send(}\textsc{client} $\cup$ \textsc{leader} $\cup$ \textsc{followers[:]},\\{\tt\hspace{40pt} attested\_msg)};\\    
    \texttt{auth\_send(}\textsc{leader}, {\tt ack)};\\    
    % \texttt{auth\_send(}\textsc{client} $\cup$ \textsc{followers[:]},\\{\tt\hspace{40pt} attested\_msg)};\\    
    \texttt{auth\_send(}\textsc{client} $\cup$ \textsc{followers[:]},\\{\tt\hspace{40pt} f\_attested\_msg)};\\    
    % \texttt{auth\_send(}\textsc{leader}{\tt, req||current\_output)};\\    
    % \texttt{auth\_send(}\textsc{client}{\tt, req||current\_output)};\\
    % {\bf if {\tt not\_seen(req)} {\sc and  sender = leader}:}\\
    % \Indp
        % {\bf for} (\textsc{followers[:]}) 
            % \texttt{auth\_send(}\textsc{followers[:]}{\tt, msg)};\\
            % \texttt{rem\_write(}\textsc{LEADERfollowers[:]}{\tt, attested\_msg)};\\
    \Indm
    \Indm
\Indm
\} \\
\vspace{-1pt}
\caption{BFT using \projecttitle{}.}
\label{algo:tnic_bft}
\end{algorithm}

%
%\dimitra{>Github code issues:
%\begin{itemize}
%    \item Line 239: has a logical bug regarding the message batching
%    \item Lines 205--210: unnecessary hash re-calcucations--- might improve performance if fixed
%    \item Continuation function needs improvement for correctness/completeness (leader should store the output for each on-going command).
%    \item For correctness, leader should have only one outstanding operation at a time.
%\end{itemize}}

%\lstinputlisting[language=C++]{codelets/pb.cc}

\subsection{Chain Replication (CR)}\label{sec:use_cases::byz_chain_rep}

\begin{algorithm}
\SetAlgoLined
\small
%\vspace{0.02cm}
\textbf{function} \texttt{head\_operation(req)} \{ \\
\Indp
 {\tt output} $\leftarrow$ \texttt{execute(req)};\\
 {\tt msg} $\leftarrow$ \texttt{req||output};\\
 \texttt{auth\_send(}\textsc{middle}{\tt,msg)};\\ \texttt{auth\_send(}\textsc{client}{\tt,msg)};\\
\Indm
\} \\

\vspace{0.15cm}

\textbf{function} \texttt{middle\_tail\_operation(msg)} \{ \\
\Indp
    {\bf assert(}\texttt{validate\_chain(msg)}{\bf)};\\
    {\tt output} $\leftarrow$ \texttt{execute(req)};\\
    {\tt chained\_msg} $\leftarrow$ \texttt{msg||output};\\
    {\bf if} (!\textsc{tail})\\
    \Indp
    \texttt{auth\_send(}\textsc{middle}{\tt,chained\_msg)};\\
    \Indm
    \texttt{auth\_send(}\textsc{client}{\tt,req||output)};\\
\Indm
\} \\
\vspace{0.15cm}
\textbf{function} \texttt{validate(msg)} \{ \\
\Indp
    \texttt{len} $\leftarrow$ \texttt{sz};\\
    \texttt{[req, out, cmt]} $\leftarrow$ \texttt{unmarshall(msg[0:len])};\\
    {\bf assert(}\texttt{memcmp(req, out)}{\bf)};\\
    {\bf assert(}\texttt{(cmt == expected\_cmt)}{\bf)};\\
    {\bf for} {\tt(i = 1; i < }{\sc node\_id}; {\tt i++)} \{\\
    \Indp
    \texttt{[out, cmt]} $\leftarrow$ \texttt{unmarshall(msg[len:len+\textsc{sz}])};\\
    {\bf assert(}\texttt{memcmp(req, out)}{\bf)};\\
    {\bf assert(}\texttt{(cmt == expected\_cmt)}{\bf)};\\
    \texttt{len} $\leftarrow$ \texttt{len} + \texttt{sz};\\
    \Indm
    {\bf return} {\tt True};\\
\Indm
\} \\
\vspace{-1pt}
\caption{Chain Replication using \projecttitle{}.}
\label{algo:tnic_chain_replication}
\end{algorithm}

We implement a Byzantine Chain Replication using \projecttitle{} that represents the replication layer of a Key-Value store. Chain Replication is a foundational protocol for building state machine replication and initially operates under the CFT model using $f+1$ nodes to tolerate up to $f$ failures. We show {\em how} to use \projecttitle{} to shield the protocol without changes to the core of the algorithm (states, rounds, etc.) while keeping the same replication factor.

\myparagraph{System model} We make the same assumptions for the system as in the previous BFT system. For error detection and reconfiguration, we assume a centralized (trusted) configuration service as in~\cite{10.1007/978-3-642-35476-2_24} that generates new configurations upon receiving reconfiguration requests from replicas. Recall that the classical Chain Replication under the CFT model relies on reliable failure detectors~\cite{chain-replication}. For liveness, we also assume that the configuration service will eventually create a configuration of correct replicas that do not intentionally issue reconfiguration requests to perform Denial-of-Service attacks. 

Clients send requests to {\tt put} or {\tt get} a value and receive the result. The replicas (e.g., head, middle, and {\em tail} nodes) are chained, and the requests flow from the head node to the tail through the intermediate middle replicas. 

Malicious primaries, i.e., the head that does not forward the message intentionally, are detected on the client's side and trigger reconfiguration~\cite{Castro:2002, minBFT}.

\myparagraph{Execution} To execute a request \texttt{req}, e.g., {\tt put}/{\tt get}, a client first obtains the current configuration from the configuration service and sends the {\tt req} to the head of the chain. The head orders and executes the request, and then it creates a {\em proof of execution message}, which is sent along the chain. The proof of execution includes the {\tt req} and the leader's action ({\tt out}) in response to that request. In our case, the leader sends the {\tt req} along with the assigned commit index. The message is then sent (signed) to the middle node that follows in the chain.

The middle node checks the message's validity by verifying that the head's output is correct, executes the {\tt req}, and forwards the request to the following replica. Similarly, every other node executes the original request, verifies the output of all previous nodes, and sends the original request and a vector of all previous outputs. A replica must construct a {\em proof of execution message} that achieves one goal. t allows the following replicas in the chain to verify all previous replicas. s such the messages is of the form < < <{\tt req}, {\tt out$_{leader}$}>$_{\sigma_0}$, {\tt out$_{middle1}$}>$_{\sigma_1}$, .., {\tt out$_{tail}$}>$_{\sigma_N}$. The tail is the last node in the chain that will execute and verify the execution of the request.

In contrast to the CFT version of the Chain Replication protocol, local operations in the tail, {\tt get} or {\tt ack} in a {\tt put} request cannot be trusted. As such, the replicas in the chain need to reply to the clients with their output after they have forwarded their proof of execution message. Clients can wait for at least $f$ replicas replies and the tail reply to collide. Clients can execute the {\tt get} requests similarly to {\tt write} requests, traversing the entire chain, or clients can consult the majority and broadcast the request to $f+1$ replicas, including the tail.

\myparagraph{Failure handling} By the protocol definition, all nodes will see and execute all messages in the same order imposed by the head node. As such, all correct replicas will always be in the same state. In addition, network partitions that may split the chain into two (or more) individual chains that operate independently cannot affect safety: the clients must verify at least $f+1$ identical replies. Suppose a correct replica or a client detects a violation (by examining the proof of execution message or having to hear for too long from a node). In that case, they can expose the faulty node and request a reconfiguration.

\noindent\fbox{\begin{minipage}{\columnwidth}
\myparagraph{System design takeaway} \projecttitle{} {\em seamlessly} shields the Chain Replication system for Byzantine settings with the same replication factor as the original CFT system.
\end{minipage}}

%\dimitra{>Github code issues:
%\begin{itemize}
%    \item check\_outputs function (L:67): Did you miss a validation step regarding transferable authentication?
%\end{itemize}}

\subsection{Accountability (PeerReview)}
\label{sec:use_cases::accountability}

We implement an accountability protocol based on the PeerReview system~\cite{bftdetection, peer-review}. Compared to the previous three BFT systems that prohibit an improper action from taking effect, accountability protocols~\cite{268272, bftdetection, peer-review} slightly weaken the system (fault) model in favor of performance and scalability. Specifically, our protocol {\em allows} Byzantine faults to happen (e.g., correct nodes might be convinced by a malicious replica to permanently delete data). Still, it guarantees that malicious actions can always be detected. Accountability protocols can be applied to different systems as generic guards that trade security for performance~\cite{peer-review}, e.g., NFS, BitTorrent, etc. 

The original version of the system did not use trusted components. t incurs a high message complexity, i.e., {\em all-to-all} communication to combat equivocation. We use \projecttitle{} to improve that message complexity.

\myparagraph{System model} We only detect faults that directly or indirectly affect a message, implying that {\em (i)} correct nodes 
can observe all messages sent and received by that node and {\em (ii)}  Byzantine faults that are not observable through the network cannot be detected. For example, a faulty storage node might report that it is out of disk space, which cannot be verified without knowing the actual state of its disks.

We further assume that each protocol participant acts according to a deterministic specification protocol. As such, detection can be accomplished even with a single correct machine, requiring only $f+1$ machines.  This does not contradict the impossibility results for agreement~\cite{FLP} because detection systems do not guarantee safety.

\myparagraph{Execution} The participants communicate through network messages generated by \projecttitle{}.  In addition, each participant maintains a {\em tamper-evident} log that stores all messages sent and received by that node as a chain. A log entry is associated with an entry index, the entry data, and an authenticator, calculated as the signed hash of the tail of the log and the current entry data. 

We frame our protocol in the context of an overlay multicast protocol~\cite{10.1145/945445.945474} widely used in streaming systems. The nodes are organized as a tree where the streaming content (e.g., audio, video) flows from a source, i.e., {\em root} node, to clients ({\em children} nodes). To support many clients, each can be a source to other clients, which will be connected as children nodes. 

In our implementation, we consider nodes in a tree topology. The tree's height equals one, comprising one source node and two client (children) nodes connected to the source. Algorithm~\ref{algo:tnic_accountability_protocol} ($\S$~\ref{sec:use_cases-appendix}) shows the operations of our implemented accountability protocol.  hen the source sends a context (executes the \texttt{root()} function), it implicitly includes a signed statement that this message has a particular sequence number (generated by \projecttitle{}). he clients execute the {\tt child()} function that validates the received message, logs the received message, executes the result, and responds to the source.

\begin{algorithm}[t]
\SetAlgoLined
\small
%\vspace{0.02cm}
\textbf{function} \texttt{root(ctx)} \{ \\
\Indp
 \texttt{auth\_send(}\textsc{child}{\tt,ctx)};\\
 {\bf upon reception of \texttt{response}:};\\
 \Indp
    {\bf assert(}\texttt{validate\_reception(response)}{\bf)};\\
    \texttt{log(response)};\\
\Indm
\Indm
\} \\

\vspace{0.15cm}

\textbf{function} \texttt{child($\alpha$||cmd||seq)} \{ \\
\Indp
    {\bf assert(}\texttt{validate\_reception($\alpha$||cmd||seq)}{\bf)};\\
    \texttt{log($\alpha$||cmd||seq)};\\
    {\tt result} $\leftarrow$ \texttt{execute(cmt)};\\
    {\tt response} $\leftarrow$ \texttt{log(result||cmd)};\\
    \texttt{auth\_send(}\textsc{root}{\tt, response)};\\
\Indm
\} \\
\vspace{0.15cm}
\textbf{function} \texttt{log\_audit()} \{ \\
\Indp
    {\bf{while}} \texttt{last\_id < log\_tail} \{\\
    \Indp

        \texttt{entry} $\leftarrow$ \texttt{validate\_log\_entry\_at(last\_id)};\\
        \texttt{last\_id++};\\
        {\bf assert(}\texttt{replay(entry)}{\bf{)}};\\
    \Indm
    \}\\
\Indm
\} \\
%\vspace{-1pt}
\caption{PeerReview using \projecttitle{}.}
\label{algo:tnic_accountability_protocol}
\end{algorithm}

Each node is assigned to a set of {\em witness} processes to detect faults. Similarly to the original system, we assume that the set of nodes and its witnesses set {\em always} contain a correct process. The witnesses audit and monitor the node's log. To detect destructive behaviors (or expose non-responsive nodes), the witnesses read the node's log and replay it to run the participant's state machine. As such, they ensure the participant's state is consistent with proper operation. 

Specifically, each witness for a participant node keeps track of n, a log sequence number, and s, the state that the participant should have been in after sending or receiving the message in log entry n. t initializes n to 0 and s to the initial state of the participant.

Whenever a witness wants to audit a node, it sends its n and a nonce (for freshness).
The participant returns an attestation of all entries between n and its current log entry using the nonce. The witness then runs the reference implementation, starting at state $s$, and progressing through all the log entries. f the reference implementation sends the same messages in the log, then the witness updates n, %\antonis{Is this the point of truncating the log?}
which is the state of the reference implementation at that point. If not, then the witness has proof it can present of the participant's failure to act correctly.

The original PeerReview system requires a receiver node to forward messages to the original sender's witnesses so they can ensure this message is {\em legitimate}, i.e., it appears in the sender's log. No other conflicting message is sent to another peer (equivocation). As such, a peer must communicate (in every round) with the witness set of any other peer, leading to a quadratic message complexity. \projecttitle{} eliminates the overhead; a participant that sends or receives a message needs to attest and append the message and its attestation in each log. A participant can process received messages only if they are accompanied by attestations generated by the sender's \projecttitle{} hardware.

\noindent\fbox{\begin{minipage}{\columnwidth}
\myparagraph{System design takeaway} \projecttitle{} can be used to optimize the message complexity in accountable systems.
\end{minipage}}

%\clearpage
% \newpage
% \fi
\bibliographystyle{ACM-Reference-Format}
\balance
\bibliography{sample}

%\theendnotes

\end{document}